\documentclass[aps,prd,superscriptaddress,nofootinbib,amsfonts,amssymb,amsmath,notitlepage]{revtex4-1}
\usepackage[utf8x]{inputenc}
\usepackage{graphicx}
\usepackage{float}
\usepackage{wrapfig}
\usepackage{bm}
\usepackage{color}
\usepackage{comment}
\usepackage{xcolor}
\usepackage[normalem]{ulem}
\usepackage{hyperref}
\usepackage{mathrsfs}
\usepackage{mhchem}
\usepackage{{booktabs}}
\hypersetup{
colorlinks=true,
citecolor=blue,
citebordercolor=red,
linktoc=all,
linkcolor=blue,
urlcolor=blue
}
\allowdisplaybreaks[1]

\catcode`\@=11
\def\lsim{\mathrel{\mathpalette\@versim<}}
\def\gsim{\mathrel{\mathpalette\@versim>}}
\def\@versim#1#2{\vcenter{\offinterlineskip
\ialign{$\m@th#1\hfil##\hfil$\crcr#2\crcr\sim\crcr } }}
\catcode`\@=12
\newcommand{\Slash}[1]{{\ooalign{\hfil/\hfil\crcr$#1$}}} 
\newcommand{\bvec}[1]{\mbox{\boldmath $#1$}}

\newcommand{\p}{\partial}

\newcommand{\al}[1]{\begin{align}#1\end{align}}
\newcommand{\bp}{\begin{pmatrix}}
\newcommand{\ep}{\end{pmatrix}}
\newcommand{\nn}{\nonumber\\}

\newcommand{\df}{\text{d}}

\newcommand{\bs}[1]{\boldsymbol}

\newcommand{\Tr}{{\rm Tr}\,}
\newcommand{\tr}{{\rm tr}\,}

\newcommand{\pmat}[1]{\begin{pmatrix}#1\end{pmatrix}}

\newcommand{\fn}[1]{\!\left(#1\right)}

\graphicspath{{./figs/}}
\newbox{\ORCIDicon}
\sbox{\ORCIDicon}{\large
                  \includegraphics[width=0.8em]{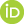}}

\begin{document}

\title{Asymptotic freedom and safety in quantum gravity}

\author{Saswato \surname{Sen}\,\href{https://orcid.org/0000-0003-2470-3483}{\usebox{\ORCIDicon}}}
\email{s.sen@thphys.uni-heidelberg.de}
\email{saswato.sen@oist.jp}
\affiliation{Institut f\"ur Theoretische Physik, Universit\"at Heidelberg, Philosophenweg 16, 69120 Heidelberg, Germany}
\affiliation{Okinawa Institute of Science and Technology Graduate University 1919-1 Tancha, Onna, Kunigami, Okinawa, Japan 904-0412}

\author{Christof \surname{Wetterich}\,\href{https://orcid.org/0000-0002-2563-9826}{\usebox{\ORCIDicon}}}
\email{c.wetterich@thphys.uni-heidelberg.de}
\affiliation{Institut f\"ur Theoretische Physik, Universit\"at Heidelberg, Philosophenweg 16, 69120 Heidelberg, Germany}

\author{Masatoshi \surname{Yamada}\,\href{https://orcid.org/0000-0002-1013-8631}{\usebox{\ORCIDicon}}}
\email{m.yamada@thphys.uni-heidelberg.de}
\affiliation{Institut f\"ur Theoretische Physik, Universit\"at Heidelberg,
Philosophenweg 16, 69120 Heidelberg, Germany}

\begin{abstract}
We compute non-perturbative flow equations for the couplings of quantum gravity in fourth order of a derivative expansion. The gauge invariant functional flow equation for arbitrary metrics allows us to extract $\beta$-functions for all couplings. In our truncation we find two fixed points. One corresponds to asymptotically free higher derivative gravity, the other is an extension of the asymptotically safe fixed point in the Einstein-Hilbert truncation or extensions thereof. 
The infrared limit of the flow equations entails only unobservably small modifications of Einstein gravity coupled to a scalar field. Quantum gravity can be asymptotically free, based on a flow trajectory from the corresponding ultraviolet fixed point to the infrared region. This flow can also be realized by a scaling solution for varying values of a scalar field. As an alternative possibility, quantum gravity can be realized by asymptotic safety at the other fixed point.
There may exist a critical trajectory between the two fixed points, starting in the extreme ultraviolet from asymptotic freedom. We compute critical exponents and determine the number of relevant parameters for the two fixed points. Evaluating the flow equation for constant scalar fields yields the universal gravitational contribution to the effective potential for the scalars.
\end{abstract}

\maketitle
\tableofcontents
\section{Introduction}
Quantum gravity can be formulated as a consistent quantum field theory for the metric if a fixed point for the flow of (generalized) couplings exists. 
If this fixed point is approached in the extreme ultraviolet, the quantum field theory is complete in the sense that it can be extrapolated to arbitrary short distances~\cite{Hawking:1979ig}. In short, one defines the microscopic theory by such a fixed point.  
The relevant parameters for small deviations from the fixed point correspond to the free parameters of the model, reflected by ``renormalizable couplings". Their number is typically finite, such that the model is predictive. The fixed point can be either a free theory--this is the case for asymptotic freedom. In contrast, the case of non-vanishing interactions at the fixed point is called asymptotic safety~\cite{Hawking:1979ig,Reuter:1996cp,Souma:1999at}.
See Refs.~\cite{Niedermaier:2006wt,Niedermaier:2006ns,Percacci:2007sz,Codello:2008vh,Reuter:2012id,Percacci:2017fkn,Reuter:2019byg,Wetterich:2019qzx,Pawlowski:2020qer,Bonanno:2020bil,Reichert:2020mja,Eichhorn:2017egq,Eichhorn:2018yfc} for reviews.
For asymptotic freedom the model is perturbatively renormalizable, while the case of asymptotic safety corresponds to a non-perturbatively renormalizable theory unless all dimensionless couplings at the fixed point are small.

For pure higher derivative gravity, based on terms quadratic in the curvature tensor $R_{\mu\nu}$ and the curvature scalar $R$, 
\al{
S=\int d^4x \sqrt{g} \left[ -\left( \frac{C}{2} + \frac{D}{3} \right)R^2 + DR_{\mu\nu}R^{\mu\nu}\right]\,,
}
Stelle has shown perturbative renormalizability for the couplings $C^{-1}$ and $D^{-1}$~\cite{Stelle:1976gc}.
(In addition, the coefficient of a term linear in the curvature scalar and a cosmological constant are renormalizable couplings.)
The perturbative renormalization flow for the couplings $C^{-1}$ and $D^{-1}$ has been computed by Fradkin and Tseytlin~\cite{Fradkin:1978yf,Fradkin:1981hx,Fradkin:1981iu}, and reads
\al{
\p_t C^{-1}&=\frac{5}{288\pi^2}C^{-2} + \frac{5}{16\pi^2}D^{-2} +\frac{5}{16\pi^2}C^{-1}D^{-1}\,,
\label{eq: C inverse pert RG eq}
\\[2ex]
\p_t D^{-1}&= -\frac{133}{160\pi^2}D^{-2}\,.
\label{eq: D inverse pert RG eq}
}
These equations show a fixed point with asymptotic freedom,
\al{
\text{SFT}:\quad C_*^{-1}=0\,,\qquad D_*^{-1}=0\,,
}
to which we refer as the SFT-fixed point.

Quantum corrections will lead to a quantum effective action $\Gamma$ which involves additional terms.
In particular, a term linear in the curvature scalar is needed for any realistic theory of gravity.
Expanding up to fourth order in derivatives (and omitting the Gauss-Bonnet topological invariant) a diffeomorphism invariant action for pure gravity takes the form
\al{
\Gamma = \int d^4x \sqrt{g}\left[V  -\frac{M_\text{p}^2}{2} R  -\left( \frac{C}{2} + \frac{D}{3} \right)R^2 + DR_{\mu\nu}R^{\mu\nu} \right]\,,
\label{eq: effective action in intro}
}
where the (reduced) Planck mass $M_\text{p}$ and the cosmological constant $V$ are additional relevant parameters beyond $C^{-1}$ and $D^{-1}$.
At the fixed point the role of $V$ and $M_\text{p}^2$ is negligible if these couplings are finite.
Flowing away from the fixed point, however, $C$ and $D$ become finite.
In this case $M_\text{p}^2$ is no longer negligible. 
For both signs of $M_\text{p}^2$ and arbitrary $V$ the effective action \eqref{eq: effective action in intro} leads to tachyons and/or ghosts, such that Minkowski space is no longer a stable approximate solution of the field equations for $V\ll M_\text{p}^4$.
As a consequence, the effective action \eqref{eq: effective action in intro} seems not to be compatible with observation.
A priori, it is not known if this is a shortcoming of the approximation to the effective action (truncation) or a basic flaw of quantum gravity based on the SFT-fixed point.
Indeed, ghosts and tachyons can be artifacts of insufficient truncations~\cite{Donoghue:2019fcb,Platania:2020knd}.

There has been recently an intense discussion of perturbative quantum gravity based on the SFT-fixed point, dubbed ``agravity"~\cite{Salvio:2014soa,Salvio:2017qkx}. See also Refs.~\cite{Anber:2011ut,Holdom:2016xfn,Salvio:2018crh}.
The conclusions are severely limited, however,  by the simple fact that the couplings $C^{-1}$ and $D^{-1}$ flow outside the perturbative domain as the renormalization scale is lowered. 
For a judgement of the fate of higher derivative gravity it seems compulsory to understand the flow of couplings in the non-perturbative domain. Only this can give an answer to the question of stability for the effective action.
Non-perturbative flow equations based on functional renormalization~\cite{Wetterich:1992yh,Reuter:1993kw,Reuter:1996cp} seem the appropriate tool for such an investigation. 

A simple ``Einstein-Hilbert truncation" of the flowing effective action or average action $\Gamma_k$ omits the quartic terms $\sim C$, $D$ in Eq.~\eqref{eq: effective action in intro}.
One finds a fixed point in the flow of the dimensionless couplings $w=M_\text{p}^2/2k^2$ and $u=V/k^4$, often called Reuter fixed point and named here R-fixed point.
The R-fixed point is associated to asymptotic safety and the gravitational interactions do not vanish at the fixed point. Its existence corresponds to non-perturbative renormalizability of quantum gravity as a quantum field theory for the metric. The R-fixed point remains present for large classes of extended truncations~\cite{Codello:2006in,Benedetti:2009rx,Benedetti:2009gn,Benedetti:2010nr,Manrique:2010mq,Manrique:2010am,Groh:2011vn,Donkin:2012ud,Christiansen:2012rx,Christiansen:2014raa,Christiansen:2015rva,Christiansen:2016sjn,Denz:2016qks,Christiansen:2017cxa,Christiansen:2017bsy,Eichhorn:2018akn,Eichhorn:2018ydy,Bonanno:2021squ,
Lauscher:2002sq,deBrito:2018jxt,
Falls:2013bv,Falls:2014tra,Falls:2017lst,Falls:2018ylp,Kluth:2020bdv,Falls:2020qhj,
Codello:2013fpa,Demmel:2014hla,Biemans:2016rvp,Gies:2016con,deBrito:2019umw,Bosma:2019aiu,Knorr:2019atm,Knorr:2021slg,Knorr:2021niv,deBrito:2020rwu,deBrito:2020xhy,deBrito:2021pmw,Ohta:2013uca,Ohta:2015zwa,Dietz:2012ic,Dietz:2013sba,Gonzalez-Martin:2017gza,Baldazzi:2021orb}.
It is also present if matter couples to gravity~\cite{Dona:2013qba,Dona:2015tnf,Percacci:2015wwa,Oda:2015sma,Eichhorn:2016esv,Meibohm:2016mkp,Biemans:2017zca,Hamada:2017rvn,deBrito:2019epw,Pawlowski:2018ixd,Wetterich:2019zdo,Wetterich:2019rsn,Eichhorn:2018nda,Alkofer:2018fxj,Alkofer:2018baq,Burger:2019upn,Hamada:2020mug,deBrito:2020dta,Eichhorn:2020kca,Eichhorn:2020sbo,Eichhorn:2021tsx,Ohta:2021bkc,Laporte:2021kyp}, as for the minimal standard model coupled to gravity.
The non-perturbative character of the fixed point implies that the number of relevant parameters is no longer determined by the canonical dimension of couplings.
This leads to an enhanced predictivity, as demonstrated by the successful  prediction of the mass of the Higgs boson~\cite{Shaposhnikov:2009pv}.
It is well possible that other parameters of the standard model may become predictable~\cite{Harst:2011zx,Harst:2011zx,Folkerts:2011jz,Eichhorn:2011pc,Eichhorn:2012va,Meibohm:2015twa,Labus:2015ska,Eichhorn:2016vvy,Christiansen:2017gtg,Eichhorn:2017lry,Eichhorn:2017eht,Eichhorn:2017ylw,Eichhorn:2017muy,Eichhorn:2017als,Eichhorn:2018whv,Eichhorn:2019dhg,Reichert:2019car,Alkofer:2020vtb,Hamada:2020vnf,Kowalska:2020zve,Kowalska:2020gie}.

In the present paper we ask what is the relation between the SFT- and R-fixed points.
We will use a truncation for which both fixed points are found. This opens the terrain for many interesting questions.
Is the SFT-fixed point viable for a definition of quantum gravity? Is there a critical trajectory from the SFT-fixed point to the R-fixed point?
What are the implications for the predictivity of quantum gravity coupled to matter?
The present paper will not yet fully answer all these questions.
It focusses on the derivation of the relevant flow equations for higher derivative gravity and provides for partial answers to these questions within the given truncation.

The minimum truncation needed for these questions takes for the effective average action the from \eqref{eq: effective action in intro}, with four running couplings $C$, $D$, $M_\text{p}^2$ and $V$ depending on the renormalization scale $k$.
Deriving these flow equations within functional renormalization is a technical challenge. There has been previous work for reproducing the perturbative $\beta$-functions for $C$ and $D$~\cite{Fradkin:1978yf,Fradkin:1981hx,Fradkin:1981iu,Julve:1978xn,Avramidi:1985ki,Avramidi:1986mj,Antoniadis:1992xu,deBerredo-Peixoto:2003jda,deBerredoPeixoto:2004if}, circumventing the technical issues by a rather special field-dependent gauge~\cite{Falls:2020qhj} or using directly expansions in small $C^{-1}$ and $D^{-1}$~\cite{Codello:2006in,Groh:2011vn}.
We aim here for an understanding of the full non-perturbative flow equations for arbitrary values of the couplings $C$, $D$, $M_\text{p}^2$ and $V$. This is needed in order to see both the SFT- and R-fixed points.
The use of the gauge invariant flow equation~\cite{Wetterich:2016vxu,Wetterich:2016ewc,Wetterich:2017aoy} constitutes an important advantage, since the contributions from the physical fluctuations in the metric can be separated from the universal ``measure contribution" of the gauge fluctuations.
Still, a major technical issue is related to mode mixing.
For particular classes of metrics, as Einstein spaces, the transverse-traceless (TT) fluctuations ($t$-fluctuations) cannot mix with the physical scalar degree in the metric ($\sigma$-fluctuations) due to symmetry.
Deriving flow equations for Einstein spaces permits a rather straightforward application of known heat kernel expansions.  
One can,  however, only obtain a flow equation for the linear combination $D+6C$ in this way.
The required flow equations for $C$ and $D$ separately require geometries beyond Einstein spaces for which the $t$- and $\sigma$-modes mix.

We display the more technical parts in various appendices and concentrate in the main part on  summaries of results and a discussion of crucial features for the derivation of the flow equations.
In Section~\ref{sec: Summary: Setup, beta functions and fixed point structure} we present the setup, the characteristic features of the flow equations and the fixed points.
Section~\ref{sec: Heat kernel method and threshold function} summarizes the heat kernel method and the flow contributions from matter fluctuations.
In Section~\ref{sec: Flow generators from metric fluctuations} we exhibit our central result on the flow contributions from the metric fluctuations. Section~\ref{sec: UV fixed point and critical exponent} discusses asymptotic safety for the R-fixed point. In Section~\ref{sec: Infrared region} we address the infrared region and conclusions are presented in Section~\ref{sec: Conclusions}.

\section{Summary: Setup, flow equations and fixed point structure}
\label{sec: Summary: Setup, beta functions and fixed point structure}
In this section, we summarize our setup, the resulting flow equations and the corresponding fixed points or scaling solutions. 
The degrees of freedom are the metric and scalar fields. For the effective action we choose a truncation with five coupling functions $U(\rho)$, $F(\rho)$, $C(\rho)$, $D(\rho)$ and $E(\rho)$, where $\rho$ is an invariant formed from the scalar fields without derivatives. The function $U(\rho)$ constitutes the effective potential for scalar fields, $F(\rho)$ is a field-dependent effective squared Planck mass, and $C(\rho)$, $D(\rho)$, $E(\rho)$ are coupling functions multiplying the gravitational invariants involving four derivatives of the metric. For a setting without a scalar field background one simply omits the dependence of $U$, $F$, $C$, $D$ and $E$ on $\rho$.
Besides the metric and scalar fields we also include the fluctuations of fermions and gauge bosons. For the matter fields (scalars, fermions, gauge bosons) we do not compute the flow of derivative terms (kinetic terms), or interaction terms as Yukawa couplings.

\subsection{Effective action for gravity}
Our ansatz for the truncated effective average action consists of a gravity part and a matter part
\al{
\Gamma_k=  \Gamma_k^\text{gravity}+\Gamma_k^\text{matter}\,.
\label{starting effective action}
}
For the gravity part, we consider the following truncated effective action,
\al{
\label{Eq: effective action for gravity part}
  \Gamma_k^\text{gravity}
  &=\int d^4 x\, \sqrt{g}
  \left[U\fn{\rho}  -\frac{F\fn{\rho}}{2} R - \frac{C\fn{\rho}}{2}R^2 +   \frac{D\fn{\rho}}{2} C_{\mu\nu\rho\sigma}C^{\mu\nu\rho\sigma} +\tilde{\mathcal L}_\text{GB} \right]
  +\Gamma_\text{gf}+\Gamma_\text{gh}\,,
}
where $R$ is the curvature scalar, $C_{\mu\nu\rho\sigma}$ is the Weyl tensor whose squared form is given by $C_{\mu\nu\rho\sigma}C^{\mu\nu\rho\sigma}=R_{\mu\nu\rho\sigma}R^{\mu\nu\rho\sigma} -2R_{\mu\nu}R^{\mu\nu}+R^2/3$.
For the computation of the universal measure contribution to the gauge invariant flow equation we use the equivalent physical gauge fixing for which $\Gamma_\text{gf}$ and $\Gamma_\text{gh}$ are the gauge fixing and the ghost action for diffeomorphisms.
These are given in Appendix~\ref{App: setups}, Eqs.~\eqref{standard gauge fixing} and \eqref{standard ghost action}.
The last term in the square brackets in Eq.~\eqref{Eq: effective action for gravity part} is the Gauss-Bonnet term which reads
\al{
\tilde{\mathcal L}_\text{GB} =E\fn{\rho}\left(R^2-4R_{\mu\nu}R^{\mu\nu}+R_{\mu\nu\rho\sigma}R^{\mu\nu\rho\sigma}\right)
=E\fn{\rho}{G_4}\,.
\label{eq: Gauss-Bonnet term}
}
For constant $E$ this is a topological invariant, while for a dynamical scalar field it contributes to the field equations. The effective action \eqref{Eq: effective action for gravity part} contains the most general diffeomorphisms invariant terms for the metric with up to four derivatives. In Appendix~\ref{sect: Basis of gravitational interactions} we present the same action in terms of different linear combinations of invariants.

The coefficients $U\fn{\rho}$, $F\fn{\rho}$, $C\fn{\rho}$, $D\fn{\rho}$ and $E\fn{\rho}$ are functions of real singlet-fields $\varphi$, $\rho=\varphi^2$, $N$-component real fields $\phi^a$, $\rho=\phi^a\phi^a/2$, or complex scalar fields $\varphi_a$, $\rho=\varphi_a^\dagger \varphi_a$. 
One can expand these coefficient functions into polynomials of $\rho$:
\al{
U(\rho)&= V + m^2 \rho +\frac{\lambda}{2}\rho^2 +\cdots\,,\\[2ex]
F(\rho)&= M_\text{p}^2 + \xi \rho + \cdots\,, \\[2ex]
C(\rho)&= C_0 + C_1 \rho +\cdots \,,\\[2ex]
D(\rho)&= D_0 + D_1 \rho +\cdots \,,\\[2ex]
E(\rho)&= E_0 + E_1 \rho +\cdots \,.
}
Here $V$ is the cosmological constant, $m^2$ is the scalar mass parameter and $\lambda$ is the quartic scalar coupling.
In the gravitational sector $M_\text{p}^2$ is the Planck mass squared for $\rho=0$ and $\xi$ is the non-minimal coupling between the scalar field and the curvature scalar. The latter plays a crucial role for the realization of Higgs inflation~\cite{Lucchin:1985ip,Futamase:1987ua,Salopek:1988qh,Cervantes-Cota:1995ehs,Bezrukov:2007ep}.

The matter part $\Gamma_k^\text{matter}$ consists of canonical kinetic terms for all matter fields. We also include gauge and Yukawa couplings, but set the effects of interactions and masses to zero in many parts of this work. We consider $N_S$ scalar bosons, $N_V$ vector bosons and $N_F$ Weyl fermions.
The explicit form of the action for the matter part is given in Section~\ref{sec: Heat kernel method and threshold function}.

\subsection{Flow equations}
The coupling functions $U$, $F$, $C$, $D$, $E$ depend on the renormalization scale $k$. Their $k$-dependence is determined by a truncation of the exact functional flow equation~\cite{Wetterich:1992yh,Tetradis:1992qt,Morris:1993qb,Tetradis:1993ts,Reuter:1993kw,Ellwanger:1993mw}.
The result of our computation can be written in the form
\al{
\p_t\Gamma_k
=\left[N_S\pi^{(S)}_k
+N_V\left(\pi_k^{(V)}-\delta_k^{(V)}\right)
+N_F\pi^{(F)}_k\right]
+\pi_k^{(t,\sigma)}-\delta_k^{(g)}\,.
\label{eq: general flow equation}
}
Here the last two terms are contributions from metric fluctuations, where $\pi^{(t,\sigma)}$ denotes contributions from the TT tensor ($t$-mode) and the physical scalar metric fluctuation ($\sigma$-mode), while contributions from the gauge modes (the longitudinal modes in metric fluctuations and the ghost fields) are included in $\delta^{(g)}$.
We employ dimensionless quantities 
\al{
&u=\frac{U}{k^4}\,,&
&w=\frac{F}{2k^2}\,, 
&\tilde\rho=\frac{\rho}{k^2}\,.
}
The flow equations at fixed $\tilde\rho$ read for $\p_t =k\p_k$
\al{
\p_t u&=\beta_U=2{\tilde \rho}\,\p_{\tilde \rho}u-4u+\frac{1}{32\pi^2}\left(N_S +2N_V-2N_F +M_U\right)\,,
\label{eq: beta function of U}
\\[2ex]
\p_t w&=\beta_F=2{\tilde \rho}\,\p_{\tilde \rho}w-2w - \frac{1}{96\pi^2}\left( N_S - 4N_V + N_F+M_F \right)\,,
\label{eq: beta function of W}
\\[2ex]
\p_t C&=\beta_C=2{\tilde \rho}\,\p_{\tilde \rho} C-\frac{1}{576\pi^2}\left( N_S+M_C\right)\,,\\[2ex]
\p_t D&=\beta_D=2{\tilde \rho}\,\p_{\tilde \rho} D+ \frac{1}{960\pi^2}\left( N_S+12N_V+3N_F +M_D \right)\,,\\[2ex]
\p_t E&=\beta_E=2{\tilde \rho}\,\p_{\tilde \rho} E -\frac{1}{5760\pi^2}\left( N_S+62N_V+\frac{11}{2}N_F +M_E \right)\,.
\label{eq: beta function of E}
}
Here $M_i$ denote the contributions from metric fluctuations.
Their explicit form is displayed in Section~\ref{sec: Physical metric fluctuations}.

The structure of the gravitational contributions to the flow equations can be understood by writing them in the form
\al{
M_U&= \frac{20}{3} L^{(t)}_U(\tilde m_t^2) + \frac{8}{5}L^{(\sigma)}_U(\tilde m_\sigma^2) -\frac{13}{2}\ell_0^4(0)\,,
\label{eq: MU in summery}
\\[2ex]
M_F&= -\frac{125}{6}  L^{(t)}_F(\tilde m_t^2) + \frac{193}{84} L^{(\sigma)}_F(\tilde m_\sigma^2) -\frac{29}{4}\ell_0^2(0)\,,
\label{eq: MF in summery}
\\[2ex]
M_C&= 20L^{(t)}_C(\tilde m_t^2)
+ \frac{9}{2} L^{(\sigma)}_C(\tilde m_\sigma^2) - L^{(t-\sigma)}_\text{mix}(\tilde m_t^2,\tilde m_\sigma^2)  -\frac{29}{2}\ell_0^0(0)\,,
\label{eq: MC in summary}
\\[2ex]
M_D&= 820 L^{(t)}_D(\tilde m_t^2) 
- \frac{51}{2} L^{(\sigma)}_D(\tilde m_\sigma^2) 
-10L^{(t-\sigma)}_\text{mix}(\tilde m_t^2,\tilde m_\sigma^2) 
+ \frac{7}{2}\ell_0^0(0)\,,
\label{eq: MD in summary}
\\[2ex]
M_E&= 1660 L^{(t)}_E(\tilde m_t^2)
- \frac{161}{2} L^{(\sigma)}_E(\tilde m_\sigma^2)  
-30 L^{(t-\sigma)}_\text{mix}(\tilde m_t^2,\tilde m_\sigma^2)  -\frac{23}{2}\ell_0^0(0)\,.
\label{eq: ME in summary}
}
The interpolating functions $L^{(t)}_i(\tilde m_t^2)$, $L^{(\sigma)}_i(\tilde m_\sigma^2)$ and $L^{(t-\sigma)}_{\rm mix}(\tilde m_t^2,\tilde m_\sigma^2)$ represent contributions from the $t$-mode, the $\sigma$-mode and the $t-\sigma$ mixing, respectively.
They are functions of the dimensionless mass terms for the $t$- and $\sigma$-modes
\al{
&\tilde m_t^2 = \frac{D}{w} -\frac{u}{w} =d-v\,,&
&\tilde m_\sigma^2 = \frac{3C}{w} -\frac{u}{4w}=3c-\frac{v}{4}\,,
\label{eq: masses of t and sigma modes}
}
where
\al{
&v=\frac{u}{w}\,,&
&c=\frac{C}{w}\,,&
&d=\frac{D}{w}\,.
\label{eq: ratios of couplins}
}
The interpolating functions are linear combinations of the ``threshold functions", with $x=\tilde m_t^2$ or $\tilde m_\sigma^2$, 
\al{
\ell_p^{2n} (x) = \frac{1}{n!}\frac{1}{(1 + x)^{p+1}}\,,
\label{eq: threshold function with Litim cutoff}
}
whose explicit form can be read off from Section~\ref{sec: Physical metric fluctuations}, Eqs.~\eqref{eq: MUTT}--\eqref{eq: MEsigma}. The explicit forms of $L_i^{(t)}$, $L_i^{(\sigma)}$ and $L_i^{t-\sigma}$ are given in Eqs.~\eqref{eq: MUTT}--\eqref{eq: mixing Me}.
The threshold functions can account for the decoupling of heavy modes for which $x\gg 1$.
The functions $L^{(t)}_i(\tilde m_t^2)$, $L^{(\sigma)}_i(\tilde m_\sigma^2)$ and $L^{(t-\sigma)}_{\rm mix}(\tilde m_t^2,\tilde m_\sigma^2)$ have poles at $\tilde m_t^2=-1$ and $\tilde m_\sigma^2=-1$.
The validity of the flow equations is restricted to $\tilde m_t^2>-1$, while the issue for $\tilde m_\sigma^2$ is more complex.

For Einstein spaces, one has $R_{\mu\nu}=(R/4)g_{\mu\nu}$ such that $C_{\mu\nu\rho\sigma}C^{\mu\nu\rho\sigma}=R_{\mu\nu\rho\sigma}R^{\mu\nu\rho\sigma}-R^2/6$ and $G_4=R_{\mu\nu\rho\sigma}R^{\mu\nu\rho\sigma}$.
For these geometries one has only two independent invariants
\al{
-\frac{C}{2}R^2 +\frac{D}{2}C_{\mu\nu\rho\sigma}C^{\mu\nu\rho\sigma} + E G_4
= -\frac{1}{2}\left(C + \frac{D}{6} \right) R^2 + \left(\frac{D}{2} + E \right) R_{\mu\nu\rho\sigma}R^{\mu\nu\rho\sigma}\,.
}
In turn, an evaluation of the flow equation on Einstein spaces, which is done in most work in the literature, only yields flow equations for $D+6C$ and $D+2E$. One finds that for these linear combinations, the mixing terms $L^{(t-\sigma)}_\text{mix}(\tilde m_t^2,\tilde m_\sigma^2)$ cancel out.
A computation of separate flow equations for $C$ and $D$ is not possible in this way.
For their extraction more general geometries have to be included, and the mixing term plays a role.

The coupling functions depend on two variables $\tilde \rho$ and $k$. We will restrict the discussion here to the case where either the $k$-dependence or the $\tilde \rho$-dependence is neglected, such that the coupling functions depend either on $\tilde\rho$ or on $k$.
The $k$-independent functions $u(\tilde \rho)$, $w(\tilde\rho)$ etc. define a ``scaling solution", that generalizes the notion of a fixed point for a finite number of couplings.
On the other hand, we may evaluate the flow with $k$ at $\tilde \rho=0$, such that the terms $\sim \tilde \rho \,\p_{\tilde\rho}u$  etc. can be omitted in Eqs.~\eqref{eq: beta function of U}--\eqref{eq: beta function of E}.
In this case, which we discuss in the following, one is left with the $\beta$-functions for five couplings.
Fixed points correspond to zeros of these $\beta$-functions. We emphasize that the flow with $k$ at $\tilde \rho=0$ can be directly mapped to the $\tilde\rho$-dependence of the scaling solution.
One simply replaces $\p_t$ by $-2\tilde\rho\,\p_{\tilde\rho}$.
Thus the flow away from fixed points at $\tilde\rho=0$ translates directly to the $\tilde\rho$-dependence of the scaling solution. 
Keeping this connection in mind, we omit in the following the term $\sim \tilde \rho\, \p_{\tilde\rho}$ etc. in the flow equations.
Then the coupling $E$ does not appear in the flow equations for the other couplings, as appropriate for a topological invariant.

\subsection{Asymptotic freedom}
The flow equations \eqref{eq: beta function of U}--\eqref{eq: beta function of E} admit the SFT-fixed point 
\al{
&C_*^{-1}=0\,,&
&D_*^{-1}=0\,.
}
Since $w$ remains finite at the fixed point, this also implies
\al{
&c_*^{-1}=0\,,&
&d_*^{-1}=0\,.
}
Finite values of $w$ and $v$ play no role precisely at the fixed point since their relative contribution to $\tilde m_t^2$ and $\tilde m_\sigma^2$ vanishes.
Close to the SFT-fixed point we can evaluate the flow equations in the limit $w\to0$, $u\to 0$ or $\tilde m_t^2\to\infty$, $\tilde m_\sigma^2\to \pm \infty$. In this limit, the beta functions for the higher derivative couplings in Eqs.~\eqref{eq: MC in summary}--\eqref{eq: ME in summary} read
\al{
\p_t C&= -\frac{1}{576\pi^2} N_S
-\frac{5}{288\pi^2}-\frac{5}{16\pi^2}\left( \frac{C^2}{D^2}+\frac{C}{D}\right)
\,,
\label{eq: beta C in perturbative}
\\[2ex]
\p_t D&=  \frac{1}{960\pi^2}\left( N_S+12N_V+3N_F \right) 
+\frac{133}{160\pi^2}\,,\\[2ex]
\p_t E &=-\frac{1}{5760\pi^2}\left( N_S+62N_V+\frac{11}{2}N_F \right)
-\frac{49}{180\pi^2}
\,.
}
These results agree with the perturbative computation \cite{deBerredoPeixoto:2004if} in higher derivative gravity. They are universal, i.e. independent of regularization and gauge parameter choice.
One infers the flow equation for the perturbative coupling $D^{-1}$
\al{
\p_t D^{-1}= -\frac{1}{960\pi^2}\left( 798+ N_S+12N_V+3N_F \right) D^{-2}\,.
\label{eq: D inverse equation}
}
For vanishing matter effects ($N_S=N_V=N_F=0$), Eq.~\eqref{eq: D inverse pert RG eq} is reproduced. For arbitrary numbers of matter particles this is of the asymptotically free form with positive $D^{-1}$ increasing as $k$ is lowered. There is unavoidably a range where $D^{-1}$ becomes large and the perturbative result is no longer valid.

For the flow equation for $C^{-1}$ one has
\al{
\p_t C^{-1} =\frac{1}{576\pi^2}(10+N_S)C^{-2} + \frac{5}{16\pi^2}D^{-2} +\frac{5}{16\pi^2}C^{-1}D^{-1}\,.
}
Setting $N_S=0$ yields Eq.~\eqref{eq: C inverse pert RG eq}.
For $C,\,D\geq 0$ all terms are positive, driving $C^{-1}$ towards smaller values as $k$-decreases while $C$ increases according to Eq.~\eqref{eq: beta C in perturbative}.
In the range of negative $C^{-1}$ the ratio 
\al{
\omega =\frac{3C}{2D}
}
has two fixed points according to the flow equation
\al{
\p_t \omega=-\frac{1}{16\pi^2 D}\left[ \frac{10+N_S}{24} +\frac{1098+N_S+12N_V+3N_F}{60}\omega + \frac{10}{3}\omega^2 \right] \,.
\label{eq: omega equation}
}
Both fixed points occur for negative $\omega$ and therefore negative $C$ for positive $D$. For pure gravity the numerical values are
\al{
\omega_*^{(\text{IR})}&=-5.467\,,&
\omega_*^{(\text{UV})}&=-0.023\,.
\label{eq: fixed point of omega}
}

At the fixed point the ratio $\omega$ is not determined since the r.h.s of Eq.~\eqref{eq: omega equation} vanishes $\sim D^{-1}$ for arbitrary values of $\omega$. For the flow trajectories away from the fixed point $\omega$ increases with decreasing $k$ for all trajectories with ``initial values" $\omega>\omega_*^{(\text{UV})}$ close to the fixed point. For initial values $\omega<\omega_*^{(\text{UV})}$ the trajectories are attracted towards $\omega_*^{(\text{IR})}$.
We conclude that the fixed points \eqref{eq: fixed point of omega} concern the behavior of trajectories away from the fixed point, but do not fix the ratio $C/D$ at the fixed point.
Starting close to the fixed point with $C<0$, $D>0$ seems to be rather natural since in this case the Euclidean action \eqref{starting effective action} is bounded from below. 
We conclude that both $C^{-1}$ and $D^{-1}$ are independent (marginally) relevant parameters. 
For the Gauss-Bonnet coupling one has at the fixed point
\al{
\theta_* = -\frac{E_*}{D_*}=0.3274\,.
}
These values agree with that found in Ref.~\cite{deBerredoPeixoto:2004if}.

At the fixed point the ratios $M_\text{p}^2/k^2$ and $V/k^4$ take finite non-zero values, given for pure gravity by
\al{
&u_*=\frac{M_U}{128\pi^2}\,,&
&w_*=-\frac{M_F}{192\pi^2}\,,&
&v_*=-\frac{3M_U}{2M_F}\,.
}
At the fixed point one finds
\al{
&M_U=\frac{93}{20}\,,&
&M_F=-\frac{1471}{60}-\frac{40}{3}\omega\,.
}
Thus $w_*$ and $v_*$ depend on $\omega$ and therefore on the particular trajectory away from the fixed point.
For $\omega=\omega_*^{\text{UV}}$ one finds the numerical values
\al{
&(\text{SFT, UV}):~
u_*=0.0069\,,&
&w_*=0.0127\,,\phantom{-}&
&v_*=0.544\,,\phantom{-}
\label{eq: AF fixed point}
}
where for $\omega=\omega_*^{(\text{IR})}$ one obtains
\al{
&(\text{SFT, IR}):~
u_*=0.0069\,,&
&w_*=-0.0257\,,&
&v_*=-0.269\,,
}
Both $u$ and $w$ correspond to relevant parameters, with critical exponents $4$ and $2$ following from Eqs.~\eqref{eq: beta function of U} and \eqref{eq: beta function of W} for $M_U$ and $M_F$ not depending on $u$ and $w$.

Taking things together the SFT-fixed point has a free undetermined parameter $\omega$. For pure gravity it has four relevant or marginal couplings, with critical exponents given by
\al{
&\theta_1=4\,,&
&\theta_2=2\,,&
&\theta_3=0\,,&
&\theta_4=0\,.
}
The Gaussian fixed point characterizes asymptotic freedom of higher derivative gravity~\cite{Fradkin:1978yf,Fradkin:1978yf,Fradkin:1981hx,Fradkin:1981iu,Julve:1978xn,Avramidi:1985ki,Avramidi:1986mj,Antoniadis:1992xu,deBerredo-Peixoto:2003jda,deBerredoPeixoto:2004if}.

Finally, we mention the limit of the Weyl invariance which is realized for $U\to 0$, $F\to 0$ and $C\to 0$ in the action \eqref{Eq: effective action for gravity part}.
For that limit in the pure gravity system, one has
\al{
&\p_t u = 2{\tilde \rho}\,\p_{\tilde \rho}u-4u+ \frac{167}{640\pi^2}\,,&
&\p_t w= 2{\tilde \rho}\,\p_{\tilde \rho}w-2w +\frac{131}{504\pi^2}\,,&
&\p_t C=-\frac{5}{288\pi^2}\,.
}
Therefore, finite values of $u$, $w$ and $C$ are induced by quantum effects and the Weyl invariance is not conserved in our current setting.

\subsection{Asymptotic safety}

Solving $\beta_U=\beta_F=\beta_C=\beta_D=0$ simultaneously, we find a further non-trivial fixed point
\al{
&(\text{R}):~
u_*=0.000281\,,&
&w_*=0.0218\,,&
&C_*=0.204\,,&
&D_*=-0.0132\,.
\label{eq: full fixed point value}
}
For this fixed point the dimensionless ratios read
\al{
& v_*=0.0129\,,&
& c_*=9.324\,,&
&d_*=-0.60479\,,&
& \omega_*=-23.1\,,
}
resulting in
\al{
&\tilde m_{t*}^2=-0.618\,,&
&\tilde m_{\sigma*}^2=28.0\,.
}
The critical exponents are given as
\al{
&\theta_1=3.1\,,&
&\theta_2=2.4\,,&
&\theta_3=10.9\,,&
&\theta_4=-88.1\,.
}
Hence, there are three relevant directions and one irrelevant one.
This result agrees with the cases of higher derivative truncations in maximally symmetric spaces,~e.g.~\cite{Falls:2018ylp}, in Einstein spaces~\cite{Benedetti:2009rx}, in an arbitrary space within a strong gravity expansion~\cite{Falls:2020qhj} and in a flat space within the vertex expansion scheme~\cite{Denz:2016qks}.

We will argue in Section~\ref{sec: UV fixed point and critical exponent} that this fixed point is the extension of the R-fixed point for one truncation. The huge absolute values of the critical exponents $\theta_3$ and $\theta_4$ are presumably artifacts of the truncation.
Similar high values are observed in a truncation which omits the term $\sim D$~\cite{Lauscher:2002sq,deBrito:2018jxt}.
Including higher order curvature invariants $\sim R^n$ the high value of $\theta_3$ is reduced to a quantity of the order one, and further critical exponents become negative, corresponding to irrelevant parameters~\cite{Falls:2013bv,Falls:2014tra,Falls:2017lst,Falls:2018ylp,Kluth:2020bdv}.
The small negative value of $D_*$ is not necessarily a cause of worry either.
Taking the four-derivative truncation at face value it would imply a tachyon in the $t$-sector and therefore an instability.
The derivative expansion yields, however, only a Taylor expansion of the inverse graviton propagator for small momenta. In the momentum range of the possible instability higher order terms can cure this issue. 
Furthermore, extended truncations could shift $D_*$ to positive values.

\subsection{Interpolating functions}
\begin{figure}
\includegraphics[width=6.2cm]{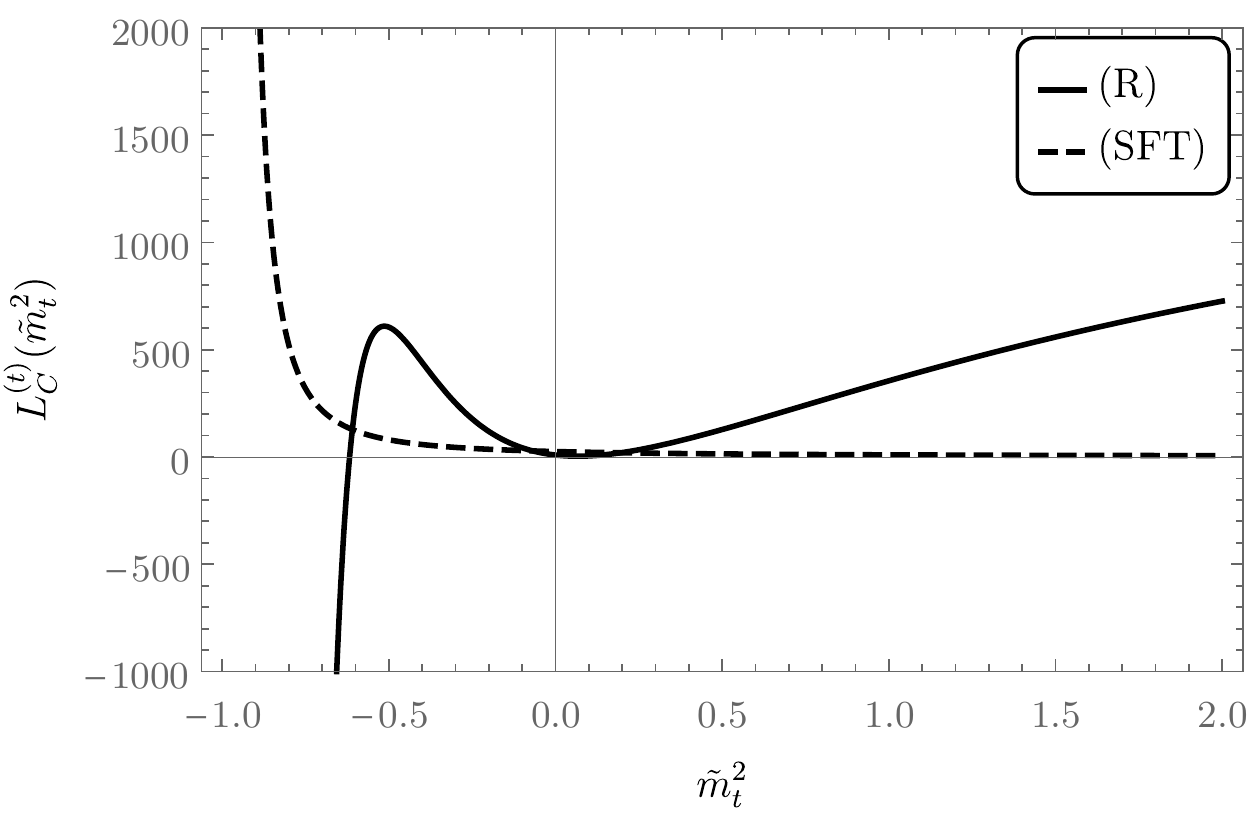}
\hspace{7ex}
\includegraphics[width=6.2cm]{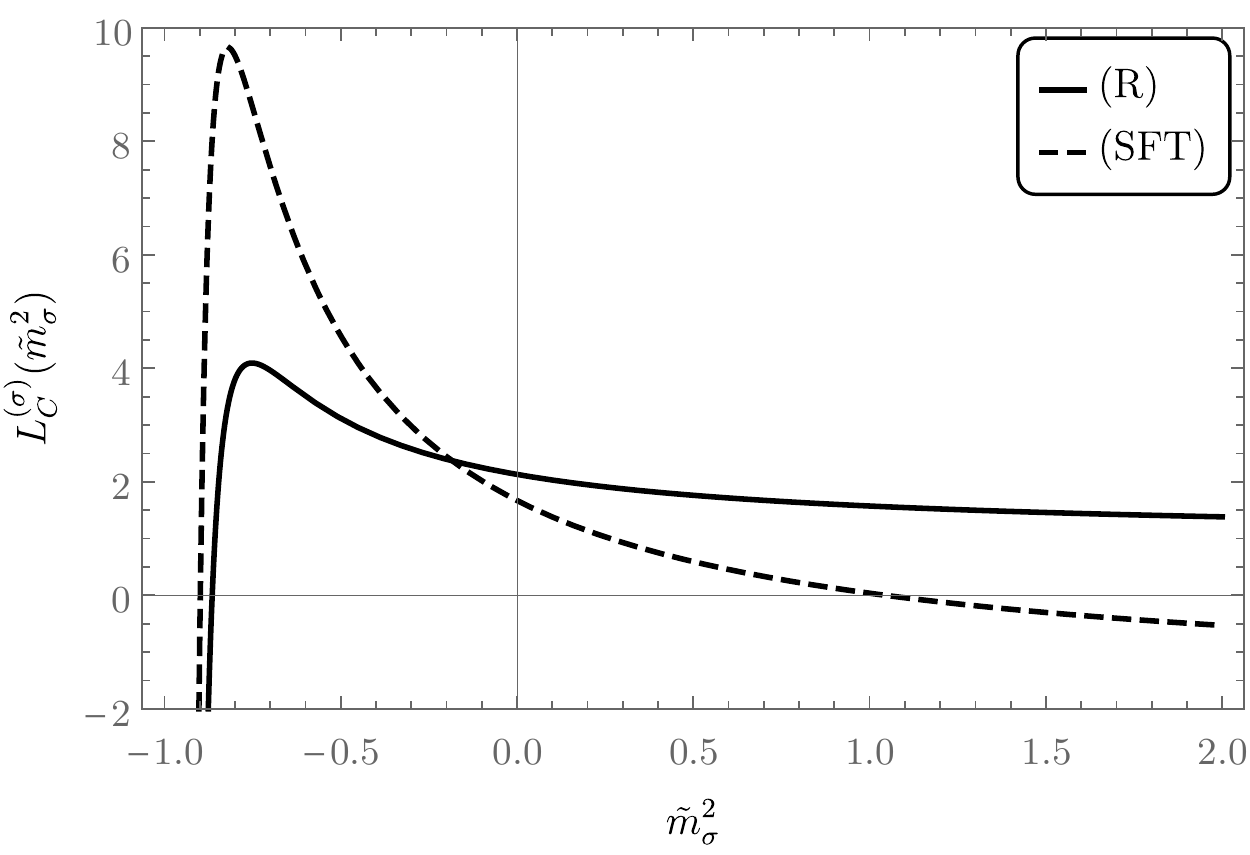}\\[2ex]
\includegraphics[width=6.2cm]{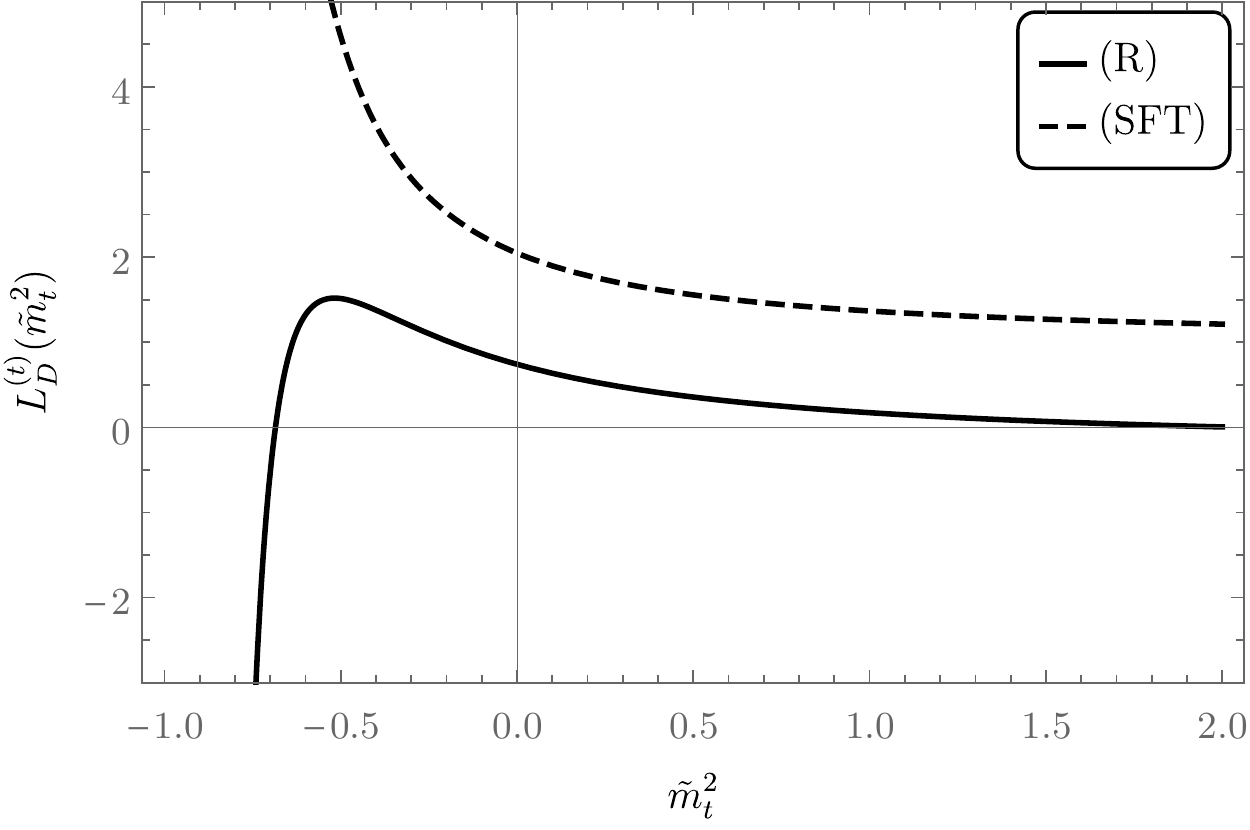}
\hspace{7ex}
\includegraphics[width=6.2cm]{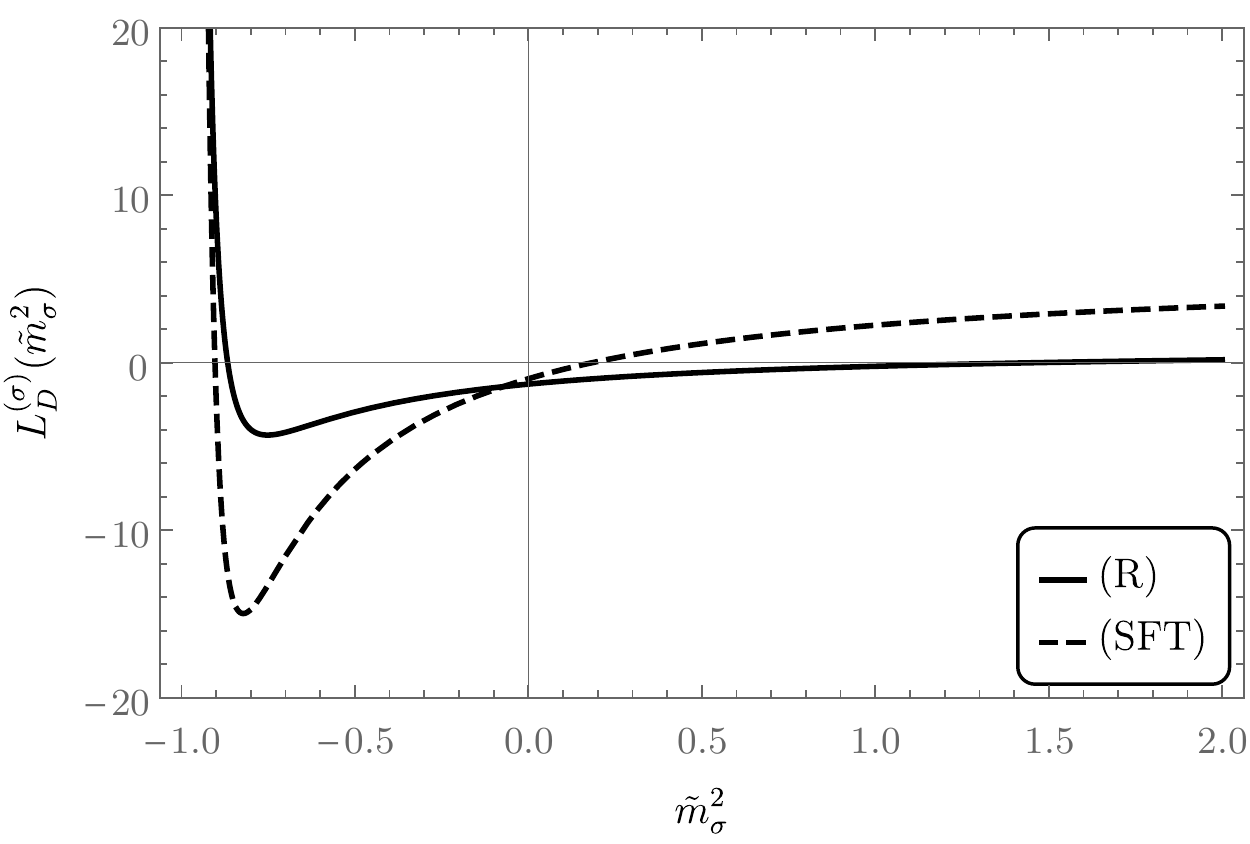}\\[2ex]
\caption{
Behavior of $L^{(t)}_i(\tilde m_t^2)$ and $L^{(\sigma)}_i(\tilde m_\sigma^2)$ as functions of the dimensionless masses $\tilde m_t^2$ and $\tilde m_\sigma^2$.
For $\omega$ and $v$ we display two cases: (R) $v=0.0128638$, $\omega=-23.1264$ and (SFT) $v=0.544229$,  $\omega=-0.0228639$.
}
\label{fig: LC LD} 
\end{figure}

\begin{figure}
\includegraphics[width=8cm]{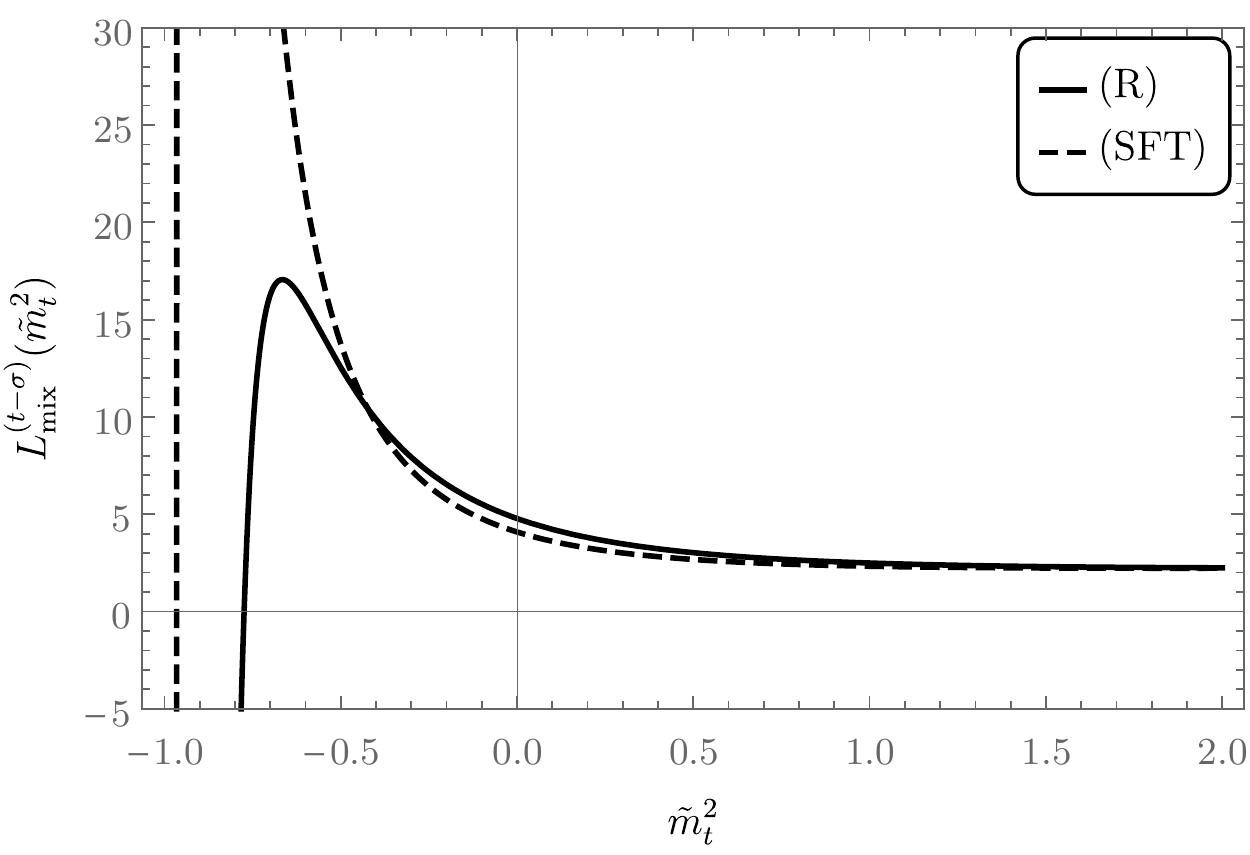}
\caption{
Behavior of $L^{(t-\sigma)}_\text{mix}(\tilde m_t^2)$ as functions of the dimensionless mass $\tilde m_t^2$.
We use the values of $v=0.0129$ in (R):~\eqref{eq: full fixed point value} and $v=0.544$ (SFT):~\eqref{eq: AF fixed point} and fix $\omega=1/2$.
}
\label{fig: Ltsigma} 
\end{figure}

For the flow away from the fixed points and for the question of a possible critical trajectory from the SFT- to the R-fixed point we need some understanding of the interpolating functions.
They depend on the couplings $c$, $d$ and $v$, or equivalently $\tilde m_t^2$, $\tilde m_\sigma^2$ and $v$.
The functions $L_i^{(t)}$ arise from the $t$-fluctuations and depend dominantly on the corresponding mass term $\tilde m_t^2$, while $L_i^{(\sigma)}$ are due to the $\sigma$-fluctuations and depend dominantly on $\tilde m_\sigma^2$. We plot in Fig.~\ref{fig: LC LD} the interpolating functions $L_D^{(t)}(\tilde m_t^2)$, $L_D^{(\sigma)}(\tilde m_\sigma^2)$, $L_C^{(t)}(\tilde m_t^2)$ and $L_C^{(\sigma)}(\tilde m_\sigma^2)$.
These functions are rather smooth until they reach the poles for $\tilde m_t^2\to -1$ or $\tilde m_\sigma^2\to -1$.
For a plot of $L_i^{(t)}(\tilde m_t^2)$ we need to fix the two other parameters that we take as $\omega$ and $v$. We show two sets, one for the values of $v$ and $\omega$ at the SFT-fixed point with $\omega=\omega_*^{(\text{UV})}$, and the other for the R-fixed point.
The same procedure is used for $L_i^{(\sigma)}(\tilde m_\sigma^2)$.
The mixing contribution $L_{\text{mix}}^{t-\sigma}(\tilde m_t^2, \tilde m_\sigma^2)$ is displayed in Fig.~\ref{fig: Ltsigma}.
The curves are shown with $v$ taken from the SFT- or R-fixed point, and $\omega=0.5$ for both curves.
Figures for the other interpolating functions can be found in Section~\ref{sec: Flow generators from metric fluctuations}.

We have chosen normalizations for the interpolating functions such that their overall size is of the order one, except for $L_\text{mix}^{(t-\sigma)}$ for the R-parameter set.
This allows for a judgement of the size of the different contributions from the prefactors in Eqs.~\eqref{eq: MU in summery}--\eqref{eq: ME in summary}. 
In particular, for the asymptotically free SFT-fixed point the $t$-fluctuations dominate, and the $t-\sigma$- mixing gives only a small contribution of around 15 percent.
In this region an omission of the $t-\sigma$- mixing contributions, which are technically the hardest part, does only lead to rather minor errors.
For pure gravity ($N_S=N_V=N_F=0$) and at $\tilde \rho=0$ one obtains at the SFT-fixed point for fixed constant $\tilde m_\sigma^2/\tilde m_t^2=3C/D$,
\al{
\beta_C&= -\frac{1}{576\pi^2}\left ( 20L^{(t)}_C(\tilde m_t^2)
+ \frac{9}{2} L^{(\sigma)}_C(\tilde m_t^2) - L^{(t-\sigma)}_\text{mix}(\tilde m_t^2,\tilde m_t^2)  -\frac{29}{2}\ell_0^0(0) \right)\Bigg|_{\tilde m_t^2 \to \infty }
\nn
&= \left[ -\frac{5}{144\pi^2} -\frac{5C^2}{16\pi ^2 D^2} -\frac{35C}{108\pi^2 D} \right]
+\left[ -\frac{1}{128\pi^2} -\frac{5D}{15552 \pi^2 C}\right]
+\left[  \frac{5C}{432\pi^2 D} + \frac{5D}{15552 \pi^2C }  \right]
+ \frac{29}{1152 \pi ^2}\nn
&= -\frac{5}{288\pi^2}-\frac{5}{16\pi^2}\left( \frac{C^2}{D^2}+\frac{C}{D}\right)
\,,\\[2ex]
\beta_D&= \frac{1}{960\pi^2}\left( 820 L^{(t)}_D(\tilde m_t^2) 
- \frac{51}{2} L^{(\sigma)}_D(\tilde m_t^2) 
-10L^{(t-\sigma)}_\text{mix}(\tilde m_t^2,\tilde m_t^2) 
+ \frac{7}{2}\ell_0^0(0)\right)\Bigg|_{\tilde m_t^2 \to \infty } \nn
&=\left[\frac{41}{48 \pi^2}  + \frac{5 C}{72\pi^2 D} \right]
+\left[-\frac{17}{640 \pi^2} +\frac{5 D}{2592 \pi^2 C}\right]
+\left[ -\frac{5C}{72\pi^2 D} - \frac{5D}{2592 \pi^2 C}\right]
+\frac{7}{1920 \pi ^2}
=\frac{133}{160\pi^2}\,.
} 
Here the first, second and third square bracket are contributions from the $t$ and $\sigma$ modes and their mixing, respectively, while the last terms are the measure contributions.
The flow of $D$ and $C$ is dominated by the $t$-fluctuations.

The coupling $D$ decreases until the negative contributions to $\beta_D$ from the mixing and $\sigma$-fluctuations get large enough to compensate the positive contribution from the $t$-fluctuations.
In our truncation this happens for negative $d$ and negative $\tilde m_t^2$.
For the R-fixed point $\tilde m_t^2$ is indeed negative, cf. Eq.~\eqref{eq: full fixed point value}.
At the R-fixed point the interpolating functions remain of the order one,
\al{
&L^{(t)}_U(\tilde m_t^2)\Big|_{\text{(R)}} =0.243\,,&
&L^{(\sigma)}_U(\tilde m_\sigma^2)\Big|_{\text{(R)}} =1.241\,,\nn
&L^{(t)}_F(\tilde m_t^2)\Big|_{\text{(R)}} =0.548\,,&
&L^{(\sigma)}_F(\tilde m_\sigma^2)\Big|_{\text{(R)}} =1.350\,,\nn
&L^{(t)}_C(\tilde m_t^2)\Big|_{\text{(R)}} =3.638\,,&
&L^{(\sigma)}_C(\tilde m_\sigma^2)\Big|_{\text{(R)}} =1.037\,,\nn
&L^{(t)}_D(\tilde m_t^2)\Big|_{\text{(R)}} = 1.194\,,&
&L^{(\sigma)}_D(\tilde m_\sigma^2)\Big|_{\text{(R)}} =0.916 \,,\nn
&L^{(t)}_E(\tilde m_t^2)\Big|_{\text{(R)}} = 1.386\,,&
&L^{(\sigma)}_E(\tilde m_\sigma^2)\Big|_{\text{(R)}} =0.920 \,,
}
with the exception of the mixing contribution,
\al{
L^{(t-\sigma)}_\text{mix}(\tilde m_t^2)|_{\text{(R)}} =97.579\,.
\label{eq: interpolating functions for mixing}
}

A fixed point with finite $D$ can only occur for negative $D$ since the decrease of $D$ with decreasing $k$ has to be stopped.
The rather large value \eqref{eq: interpolating functions for mixing}, together with a rather large negative $\tilde m_t^2$, may cast same doubts on the robustness of the R-fixed point with respect to extended truncations.

\subsection{Gravity contributions to the flow of the effective potential for scalars}
The flow equations \eqref{eq: beta function of U}--\eqref{eq: beta function of E} are valid for arbitrary constant scalar fields, or arbitrary constant $\tilde \rho$.
In particular, Eq.~\eqref{eq: beta function of U} describes the flow of the scalar potential. 
In principle, $N_S$, $N_V$ and $N_F$ depend on $\tilde \rho$, reflecting the flow-contributions from matter loops~\cite{Wetterich:2019zdo}.
We focus here on the gravitational contributions encoded in $M_U(\tilde\rho)$. 
The central quantity is the gravity induced scalar anomalous dimension $A$. 
For $A>0$ the quartic scalar couplings become irrelevant parameters.
Together with the assumption that the flow of couplings below the Planck scale does not deviate much from the one of the standard model this leads to the successful prediction of the mass of the Higgs boson~\cite{Shaposhnikov:2009pv}, or more precisely for the mass ratio between top quark and Higgs boson~\cite{Eichhorn:2017ylw,Wetterich:2019qzx}.
For $A>2$ also the scalar mass terms become irrelevant couplings, driving all scalar masses rapidly to zero.
This is the basis for a possible solution of the gauge hierarchy problem by the running of couplings~\cite{Wetterich:2016uxm,Wetterich:1981ir}.

The metric-fluctuation induced scalar anomalous dimension is given by the derivative of $M_U$ with respect to $u$ for $\tilde \rho\to0$,
\al{
A=\frac{1}{32\pi^2}\frac{\p M_U}{\p u} \bigg|_{\tilde\rho\to 0}
= \frac{1}{48\pi^2\tilde M_\text{p}^2}\left[ \frac{20( 1+\frac{3}{2}d)}{(1 + \tilde m_t^2)^2} + \frac{\frac{9}{20}(1+ 5c)}{(1+\tilde m_\sigma^2)^2} \right]\bigg|_{\tilde\rho\to 0}\,.
\label{eq: metric-fluctuation induced anomalous dimension}
}
Here the first and second term correspond to contributions from the $t$-mode (graviton) and $\sigma$-mode in the metric, respectively.
In our previous papers \cite{Pawlowski:2018ixd,Wetterich:2019zdo} within the Einstein-Hilbert truncation ($C\to 0$, $D\to 0$ and $E\to 0$), we have found
\al{
A_\text{EH} = \frac{1}{48\pi^2\tilde M_\text{p}^2}\left[ \frac{20}{(1 - v_0)^2} + \frac{1}{(1- v_0/4)^2} \right]\,.
}
The dominant graviton (the $t$-mode) contribution agrees with Eq.~\eqref{eq: metric-fluctuation induced anomalous dimension} for $D\to0$, while the part of the scalar mode has a difference of a factor $9/20$ for $C\to 0$. This difference arises from the different regularization scheme for the propagator of the scalar mode.

At the SFT-fixed point the gravity induced scalar anomalous dimension vanishes according to 
\al{
&\text{(SFT)}:\quad
A=\frac{1}{48\pi^2 w}\left( \frac{15}{d} + \frac{2}{c}\right)\to 0\,.
}
On the other hand, for the R-fixed point one finds 
\al{
&\text{(R)}:\quad
A=0.618\,.
}
This is positive and below two.
Adding contributions from matter fluctuations can change the value of $A$. 
In particular, a decrease of the fixed point value for $w$ leads to an increase of $A$.
Our investigation gives further support to the prediction for the mass of the Higgs boson, while the question of the gauge hierarchy will depend on the precise matter content.

\section{Flow generator and heat kernel method}
\label{sec: Heat kernel method and threshold function}
We compute the flow equations~\eqref{eq: beta function of U}--\eqref{eq: beta function of E} by the heat kernel method.
For this purpose we consider arbitrary geometries close to flat space.
Geometries beyond Einstein spaces are needed for the extraction of the gravitational contributions to the individual coupling functions for the higher derivative terms in the gravitational effective action.
The evaluation of the heat kernels for the differential operators needed in this context is technically new terrain  and will be described in the next section.
In this section we introduce the method and present the flow-contributions of matter fluctuations that can be obtained by more standard techniques.

We want to evaluate the flow generator
\al{
\zeta =\p_t \Gamma_k =\sum_i\zeta_i\,,
}
as a general functional of the (``background")-metric field. 
The sum $\sum_i$ is over fluctuation contributions of degrees of freedom that do not mix.
For our application the metric is closed to flat space, but not restricted otherwise.
For the heat kernel method $\zeta_i$ is represented as a trace over suitable differential operators, $\zeta_i=\tr W\fn{z=\Delta_i}$ where $\Delta_i$ is an appropriate Laplacian acting on the degree of freedom $i$.  
The flow generator can be expanded as
\al{
\zeta_i=\frac{1}{2}\tr_{(i)} W\fn{z}
= \frac{1}{2(4\pi)^2}\int_x\sqrt{g}\left[
Q_{2}[W] c_0^{(i)}+Q_1[W] c_2^{(i)}+Q_0[W] c_4^{(i)}+\cdots
\right]\,.
\label{eq: heat kernel expansion}
}
Here $\tr_{(i)}$ acts on the space for the degree of freedom $i$, and the coefficients $c_0^{(i)}$, $c_2^{(i)}$ and $c_4^{(i)}$ are given by
\al{
&c_0^{(i)}=b_0^{(i)}\,,&
&c_2^{(i)}=b_2^{(i)}R\,,&
&c_4^{(i)}=b_4^{(i)} R^2+{\hat b}_4^{(i)} R_{\mu\nu}R^{\mu\nu}+{\tilde b}_4^{(i)} R_{\mu\nu\rho\sigma}R^{\mu\nu\rho\sigma}\,,
\label{eq: heat kernel coefficients up to 4th}
}
where the values of the heat kernel coefficients $b_n^{(i)}$ depend on the degrees of freedom of a field on which the Laplacian acts.
The coefficients of $(1,\,-\frac{1}{2}R,\,-\frac{1}{2}R^2,\,\frac{1}{2}C_{\mu\nu\rho\sigma}C^{\mu\nu\rho\sigma},\, G_4)$ yield $(\p_t U,\,\p_t F,\,\p_t C,\,\p_t D,\,\p_t E)$ at fixed $\rho$.
Switching to dimensionless couplings yields the terms $-4u$ and $-2w$, and the change to the dimensionless scalar invariant $\tilde\rho$ induces the terms $\sim 2\tilde\rho \p_{\tilde \rho}$ in the flow equations~\eqref{eq: beta function of U}--\eqref{eq: beta function of E}.

The detailed steps of these calculations are displayed in the Appendices~\ref{App: setups}--\ref{App: Contributions from metric fluctuations}.
In the main text we summarize the most important points.
For the evaluation of Eq.~\eqref{eq: heat kernel expansion} the explicit heat kernel coefficients are summarized in Appendix~\ref{App: heat kernel coefficients}.
The threshold functions $Q_n$ are given by
\al{
&Q_n[W]=\frac{1}{\Gamma\fn{n}}\int^\infty_0 \df z\,z^{n-1}W\fn{z}
\qquad
\text{for $n\geq 1$} \,,&
&Q_{-n}[W]=(-1)^n\frac{\p^n W}{\p z^n}\Bigg|_{z=0}
\qquad
\text{for $n\geq 0$} \,.
}
The contributions to the flow generator take typically the form 
\al{
W_{p,\epsilon}(z)=\frac{\p_t {\mathcal R}_k(z)}{(ak^2 z^{1+2\epsilon}+ b z^{2+2\epsilon} +ck^4 z^{2\epsilon}+{\mathcal R}_k(z))^{p+1}}\,,
}
where $a$, $b$, $c$ are dimensionless couplings and $\mathcal R_k$ constitutes the infrared cutoff.
The quantity $\epsilon$ depends on the normalization of the fields. For contributions from the $t$-mode and matter modes, one has $\epsilon=0$, while $\epsilon=1$ is used for the $\sigma$-mode contributions.
For the cutoff function $\mathcal R_k(z)$ we employ a generalization of the Litim cutoff~\cite{Litim:2001up} such that $z$ in the denominator of $W(z)$ is replaced by $P_k(z)=z+R_k(z)=z+(k^2-z)\theta(k^2-z)$.
This yields
\al{
Q_n[p;\epsilon;a,b,c]&\equiv Q_n[W_{p,\epsilon}]\nn
&= \frac{(1+2\epsilon)(n+2+2\epsilon)}{n+1 +2\epsilon}a^{-p}  \Bigg( 1+ 2\frac{(1+\epsilon)(n+1 +2\epsilon)}{(1+2\epsilon)(n+2+2\epsilon)}\frac{b}{a}  +  2\frac{\epsilon \left( n+2+2\epsilon \right)(n+1 +2\epsilon)}{(n+2\epsilon)(1+2\epsilon)(n+2+2\epsilon)}\frac{c}{a}\Bigg) \nn
&\quad \times \left[2k^{2n-4p-4p\epsilon}\ell_{p}^{2n} (b/a+c/a) \right] \qquad (n\geq 0)\,,
}
where we neglected the anomalous dimensions, i.e. $\p_t a=\p_t b=\p_t c=0$.
The full expressions for $Q_n[W]$ with the anomalous dimensions can be found in Eq.~\eqref{App: threshold functions general} in Appendix~\ref{App: heat kernel coefficients}. 
For more details on the heat kernel technique we refer to Refs.~\cite{Codello:2008vh,Wetterich:2019zdo}.
Our general setting will become more explicit once we evaluate next the individual contributions from the fluctuations of free and massless scalar fields, fermions and gauge bosons.

\subsection{Scalar bosons}
We start by evaluating contributions from a scalar field whose effective action is given by ($\rho=\varphi^2$)
\al{
\Gamma_k^{(S)}=\int_x \sqrt{g}\left[ 
\frac{1}{2}(\p_\mu \varphi)^2 +U\fn{\rho} 
\right]\,.
}
The second functional derivative with respect to $\varphi$ yields
\al{
\label{eq: second derivative term by scalar fields}
\Gamma_k^{(S,2)}=\sqrt{g}\left( \Delta_{S} +m_\varphi^2\fn{\varphi}  \right)\,,
}
where $\Delta_{S}=-D^2$ is the Laplacian acting on a spin-0 scalar field and we define the mass term,
\al{
m_\varphi^2\fn{\varphi}=\frac{\p^2 U}{\p \varphi^2}\,.
}

In order to compute the flow-contributions, an appropriate IR cutoff function $R_k\fn{\Delta_{S}}$ is added so that the Laplacian $\Delta_{S}$ in Eq.~\eqref{eq: second derivative term by scalar fields} is replaced to $P_k=\Delta_{S}+R_k\fn{\Delta_{S}}$.
The flow equation for the scalar contribution reads
\al{
\p_t \Gamma_k=\pi^{(S)}_k=\frac{1}{2}\tr_{(0)} W^{(S)}\fn{\Delta_{S}}\,,
}
where the flow kernel is
\al{
&W^{(S)}(\Delta_{S})=\frac{\p_t R_k(\Delta_{S})}{\Delta_{S}+ R_k(\Delta_{S})+m_\varphi ^2}
=\frac{\p_t R_k(\Delta_{S})}{P_k(\Delta_{S})+m_\varphi ^2}
\,.
}
Using the heat kernel expansion \eqref{eq: heat kernel expansion} with the corresponding heat kernel coefficients to $\Delta_{S}$ (see Table~\ref{hkcs} in Section~\ref{App: hear kernel technique}), one obtains
\al{
\pi^{(S)}_k
=\frac{1}{16\pi^2}\int_x\sqrt{g}\left[ 
k^4 \ell_0^4\fn{\tilde m_\varphi^2} +\frac{1}{6}k^2\ell_0^2\fn{\tilde m_\varphi^2} R +\frac{1}{180}\left(\frac{5}{2}R^2-R_{\mu\nu}R^{\mu\nu} +R_{\mu\nu\rho\sigma}R^{\mu\nu\rho\sigma}  \right) \ell^0_0\fn{\tilde m_\varphi^2} 
\right]\,,
}
where $\tilde m_\varphi^2=m_\varphi^2/k^2$.
For the Litim cutoff the threshold functions are given by Eq.~\eqref{eq: threshold function with Litim cutoff}.
For $N_S$ massless scalars ($m_\varphi^2=0$) this yields in Eqs.~\eqref{eq: beta function of U}--\eqref{eq: beta function of E} the contributions $\sim N_S$.

\subsection{Gauge bosons}
We employ the effective action for a gauge theory as
\al{
\Gamma_k^{(V)}=\frac{1}{4}\int_x\sqrt{g}\,F_{\mu\nu}^aF^{a\mu\nu}+\Gamma_\text{gf}^{(V)}+\Gamma_\text{gh}^{(V)}\,,
\label{effective action for gauge theory}
}
where $F_{\mu\nu}^a$ is the field strength of the gauge field $A_\mu^a$.
Here, $\Gamma_\text{gf}^{(V)}$ and $\Gamma_\text{gh}^{(V)}$ are the actions of the gauge fixing and ghost fields ($c_a$, $\bar c_a$) associated with the gauge field $A_\mu^a$ and are given respectively by
\al{
&\Gamma_\text{gf}^{(V)} =\frac{1}{2\alpha_V}\sum_{a} \int \df^4 x \sqrt{g}\, \Big( (D_\mu A^\mu)_a\Big)^2\,,&
&\Gamma_\text{gh}^{(V)} = \int \df^4 x \sqrt{g}\, \bar c_a\, \p_\mu (D^\mu c)_a\,,
}
with $\alpha_V$ the gauge fixing parameter.

The Hessian of $A_\mu^a$, i.e. the second-order functional derivative of Eq.~\eqref{effective action for gauge theory} with respect to $A_\mu^a$, is computed as
\al{
\left(\Gamma_k^{(V,2)} \right)^{\mu\nu}= \sqrt{g} \left[ g^{\mu\nu}D^2 - \left(1-\frac{1}{\alpha_V} \right) D^\mu D^\nu \right]\,.
}
In the Landau gauge $\alpha_V\to 0$, the flow equation for Eq.~\eqref{effective action for gauge theory} reads
\al{
\p_t \Gamma_k=\zeta^{(V)}_k=\pi_k^{(V)}-\delta_k^{(V)}
= \frac{1}{2}\tr_{(1)} W^{(V)}\fn{z} - \frac{1}{2}\tr_{(0)} W^{(V)}\fn{z}
\,.
\label{eq: contributions from gauge fields}
}
For vanishing gauge couplings the r.h.s. of Eq.~\eqref{eq: contributions from gauge fields} becomes the sum of contributions from individual gauge bosons. For a single gauge boson the flow kernel is given by
\al{
W^{(V)}(\Delta_{L1})=\frac{\p_t R_k(\Delta_{L1})}{\Delta_{L1}+R_k(\Delta_{L1})}=\frac{\p_t R_k(\Delta_{L1})}{P_k(\Delta_{L1})}\,,
}
where the regulator $R_k(z)$ replaces the Lichnerowicz Laplacian $z=\Delta_{L1}$ (defined in Eq.~\eqref{App: Lichnerowicz Laplacian for spin1}) to $P_k(z)=z+R_k(z)$.
The use of the Litim cutoff and the heat kernel expansion yields the contribution from the physical mode in $A_\mu^a$ as 
\al{
\pi_k^{(V)}
=\frac{1}{2}\tr_{(1)} W^{(V)}\fn{\Delta_{L1}}
=\frac{1}{16\pi^2}\int_x\sqrt{g}\left[
3k^2 \ell_0^4(0) -\frac{1}{2}\ell_0^2(0)k^2R +\frac{1}{120}\left( -15 R^2 +58R_{\mu\nu}R^{\mu\nu} -8R_{\mu\nu\rho\sigma}R^{\mu\nu\rho\sigma}  \right)\ell_0^0(0)
\right]\,.
}
The contribution of the gauge mode and ghost yields for the ``physical" Landau gauge the universal measure contribution,
\al{
\delta_k^{(V)}
=\frac{1}{2}\tr_{(0)} W^{(V)}\fn{\Delta_{L1}}
=\frac{1}{16\pi^2}\int_x\sqrt{g}\left[ 
k^4 \ell_0^4\fn{0} +\frac{1}{6}k^2\ell_0^2\fn{0} R +\frac{1}{180}\left(\frac{5}{2}R^2-R_{\mu\nu}R^{\mu\nu} +R_{\mu\nu\rho\sigma}R^{\mu\nu\rho\sigma}  \right) \ell^0_0\fn{0} 
\right].
}
The two terms sum up to
\al{
\zeta^{(V)}_k&=\pi_k^{(V)}-\delta_k^{(V)} \nn
&=\frac{1}{16\pi^2}\int_x\sqrt{g}\left[
 2k^4 \ell_0^4\fn{0} -\frac{2}{3}k^2\ell_0^2\fn{0} R+\frac{1}{180}\left( -25 R^2 +88R_{\mu\nu}R^{\mu\nu} -13R_{\mu\nu\rho\sigma}R^{\mu\nu\rho\sigma}  \right)\ell_0^0(0)
\right].
}
For $N_V$ gauge bosons this yields in Eqs.~\eqref{eq: beta function of U}--\eqref{eq: beta function of E} the contributions $\sim N_V$.

\subsection{Weyl fermions}
We next consider the contributions from Weyl fermions to the flow of the gravitational interactions. Spinor fields in curved spacetime involve the vierbein fields $e_\mu^m$.
The covariant derivative $D_\mu$ involves the spin connection constructed from $e_\mu^m$.
With $\Slash D=\gamma^\mu D_\mu$ and $e=\det (e_\mu^m)$ the effective action with a Yukawa coupling to a scalar field $\varphi$ reads
\al{
\Gamma_k^{(F)}=\int_x \sqrt{e}\left[ 
\bar\psi i{\Slash D}\psi + y \varphi   \bar\psi \gamma^5\psi
\right]\,.
}
One obtains the Hessian as
\al{
\Gamma_k^{(F,2)} =\frac{\overrightarrow \delta}{\delta \bar\psi}\Gamma_k^{(F)}\frac{\overleftarrow \delta}{\delta \psi}
= \sqrt{e} \left[ i\Slash D + y\varphi \gamma^5  \right].
}
The squared covariant derivative becomes $-\Slash D^2=-D^2 + R/4=\Delta_{L\frac{1}{2}}$, which is the  Lichnerowicz Laplacian acting on a spinor field. Thus, we regulate $z=\Delta_{L\frac{1}{2}}$ by employing the regular $R_k(z)$ to obtain the flow generator
\al{
\pi^{(F)}_k= \frac{1}{2} \tr_{(\frac{1}{2})} W^{(F)}(\Delta_{L\frac{1}{2}}) 
=\frac{1}{2} \tr_{(\frac{1}{2})} \frac{\p_t R_k(\Delta_{L\frac{1}{2}})}{\Delta_{L\frac{1}{2}}+ R_k(\Delta_{L\frac{1}{2}})}
=\frac{1}{2} \tr_{(\frac{1}{2})} \frac{\p_t R_k(\Delta_{L\frac{1}{2}})}{P_k(\Delta_{L\frac{1}{2}})}\,.
}
From the heat kernel technique, one finds 
\al{
\pi^{(F)}_k
=-\frac{1}{16\pi^2}\int_x\sqrt{g}\left[
 2k^4 \ell_0^4\fn{m_\psi^2} -\frac{1}{6}k^2\ell_0^2\fn{m_\psi^2} R +
\frac{1}{720}\left(
5R^2 -8R_{\mu\nu}R^{\mu\nu} -7R_{\mu\nu\rho\sigma}R^{\mu\nu\rho\sigma}
\right)\ell^0_0\fn{m_\psi^2}
\right],
}
with $m_\psi^2=y^2\rho/k^2$.
The terms $\sim N_F$ in Eqs.~\eqref{eq: beta function of U}--\eqref{eq: beta function of E} account for the contributions form $N_F$ massless Weyl fermions ($m_\psi^2=0$).

\section{Flow generators from metric fluctuations}
\label{sec: Flow generators from metric fluctuations}
This section summarizes the main technical achievement of this work, namely the computation of the metric fluctuations to the functional flow of all coupling functions of higher derivative gravity in fourth order.
To this end, we use the heat kernel method again. The two-point function (Hessian) of the ghost fields can be written in terms of the form in the so-called non-minimal operator ${\mathcal D}_1=\Delta_V \delta^\mu_\nu -D^\mu D_\nu -R^\mu{}_\nu$ for which we use the heat kernel coefficients obtained in Refs.~\cite{Endo:1984sz,Gusynin:1988zt,Gusynin:1997dc}. On the other hand, the Hessian for the metric field becomes complicated and cannot be simplified to be in such a non-minimal operator. Therefore, we use the off-diagonal heat kernel expansion introduced in Ref.~\cite{Benedetti:2010nr} to evaluate the heat kernel coefficients for e.g. $\tr [e^{-s\Delta} R^{\mu\nu}D_\mu D_\nu]$ appearing in the flow kernel for metric fluctuations.

\subsection{Physical metric decomposition and gauge invariant flow}
Our starting point for the derivation of the flow equations is to employ the ``physical decomposition"~\cite{Wetterich:2016vxu,Wetterich:2016ewc,Wetterich:2017aoy} of the metric fluctuations.
\al{
\label{physical metric decomposition}
&g_{\mu\nu}=\bar g_{\mu\nu} + h_{\mu\nu}\,,&
&h_{\mu\nu}=f_{\mu\nu} + a_{\mu\nu} \,, &
&a_{\mu\nu}=D_\mu a_\nu +D_\nu a_\mu \,,
} 
with $\bar g_{\mu\nu}$ the argument of the effective action (often referred to as ``background field"). Hereafter we omit the bar on the background field. 
The physical metric fluctuation $f_{\mu\nu}$ satisfies the transverse condition, i.e. $D^\mu f_{\mu\nu}=0$. 
The ``gauge fluctuations" $a_{\mu\nu}$ denote the directions in field space generated by infinitesimal gauge transformations (diffeomorphisms).
In a physical gauge they decouple from the physical fluctuations.

In second order in $f_{\mu\nu}$ the  expansion of the effective action \eqref{Eq: effective action for gravity part} yields
\al{
\label{Eq: two-point function for f}
\Gamma_{(ff)}  &=\frac{1}{4}\int_x\sqrt{g}\, f^{\mu\nu}\left[ 
 (\widetilde{\mathcal D}_f)_{\mu\nu}{}^{\rho\tau}
-U\left( T_{\mu\nu}{}^{\rho\tau}-I_{\mu\nu}{}^{\rho\tau} \right)
 +\Big({\mathbb M}( R, \Delta_T)\Big)_{\mu\nu}^{\phantom{\mu\nu}\rho\tau}
\right]f_{\rho\tau}\,,
}
where the covariant derivative operator acting on the physical metric fluctuations $f$ is given by
\al{
(\widetilde{\mathcal D}_f)_{\mu\nu}{}^{\rho\tau}&= 
\left[ \frac{F}{2}\Delta_T+D\Delta_T^2\right] T_{\mu\nu}{}^{\rho\tau}
-3\left[ \frac{F}{2}\Delta_T +\frac{8}{3}C\Delta_T^2 +\frac{1}{9}D\Delta_T^2 \right]I_{\mu\nu}{}^{\rho\tau}\,.
\label{differential operator for f}
}
The ``interaction piece" ${\mathbb M}(\bar R,\bar \Delta_T)$ is a tensor depending on curvature tensors and the covariant derivatives.
With 
\al{
&T_{\mu\nu}{}^{\rho\tau}=E_{\mu\nu}{}^{\rho\tau}-I_{\mu\nu}{}^{\rho\tau}\,,&
&E_{\mu\nu}{}^{\rho\tau}=\frac{1}{2}(\delta_\mu^\rho \delta_\nu^\tau +\delta_\mu^\tau \delta_\nu^\rho)\,,&
&I_{\mu\nu}{}^{\rho\tau}=\frac{1}{4}g_{\mu\nu}g^{\rho\tau}\,,
}
one sees that $T$ is orthogonal to $I$, namely $T_{\mu\nu}{}^{\rho\tau}I_{\rho\tau}{}^{\alpha\beta}=0$.
The part $\sim T$ is the kinetic part of the inverse graviton propagator.
The flow equation for the system (with the physical gauge fixing action \eqref{standard gauge fixing} for $\alpha\to0$ and $\beta=-1$) takes the form
\al{
\p_t \Gamma_k=\zeta_k=\pi_k^{(f)}-\delta_k^{(g)}\,,
}
where $\pi_k^{(f)}$ is contributions from the physical modes $f_{\mu\nu}$, whereas $\delta_k^{(g)}$ contains contributions from the gauge modes $a_{\mu\nu}$ and the ghost fields.

\subsection{Physical metric fluctuations}
\label{sec: Physical metric fluctuations}
The physical metric fluctuations $f_{\mu\nu}$ can be further decomposed as
\al{
f_{\mu\nu}= t_{\mu\nu} + \hat S_{\mu\nu}\sigma\,.
}
Here $t_{\mu\nu}$ is the TT tensor, i.e. satisfies $D^\mu t_{\mu\nu}=0$ and $g^{\mu\nu}t_{\mu\nu}=0$, and $\sigma$ is the scalar physical fluctuation of the metric defined by
\al{
\sigma=g^{\mu\nu}f_{\mu\nu}\,.
}
The tensor $\hat S_{\mu\nu}$ obeys
\al{
\hat S_{\mu\nu} =(g_{\mu\nu}\Delta_S + D_\mu D_\nu -R_{\mu\nu})(3\Delta_S -R)^{-1},
}
such that $D^\mu \hat S_{\mu\nu}=0$ and $g^{\mu\nu} \hat S_{\mu\nu}=1$. With $\sigma\hat S^{\mu\nu} t_{\mu\nu}=-\sigma(3\Delta_S -R)^{-1} R^{\mu\nu} t_{\mu\nu}$ one has, in a general spacetime, mixing effects between $t_{\mu\nu}$ and $\sigma$. Only in an Einstein spacetime, $R_{\mu\nu}=(R/4)g_{\mu\nu}$, the TT tensor mode ($t$-mode) decouples from the scalar mode ($\sigma$-mode). 

The Hessian for the physical metric fluctuations takes the following structure:
\al{
\left(\Gamma^{(2)}_{(ff)}\right)_{\mu\nu}^{\phantom{\mu\nu}\rho\sigma}&=\frac{1}{2}\Bigg[
 {\mathbb K}_{(t)} (P_t)_{\mu\nu}{}^{\rho\tau}  +\Big({\mathbb M}_{(t)}( R, \Delta_T)\Big)_{\mu\nu}^{\phantom{\mu\nu}\rho\sigma} 
 -\frac{8}{3}\Big( {\mathbb K}_{(\sigma)}  +{\mathbb M}_{(\sigma)}( R, \Delta_S) \Big) I_{\mu\nu}{}^{\rho\tau} \nn
&\qquad
+\Big({\mathbb M}_{(t \sigma)}( R, \Delta_T)\Big)_{\mu\nu}  g^{\rho\sigma}
+ g_{\mu\nu} \Big({\mathbb M}_{(\sigma t)}( R, \Delta_T)\Big)^{\rho\sigma}
\Bigg]\,,
\label{Hessian for physical metric}
}
where $\Delta_T=-D^2$ is the Laplacian acting on tensor fields. The kinetic parts are defined by
\al{
&{\mathbb K}_{(t)}=\frac{F}{2}\Delta_T+D\Delta_T^2 -U\,,&
&{\mathbb K}_{(\sigma)}=\frac{F}{2}\Delta_S +3C\Delta_S^2-\frac{U}{4}\,.
\label{kinetic terms of t and sigma}
} 
They correspond to the inverse propagators of the $t$-mode and $\sigma$-mode.
The interaction parts ${\mathbb M}$ are lengthy. Their explicit forms are shown in Appendix~\ref{sect: Inverse two-point functions}: See Eqs.~\eqref{App: tt mode interaction}--\eqref{App: mixing term of Hessian}.
The last terms involve the mixing between the $t$- and $\sigma$-mode. This vanishes for Einstein spaces due to the different transformation of these modes under rotation symmetry.

For the particular case of flat spacetime we can perform a Fourier transformation. The propagators of the graviton and $\sigma$-mode are given respectively by 
\al{
G_g(q^2)&= \frac{1}{\frac{F}{2}q^2 + D q^4 -U} = \frac{D^{-1}}{q^2( q^2 + M_t^2)-U/D}
\sim \frac{1}{q^2} -\frac{1}{q^2 + M_t^2} \qquad (U\to 0) \,,\\
G_\sigma(q^2)&=\frac{1}{\frac{F}{2}q^2 + 3C q^4 -U} = \frac{(3C)^{-1}}{q^2( q^2 + M_\sigma^2)-U/(3C)}
\sim \frac{1}{q^2} - \frac{1}{q^2+ M_\sigma^2} \qquad (U\to 0)\,,
}
with $M_t^2=F/(2D)$ and $M_\sigma^2=F/(6C)$.
The last terms are taken for $U\to 0$, where we observe the usual ghost for the $t$-mode of fourth-order gravity.
We also note the negative sign of the kinetic term for the $\sigma$-mode in Eq.~\eqref{Hessian for physical metric}.
Functional renormalization deals with these well known problematic feature by introducing an infrared cutoff.

For the two-point functions, we employ the regulator such that $\Delta_{i}$ is replaced by $P_k\fn{\Delta_{i}}$ in the kinetic terms ${\mathbb K}_{(t)}$ and ${\mathbb K}_{(\sigma)}$ in Eq.~\eqref{kinetic terms of t and sigma}. 
The contribution to the flow generator from the physical metric fluctuations reads
\al{
\label{eq: flow generator for ff mode}
\pi_k^{(f)}&= \frac{1}{2}\tr_{(2)}\frac{\p_t \mathcal R_k}{\Gamma_k^{(2)}+\mathcal R_k}\bigg|_{ff}
\nn[2ex]
&=\frac{1}{2}\tr_{(2)}\frac{\p_t \mathcal R_k}{\Gamma_k^{(2)}+\mathcal R_k}\bigg|_{tt}
+\frac{1}{2}\tr_{(0)}\frac{\p_t \mathcal R_k}{\Gamma_k^{(2)}+\mathcal R_k}\bigg|_{\sigma\sigma}
+J_{\text{grav0},k}=\pi_k^{(t)}+\pi_k^{(\sigma)}+J_{\text{grav0},k}\,.
}
Due to the regulator functions for the $t$-mode and $\sigma$-mode the Laplacians ($\Delta_T$ and $\Delta_S$) are replaced to $P_k(\Delta_i)=\Delta_i+R_k(\Delta_i)$.
The last term in Eq.~\eqref{eq: flow generator for ff mode} arises from the regulated Jacobian which accounts in the functional integral for the decomposition of the metric fluctuations into physical modes and gauge modes.
Computations are given in Appendix~\ref{App: Contributions from metric fluctuations}.
Here, we list the explicit forms of contributions from the $t$- and $\sigma$-modes to the beta functions~\eqref{eq: beta function of U}--\eqref{eq: beta function of E}: One obtains from $\pi_k^{(t)}$ 
\al{
&M^{(t)}_U=\frac{20}{3}(3d+2)\ell_{0}^4(\tilde m_t^2)\nn
&\phantom{M^{(t)}_U}
\equiv \frac{20}{3}L_U^{(t)}(\tilde m_t^2)\,,
\label{eq: MUTT}
\\[2ex]
&M^{(t)}_F=-\bigg[ \frac{5}{2}(4d+3) \ell_0^2(\tilde m_t^2)
+\frac{40}{3}(3d+2) \ell_1^4(\tilde m_t^2)
+\frac{15}{2}(2c+d)(5+8d)\ell_1^6(\tilde m_t^2) \bigg] \nn
&\phantom{M^{(t)}_F}
\equiv -\frac{125}{6}L_F^{(t)}(\tilde m_t^2)\,,
\label{eq: MFTT}
\\[2ex]
&M^{(t)}_C=
-\frac{770}{9}(d+1)\ell_0^0(\tilde m_t^2)
+\frac{410}{9} (4 d+3) \ell_1^2(\tilde m_t^2) 
+\frac{40}{27} (3 d+2) (3 c+7 d) \ell_1^4(\tilde m_t^2)
+\frac{1120}{9} (3 d+2) \ell_2^4(\tilde m_t^2) \nn
&\qquad\qquad
+\frac{80}{3} (8 d+5) (9 c+5 d) \ell^6_2(\tilde m_t^2)
+432 (5 d+3) \left(2 c^2+2 c d+d^2\right) \ell^8_2(\tilde m_t^2)
\nn
&\phantom{M^{(t)}_C}
\equiv 20L_C^{(t)}(\tilde m_t^2)
\,,
\label{eq: MCTT}
\\[2ex]
&M^{(t-\sigma)}_C=
\frac{5 d+3}{50(1+\tilde m_\sigma^2)}\ell_1^8(\tilde m_t^2)
+\frac{d (12 d+7) }{15 (1+\tilde m_\sigma^2)}\ell_1^{10}(\tilde m_t^2)
-\frac{400 (7 d+4) (d-6 c)^2}{21 (1+\tilde m_\sigma^2)} \ell_1^{12}(\tilde m_t^2)\,,
\\[2ex]
&M^{(t)}_D=
\frac{1030}{9} (d+1) \ell_0^0(\tilde m_t^2)
+\frac{500}{9} (4 d+3) \ell_1^2(\tilde m_t^2)
+\frac{200}{27} (3 d+2) (6 c-13 d) \ell_1^4(\tilde m_t^2)
+\frac{2800}{9} (3 d+2) \ell_2^4 (\tilde m_t^2)\nn
&\qquad\qquad
+\frac{1600}{3} d (8 d+5) \ell_2^6 (\tilde m_t^2)
+4080 d^2 (5 d+3) \ell_2^8 (\tilde m_t^2)
\nn
&\phantom{M^{(t)}_D}
\equiv 820L_D^{(t)}(\tilde m_t^2)\,,
\label{eq: MDTT}
\\[2ex]
&M^{(t-\sigma)}_D=
\frac{5 d+3}{5(1+  \tilde m_\sigma^2)} \ell_1^8 (\tilde m_t^2)
+\frac{2d (12 d+7) \ell_1^{10} (\tilde m_t^2)}{3(1+  \tilde m_\sigma^2)}
-\frac{4000 (7 d+4) (d-6 c)^2 }{21(1+ \tilde m_\sigma^2)}\ell_1^{12} (\tilde m_t^2)\,,
\\[2ex]
&M^{(t)}_E=
\frac{430}{3} (d+1) \ell_0^0 (\tilde m_t^2) 
+\frac{500}{3} (4 d+3) \ell_1^2 (\tilde m_t^2)
+\frac{200}{9} (3 d+2) (6 c+5 d) \ell_1^4 (\tilde m_t^2)
+\frac{1600}{3} (3 d+2) \ell_2^4 (\tilde m_t^2) \nn
&\qquad\qquad
+400 d (8 d+5) \ell_2^6 (\tilde m_t^2)
+6480 d^2 (5 d+3) \ell_2^8 (\tilde m_t^2)
\nn
&\phantom{M^{(t)}_E}
\equiv 1660L_E^{(t)}(\tilde m_t^2)\,,
\label{eq: METT}
\\[2ex]
&M^{(t-\sigma)}_E=
\frac{3 (5 d+3) }{5(1+ \tilde m_\sigma^2)}\ell_1^8 (\tilde m_t^2)
+\frac{2 d (12 d+7) }{1+ \tilde m_\sigma^2}\ell_1^{10} (\tilde m_t^2)
-\frac{4000 (7 d+4) (d-6 c)^2 }{7(1+ \tilde m_\sigma^2)}\ell_1^{12} (\tilde m_t^2)
-360 (d+1) \frac{N}{\chi_E} \ell_0^0 (\tilde m_t^2)\,,
}
and from $\pi_k^{(\sigma)}+J_{\text{grav0},k}$,
\al{
&M^{(\sigma)}_U= \frac{3}{10} (24 +80c -5v) \ell_0^4(\tilde m_\sigma^2)  \nn
&\phantom{M^{(\sigma)}_U}
\equiv \frac{8}{5}L_U^{(\sigma)}(\tilde m_\sigma^2)\,,
\label{eq: MUsigma}
\\[2ex]
&M^{(\sigma)}_F= \frac{1}{12}(144c-10v+45)\ell_0^2(\tilde m_\sigma^2)
-\frac{v}{20}(120c-7v+35)\ell_1^6(\tilde m_\sigma^2)
+\frac{1}{7}(126c-7v+36)\ell_1^8(\tilde m_\sigma^2)\nn
&\qquad\qquad
+\frac{27c}{7}(224c-12v+63)\ell_1^{10}(\tilde m_\sigma^2)
-\frac{5}{3}\ell_0^2(0)\nn
&\phantom{M^{(\sigma)}_F}
\equiv \frac{193}{84}L_F^{(\sigma)}(\tilde m_\sigma^2)\,,
\label{eq: MFsigma}
\\[2ex]
&M^{(\sigma)}_C=
(12 c-v+4) \ell_0^0 (\tilde m_\sigma^2)
-\frac{7}{20} v (80 c-5 v+24) \ell_1^4 (\tilde m_\sigma^2)
+\frac{11}{15} (120 c-7 v+35) \ell_1^6 (\tilde m_\sigma^2)
\nn
&\qquad\qquad
-\frac{8}{63} (183 c-10 d) (126 c-7 v+36) \ell_1^8 (\tilde m_\sigma^2)
+\frac{20}{21} (21 c-d) (126 c-7 v+36) \ell_1^8 (\tilde m_\sigma^2)
 \nn
&\qquad\qquad
+\frac{5}{63} v^2 (126 c-7 v+36) \ell_2^8 (\tilde m_\sigma^2)
-\frac{2}{7} v (224 c-12 v+63) \ell_2^{10} (\tilde m_\sigma^2)
+\frac{25}{36} (288 c-15 v+80) \ell_2^{12} (\tilde m_\sigma^2) \nn
&\qquad\qquad
-\frac{40}{3} c v (288 c-15 v+80) \ell_2^{12} (\tilde m_\sigma^2)
+\frac{80}{3} c (1080 c-55 v+297) \ell_2^{14} (\tilde m_\sigma^2) 
+\frac{60480}{11} c^2 (220 c-11 v+60) \ell_2^{16} (\tilde m_\sigma^2) 
\nn
&\phantom{M^{(\sigma)}_C}
\equiv \frac{9}{2}L_C^{(\sigma)}(\tilde m_\sigma^2)\,,
\label{eq: MCsigma}
\\[2ex]
&M^{(\sigma-t)}_C=
\frac{(126 c-7 v+36) }{210 (1+\tilde m_t^2)}\ell_1^8(\tilde m_\sigma^2) 
+\frac{(126 c-7 v+36) }{210 (1+\tilde m_t^2)}\ell_1^8 (\tilde m_\sigma^2)
+\frac{3 d (224 c-12 v+63) }{140 (1+\tilde m_t^2)}\ell_1^{10}(\tilde m_\sigma^2) \nn
&\qquad\qquad
-\frac{25 (d-6 c)^2 (288 c-15 v+80) }{9 (1+\tilde m_t^2)}\ell_1^{12} (\tilde m_\sigma^2)
-\frac{17}{18}\ell_0^0(0)\,,
\\[2ex]
&M^{(\sigma)}_D=
(12 c-v+4) \ell_0^0 (\tilde m_\sigma^2)
+\frac{5}{4} v (-80 c+5 v-24) \ell_1^4(\tilde m_\sigma^2) 
+ \frac{13}{3} (120 c-7 v+35) \ell_1^6(\tilde m_\sigma^2)
 \nn
&\qquad\qquad
-\frac{20}{63} (183 c-10 d) (126 c-7 v+36) \ell_1^8 (\tilde m_\sigma^2)
+\frac{5}{14} v (224 c-12 v+63) \ell_2^{10} (\tilde m_\sigma^2)
+\frac{25}{36} (288 c-15 v+80) \ell_2^{12}(\tilde m_\sigma^2)\nn
&\qquad\qquad
+\frac{5}{63} v^2 (126 c-7 v+36) \ell_2^8(\tilde m_\sigma^2) 
+\frac{50}{3} c v (288 c-15 v+80) \ell_2^{12}(\tilde m_\sigma^2) 
+\frac{80}{3} c (1080 c-55 v+297) \ell_2^{14}(\tilde m_\sigma^2)\nn
&\qquad\qquad
+\frac{60480}{11} c^2 (220 c-11 v+60) \ell_2^{16}(\tilde m_\sigma^2) 
\nn
&\phantom{M^{(\sigma-t)}_D}
\equiv -\frac{51}{2}L_D^{(\sigma)}(\tilde m_\sigma^2)\,,
\label{eq: MDsigma}
\\[2ex]
&M^{(\sigma-t)}_D=
\frac{(126 c-7 v+36) }{21 (1+\tilde m_t^2)}\ell_1^8 (\tilde m_\sigma^2)
+\frac{3 d (224 c-12 v+63) }{14 (1+\tilde m_t^2)}\ell_1^{10} (\tilde m_\sigma^2) 
-\frac{250 (d-6 c)^2 (288 c-15 v+80) }{9 (1+\tilde m_t^2)}\ell_1^{12} (\tilde m_\sigma^2)\nn
&\qquad\qquad
-\frac{341}{18}\ell_0^0(0)\,,
\\[2ex]
&M^{(\sigma)}_E=
(12 c-v+4) \ell_0^0(\tilde m_\sigma^2)
+\frac{15}{4} v (-80 c+5 v-24) \ell_1^4 (\tilde m_\sigma^2)
+13 (120 c-7 v+35) \ell_1^6 (\tilde m_\sigma^2) \nn
&\qquad\qquad
-\frac{20}{21} (183 c-10 d) (126 c-7 v+36) \ell_1^8(\tilde m_\sigma^2)
+\frac{15}{14} v (224 c-12 v+63) \ell_2^{10} (\tilde m_\sigma^2)
+\frac{25}{12} (288 c-15 v+80) \ell_2^{12} (\tilde m_\sigma^2) \nn
&\qquad\qquad
+\frac{5}{21} v^2 (126 c-7 v+36) \ell_2^8 (\tilde m_\sigma^2)
+50 c v (288 c-15 v+80) \ell_2^{12} (\tilde m_\sigma^2)
+80 c (1080 c-55 v+297) \ell_2^{14} (\tilde m_\sigma^2)\nn
&\qquad\qquad
+\frac{181440}{11} c^2 (220 c-11 v+60) \ell_2^{16} (\tilde m_\sigma^2)
\nn
&\phantom{M^{(\sigma)}_E}
\equiv L_E^{(\sigma)}(\tilde m_\sigma^2)\,,\\[2ex]
&M^{(\sigma-t)}_E=
\frac{(126 c-7 v+36) }{7 (1+\tilde m_t^2)}\ell_1^8(\tilde m_\sigma^2)
+\frac{9 d (224 c-12 v+63) }{14 (1+\tilde m_t^2)}\ell_1^{10}(\tilde m_\sigma^2)
-\frac{250 (d-6 c)^2 (288 c-15 v+80) }{3 (1+\tilde m_t^2)}\ell_1^{12}(\tilde m_\sigma^2)\nn
&\qquad\qquad
-\frac{317}{6}\ell_0^0(0)\,.
\label{eq: MEsigma}
}
We also define the interpolating functions for the mixing effects as
\al{
&M_C^{(t-\sigma)}+ M_C^{(\sigma-t)}= -L_\text{mix}^{t-\sigma}(\tilde m_t^2,\tilde m_\sigma^2) \,,\\[2ex]
&M_D^{(t-\sigma)}+ M_D^{(\sigma-t)}= -10L_\text{mix}^{t-\sigma}(\tilde m_t^2,\tilde m_\sigma^2)\,,\\[2ex]
&M_E^{(t-\sigma)}+ M_E^{(\sigma-t)}= -30L_\text{mix}^{t-\sigma}(\tilde m_t^2,\tilde m_\sigma^2)\,.
\label{eq: mixing Me}
}
In Eq.~\eqref{eq: METT}, the Euler characteristic and the number of Killing vectors and conformal ones are denoted by $\chi_E$ and $N$, respectively.
For geometries with a topology of Euclidean flat space they are given by $\chi_E=1$, $N=d(d+1)/2$.

The expressions $(1+\tilde m_t^2)^{-1}$ and $(1+\tilde m_\sigma^2)^{-1}$, with $\tilde m_t^2=d-v$, $\tilde m_\sigma^2=3c-v/4$ as in Eq.~\eqref{eq: masses of t and sigma modes}, correspond to the regularized propagators of the $t$- and $\sigma$-modes, multiplied with $Dk^4$ and $3Ck^4$, respectively.
For the particular Litim-type regulator employed here the combination $P_k(\Delta_i)$ is effectively replaced by $k^2$.
By virtue of the gauge invariant formulation, or equivalent physical gauge fixing, the contribution of the different physical modes is well visible and separated from the sector of the gauge modes.
Despite the somewhat lengthy expressions the structure of the contributions to the flow generator is well visible.
The terms involving both factors of $(1+\tilde m_t^2)^{-1}$ and $(1+\tilde m_\sigma^2)^{-1}$ are due to the mixing of the $t$- and $\sigma$-modes for geometries that do not exhibit rotation symmetry. 

\begin{figure}
\includegraphics[width=6.2cm]{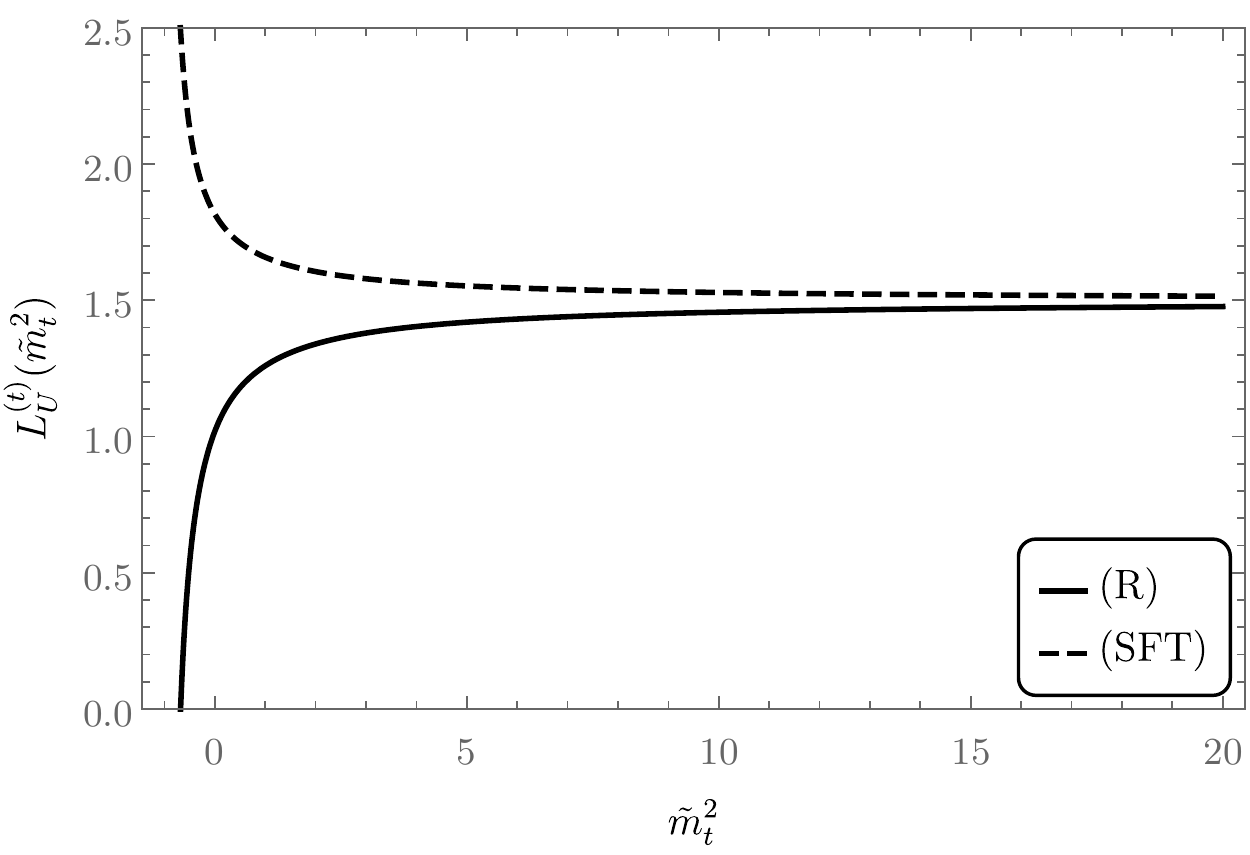}
\hspace{7ex}
\includegraphics[width=6.2cm]{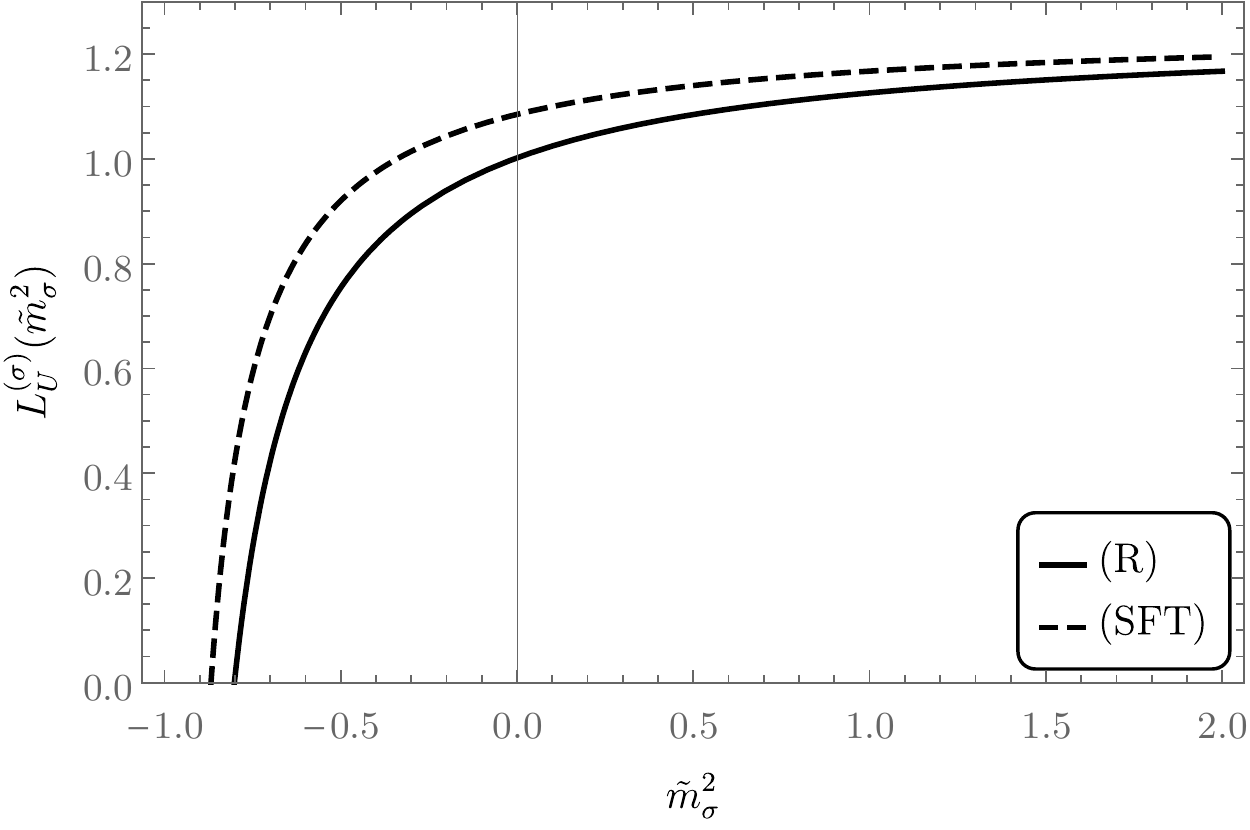}\\[2ex]
\includegraphics[width=6.2cm]{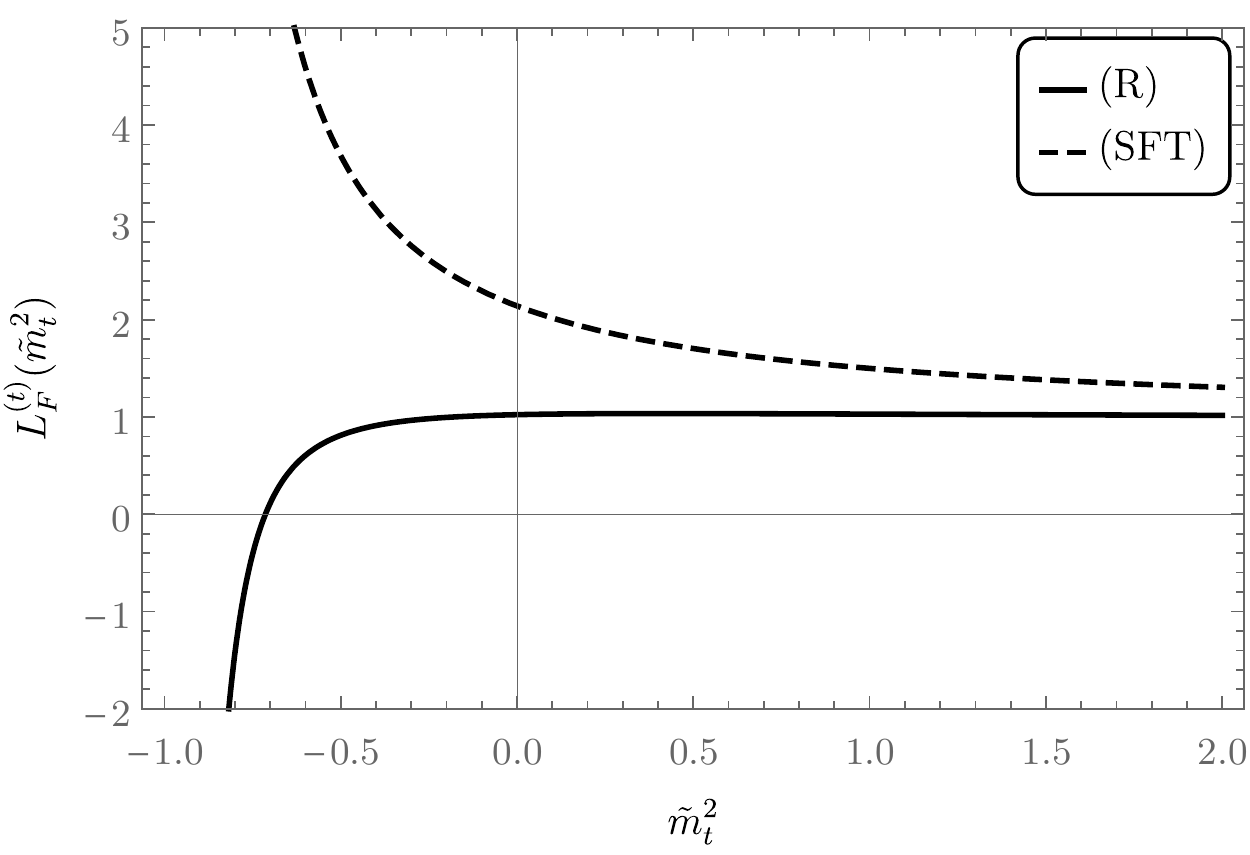}
\hspace{7ex}
\includegraphics[width=6.2cm]{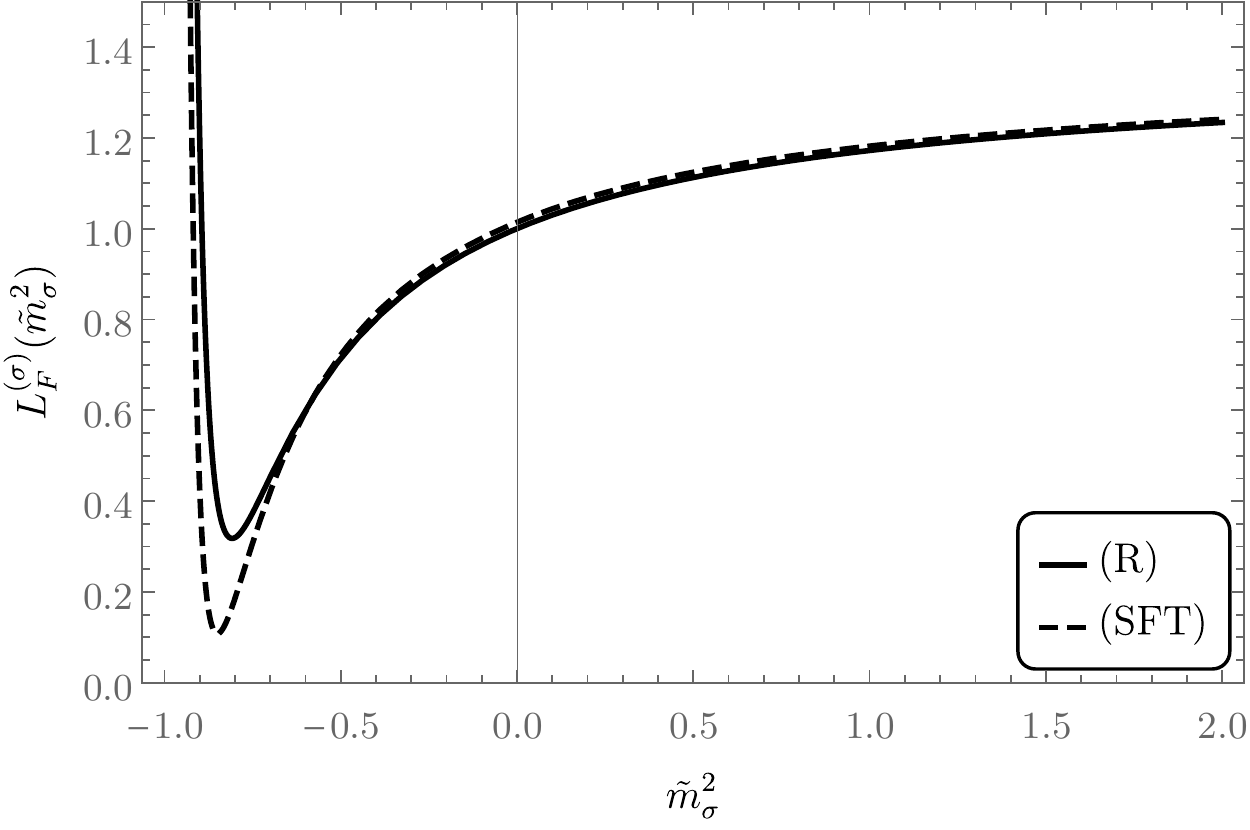}\\[2ex]
\caption{
Behavior of $L^{(t)}_i(\tilde m_t^2)$ and $L^{(\sigma)}_i(\tilde m_\sigma^2)$ as functions of the dimensionless masses $\tilde m_t^2$ and $\tilde m_\sigma^2$.
For $\omega$ and $v$ we display two cases: (R) $v=0.0128638$, $\omega=-23.1264$ and (SFT) $v=0.544229$,  $\omega=-0.0228639$.
}
\label{fig: LU LF} 
\end{figure}

\begin{figure}
\includegraphics[width=5cm]{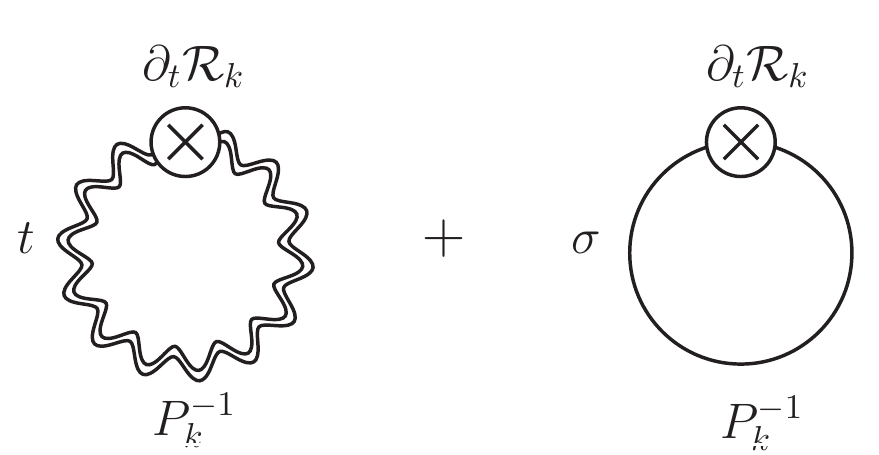}
\caption{
One loop diagram for the gravitational contribution to the flow of the effective potential.
}
\label{fig: one_loop diag} 
\end{figure}

The contributions from the $t-\sigma$-mixing correspond to $L_\text{mix}^{(t-\sigma)}$ in Eqs.~\eqref{eq: MC in summary}--\eqref{eq: ME in summary}, which can be read off directly from the explicit expressions in Eqs.~\eqref{eq: MCTT}--\eqref{eq: METT} and \eqref{eq: MCsigma}--\eqref{eq: MEsigma}.
The remaining parts in these equations define the interpolating functions $L_C^{(i)}$, $L_D^{(i)}$, $L_E^{(i)}$ that we show for particular values of the couplings in Fig.~\ref{fig: LC LD}.
Similarly, Eqs.~\eqref{eq: MUTT}, \eqref{eq: MFTT}, \eqref{eq: MUsigma}, \eqref{eq: MFsigma} define the interpolating functions $L_U^{(i)}$ and $L_F^{(i)}$ in Eqs.~\eqref{eq: MU in summery} and \eqref{eq: MF in summery}.
We show these interpolating functions in Fig.~\ref{fig: LU LF} for the same parameter sets as for Fig.~\ref{fig: LC LD}.
The flow generator for the effective scalar potential $\sim M_U$ is particularly simple and has a simple expression in terms of one loop diagrams in Fig.~\ref{fig: one_loop diag}. 
It can be equivalently evaluated in flat space, with propagators taken for constant scalar fields.
For the other expressions the $t$- and $\sigma$-propagators in Fig.~\ref{fig: one_loop diag} have to be evaluated in a curved background, and there is propagator mixing in the absence of rotation symmetry.
A perturbative computation would expand the propagators in deviations from flat space, introducing external legs for metric fluctuations in Fig.~\ref{fig: one_loop diag}.
The increasing number of legs needed for the flow of higher derivative couplings is directly related to the increasing complexity of the expressions in Eqs.~\eqref{eq: MUTT}--\eqref{eq: MEsigma}.
A Taylor expansion of the functional flow equation in the number of external fluctuation fields produces the vertex expansion for the flow equations, which has been investigated extensively for quantum gravity~\cite{Christiansen:2012rx,Christiansen:2014raa,Christiansen:2015rva,Christiansen:2016sjn,Denz:2016qks,Christiansen:2017cxa,Christiansen:2017bsy,Eichhorn:2018akn,Eichhorn:2018ydy,Bonanno:2021squ}.
In comparison to this work the gauge invariant flow equation has additional contributions from the field dependence of the infrared regulator and requires a particular physical gauge fixing.
A more detailed discussion of these issues can be found in Ref.~\cite{Wetterich:2017ixo,Wetterich:2018qsl}. 


\subsection{Measure contribution}
The measure contribution includes the spin-1 gauge modes $a_\mu$ in the metric fluctuations $(a_{\mu\nu}= D_\mu a_\nu + D_\nu a_\mu)$ and the ghost modes $C_\mu$, $\bar C_\mu$, as well as the Jacobian arising from the decomposition \eqref{physical metric decomposition}.
The gauge invariant formulation tells us that there is a simple relation~\cite{Wetterich:2016ewc,Wetterich:2017aoy} between the metric fluctuations $\delta_k^{(g)}$ and the ghost ones $\epsilon_k^{(g)}$ such that $\delta_k^{(g)}=2\epsilon_k^{(g)}$.
Thanks to these relations, the total measure contribution takes the simple form 
\al{
\eta_k^{(g)}=
-\delta_k^{(g)}
=-\frac{1}{2}\text{Tr}_{(1)}\frac{\p_t P_k({\mathcal D}_1)}{P_k({\mathcal D}_1)}\,,
 \label{eq: measure contributions}
}
with the differential operator acting on a vector field,
 \al{
 \label{Eq: differential operators for measure contributions}
\left({\mathcal D}_1\right)^\mu{}_\nu=\Delta_V \delta^\mu_\nu -D^\mu D_\nu -R^\mu{}_{\nu}\,.
}
Here $\Delta_V=-D^2$ is the Laplacian acting on vector fields.
In Eq.~\eqref{eq: measure contributions}, we employ a regulator which replaces ${\mathcal D}_1$ by $P_k$. 
More explicit calculations using the heat kernel method are presented in Appendix~\ref{App: measure contributions}.
Here we show only the result:
\al{\label{eq: result of measure contributions}
\delta^{(g)}_k&=\frac{1}{16\pi^2} \int_x\sqrt{g} \left[ \frac{13}{4}k^4 \ell_0^4\fn{0}  + \frac{29}{24} k^2 \ell_0^2\fn{0}{R} + \left(  \frac{23}{144}{R}^2+ \frac{67}{360}{R}_{\mu\nu}{R}^{\mu\nu}  -\frac{11}{180}{R}_{\mu\nu\rho\sigma}{R}^{\mu\nu\rho\sigma} \right)\ell_0^0\fn{0}  \right]\,.
}
As an important advantage of the gauge invariant formulation the gauge sector decouples from the physical sector.
As a consequence, this measure contribution is universal in the sense that it does not depend on the couplings in the physical sector.
The expression \eqref{eq: result of measure contributions} does not involve the coupling function $u$, $w$, $c$, $d$.

\section{R-fixed point and critical exponents}
\label{sec: UV fixed point and critical exponent}
The fixed points of the flow equations~\eqref{eq: beta function of U}--\eqref{eq: beta function of E} obtain by setting the terms $\sim \p_t$ and $\sim \tilde \rho\p_{\tilde \rho}$ to zero.
For pure gravity, with $N_S=N_F=N_V=0$, the fixed point conditions are
\al{
&M_C=M_D=0\,,&
&u=\frac{M_U}{128\pi^2}\,,&
&w=-\frac{M_F}{192\pi^2}\,,&
&v=-\frac{3M_U}{2M_F}\,.
}
These are four non-linear equations for the four couplings $u$, $v$, $c$ and $d$ that we have solved numerically. The numerical investigation finds several fixed points.
Part of them are outside the validity of our truncation, occurring for example at $\tilde m_t^2<-1$.
We have selected the R-fixed point \eqref{eq: full fixed point value}.
We will next argue that this fixed point corresponds to the fixed point of asymptotic safety found earlier for other truncations of the effective average action.
For this purpose we compute these truncations in our setting for the gauge invariant flow equation and the particular infrared regulator employed in this paper.
This allows for comparison with earlier results, and shows how the R-fixed point depends on the truncation.

\subsection{Einstein-Hilbert truncation}
We first show the result of the Einstein-Hilbert truncation ($C=D=E=0$) in our scheme without matter effects $N_S=N_V=N_F=0$. 
Solving $\beta_U=\beta_F=0$, we find a non-trivial fixed point
\al{
&u_*=0.0061\,,&
&w_*=0.022\,,&
&v_*=0.279 \,,&
&\tilde m_{t*}^2 = -0.28\,,&
&\tilde m_{\sigma*}^2= -0.070\,,
\label{eq: EH truncation non-trivial fixed point}
}
at which the critical exponents read
\al{
\theta_{1,2}=2.63 \pm 1.39i\,.
}
These values are in a range found for the R-fixed point within earlier truncations. 
The value for $w_*$ is similar to Eq.~\eqref{eq: full fixed point value}, while $u_*$ is substantially larger.
This is due to the difference of the values for $\tilde m_t^2$ and $\tilde m_\sigma^2$ between Eqs.~\eqref{eq: full fixed point value} and \eqref{eq: EH truncation non-trivial fixed point}.
For the values in Eq.~\eqref{eq: EH truncation non-trivial fixed point} the $\beta$-function for $D$ remains positive, ``destabilizing" the fixed point \eqref{eq: EH truncation non-trivial fixed point} once the higher derivative terms are included.
As we have argued before, substantially more negative values for $\tilde m_t^2$ are needed in order to find a fixed point for $D$. This is the main reason for the shift in the fixed point values between Eqs.~\eqref{eq: full fixed point value} and \eqref{eq: EH truncation non-trivial fixed point}.
The fixed point $w_*$ in Eq.~\eqref{eq: EH truncation non-trivial fixed point} corresponds to the dimensionless Newton constant $g_{N*}=1/(16\pi w_*)=0.92$.

Let us finally look at the flow equation for $E$ evaluated for the fixed point \eqref{eq: EH truncation non-trivial fixed point}. 
Since the Gauss-Bonnet term is topological, its coefficient $E$ does not contribute to any beta functions, whereas the beta function of $E$ depends on other couplings. It takes into account the number of (normal and conformal) Killing vectors, and therefore monitors topological features of the background geometry. We find a vanishing $\beta_E$ for $N_G=3.1\chi_E$, but we will not impose $\beta_E=0$ for the search of fixed points.

\subsection{$R^2$ truncation}
Next, we analyze the $R^2$ truncation for pure gravity. The earlier works on that truncation have been done in the Einstein spacetime $R_{\mu\nu}=(R/4) g_{\mu\nu}$~\cite{Benedetti:2009rx,Benedetti:2009gn,Hamada:2017rvn} or the maximally symmetric spacetime $R_{\mu\nu\rho\sigma}=(R/12)(g_{\mu\rho}g_{\nu\sigma} - g_{\mu\sigma}g_{\nu\rho})$~\cite{Falls:2017lst,deBrito:2018jxt,Falls:2018ylp,Kluth:2020bdv}.
In the present work, we do not assume such a special spacetime background. We first consider the $R^2$ truncation defined by setting $D=E=0$ and then solving $\beta_U=\beta_F=\beta_C=0$.
We find a fixed point for
\al{
&u_*= 0.00018\,,&
&w_*=0.0095\,,&
&C_*=-0.00292\,,
&v_*= 0.0187\,,&
&c_* = -0.307\,,\nn[1ex]
&&&\tilde m_{t*}^2=-0.0187\,,&&
&\tilde m_{\sigma*}^2 = -0.93\,,
\label{eq: FP in R2}
}
at which we have the critical exponents
\al{
&\theta_{1,2}=2.48 \pm 0.342i\,,&
&\theta_3=306.6\,.
\label{eq: CE in R2}
}
The critical exponents for $F$ and $U$ are close to those of the Einstein-Hilbert truncation, whereas the $R^2$ coupling has a huge value of the critical exponent. Such a situation has been actually seen in the previous works and indicates the insufficiency of the truncation. By improving the truncation, i.e. including higher order operators such as $R^3$, $R^4$, etc., the critical exponents converge to reasonable values~\cite{Falls:2018ylp,Kluth:2020bdv}.

As an alternative $R^2$-truncation, we can look at the flow of the linear combination $\tilde C=C+D/6$, setting $C=\tilde C$ for the flow generators.
This corresponds to the previous investigations for Einstein spaces. For this procedure we find a fixed point at
\al{
u_*=0.00098\,,&
&w_*=0.011\,,&
&C_*=-0.00342\,,&
&v_*=0.0859\,,&
&c_*=-0.300\,.
}
This fixed point value gives the critical exponents
\al{
&\theta_{1,2}=2.47 \pm 0.709i\,,&
&\theta_3=188.4\,.
}
Comparison with Eq.~\eqref{eq: FP in R2} demonstrates that the fixed point values depend substantially on the choice of the precise truncation, while the first two critical exponents are more robust.
The extraction of the flow of $C$ is ambiguous and needs a specification of assumptions on the ratio $D/C$. 

\subsection{Vanishing cosmological constant}
Another interesting approximation or truncation neglects the scalar potential or cosmological constant.
A numerical search for simultaneous zeros of the beta functions $\beta_F$, $\beta_C$, $\beta_D$ for $u=v=0$ and $N_S=N_V=N_F=0$ shows three fixed points
\al{
&\text{FP}_1:~~w_*=0.0198\,,&
&C_*=0.174\,,&
&D_*=-0.012\,,&
&c_*=8.82 \,,&
&d_*=-0.605 \,,\nn[1ex]
&&&\tilde m_{t*}^2=-0.605\,,&&&
&\tilde m_{\sigma*}^2 = 26.4\,,
\label{eq: u=0 fixed point}
\\[2ex]
&\text{FP}_2:~~w_*=0.0033\,,&
&C_*=-0.00101\,,&
&D_*=-0.00242\,,&
&c_*=-0.304 \,,&
&d_*=-0.724\,,\nn[1ex]
&&&\tilde m_{t*}^2=-0.725\,,&&&
&\tilde m_{\sigma*}^2 = -0.911\,,
\\[2ex]
&\text{FP}_3:~~w_*=0.0033\,,&
&C_*=0.00085\,,&
&D_*=-0.0022\,,&
&c_*=0.259\,,&
&d_*=-0.674 \,,\nn[1ex]
&&&\tilde m_{t*}^2=-0.674\,,&&&
&\tilde m_{\sigma*}^2 = 0.778\,.
}
For these fixed points the critical exponents are found as
\al{
&\text{FP}_1:~~
\theta_1=2.10\,,&
&\theta_2=9.51\,,&
&\theta_3=-87.8\,,
\label{eq: critical exponent in u=0}
\\[2ex]
&\text{FP}_2:~~
\theta_1=2.33\,,&
&\theta_2=357.0\,,&
&\theta_3=-1700.2\,,\\[2ex]
&\text{FP}_3:~~
\theta_1=1.89\,,&
&\theta_2=-20.3\,,&
&\theta_3=-588.7\,.
}
The fixed point $\text{FP}_1$ is close to the R-fixed point in Eq.~\eqref{eq: full fixed point value}, while the critical exponents are sensitive to the omission of $u$.

\subsection{Full system}
\begin{figure}
\includegraphics[width=8.5cm]{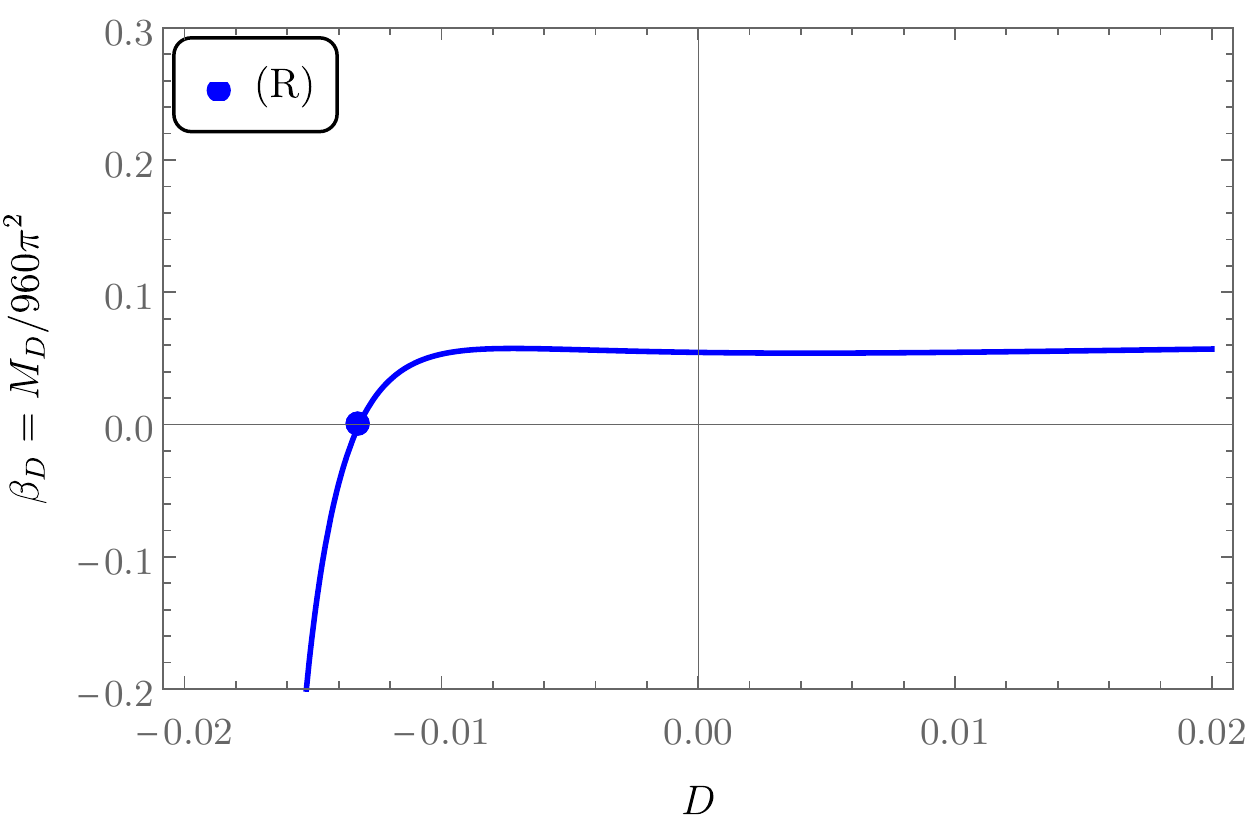}
\caption{
The behavior of the beta function of $D$ as a function of $D$. For the other couplings we use the values at the R-fixed point in Eq.~\eqref{eq: full fixed point value}. The blue point is just its fixed point value of $D_*$.
}
\label{fig: MD as a function of D} 
\end{figure}

For a numerical search of the fixed points for the full system, we look for solutions $\beta_U=\beta_F=\beta_C=\beta_D=0$.
This yields the R-fixed point in Eq.~\eqref{eq: full fixed point value}. While $\tilde m_t^2$ takes substantial negative values as necessary for a stop of the flow of $D$, there is no sign of a problematic behavior.
This is exemplified by the flow of $D$ for which we show the behavior of $\beta_D=M_D/960\pi^2$ as a function of $D$ in Fig.~\ref{fig: MD as a function of D}. For all couplings except for $D$, we use the fixed point value \eqref{eq: full fixed point value}. The location of the fixed point value $D_*$ is shown by a fat blue dot.
While the vicinity of the pole at $\tilde m_t^2=-1$ yields the needed substantial negative contributions to $\beta_D$, the effect is not dramatic. 
In the current setup we have found no other fixed point with $\tilde m_t^2>-1$.
The fixed points $\text{FP}_2$ and $\text{FP}_3$ in the truncation for $u=0$ may therefore be considered as artifacts due to the truncation.

\section{Infrared region}
\label{sec: Infrared region}
The coupling $w$ corresponds to a relevant parameter, both at the SFT- and R-fixed points.
Away from the fixed points $w$ typically increases and one enters the infrared region of large $w$.
In this region the Einstein-Hilbert action becomes a very good approximation.
We will discuss this issue by investigating the scaling solution in the region of large $\tilde \rho$.
Equivalently, the infrared region can be discussed at $\rho=0$ for the $k$-dependence of the coupling constants.

The scaling solution for large $\tilde \rho$ is characterized by $w\sim \tilde \rho$, i.e.
\al{
&w=\frac{1}{2}\xi_\infty \tilde \rho + w_\infty\,,&
&\xi_\infty>0\,,
}
with constant $\xi_\infty$ and $w_\infty$.
If for $\tilde \rho\to \infty$ the coupling functions $u$, $C$, $D$ remain finite or increase only slowly with $\tilde \rho$, the leading behavior in the infrared region is given by 
\al{
&v=0\,,&
&c=0\,,&
&d=0\,.
}
In this limit the gravitational contributions to the flow generators simplify considerably since $\tilde m_t^2$ and $\tilde m_\sigma^2$ vanish.
One finds (with $M_D^{(m)}$ the measure contribution) 
\al{
&M_D^{(t)}=\frac{5270}{9}\,,&
&M_D^{(\sigma)}=-\frac{19589}{2268}\,,&
&M_D^{(m)}=\frac{7}{2}\,,
}
such that $M_D$ and $\beta_D$ take constant values
\al{
&M_D=\frac{1316389}{2268}\approx 580.4\,,&
&\tilde \beta_D=2A_D=\frac{1316389}{2177280\pi^2}\approx 0.061\,.
}
Here $\tilde \beta_i$ obtains from $\beta_i$ by omitting the term $\sim \tilde\rho \p_{\tilde\rho}$.
Similarly, one obtains
\al{
&M_C^{(t)}=\frac{1574}{9}\,,&
&M_C^{(\sigma)}=\frac{12487}{2268}\,,&
&M_C^{(m)}=-\frac{29}{2}\,,
\nonumber
}
\al{
&M_C=\frac{376249}{2268}\approx 165.9\,,&
&\tilde\beta_C=-2A_C=-\frac{376249}{1306368\pi^2}\approx -0.029\,.
\label{eq: IR values of C}
}
and
\al{
&M_F=-\frac{361}{14}\approx 25.8\,,&
&M_U=\frac{301}{60}\approx 5.02\,,\nn[2ex]
&\tilde\beta_F+2w=2w_\infty=\frac{361}{1344\pi^2}\approx 0.027\,,&
&\tilde\beta_U+4u=4u_\infty=\frac{301}{1920\pi^2}\approx 0.016\,.
}

The scaling solution exists for a constant value of $u=u_\infty$, such that $v=u/w\sim \tilde\rho^{-1}$ vanishes indeed for $\tilde\rho\to\infty$. The coupling $C$ increases logarithmically,
\al{
C(\tilde\rho) = \bar C + A_C \ln \tilde\rho\,,
\label{eq: asymptotic behavior of C}
}
with constant $\bar C$ adjusted such that the asymptotic behavior \eqref{eq: asymptotic behavior of C} for large $\tilde \rho$ matches the behavior at the SFT-fixed point for $\tilde\rho\to 0$.
This increase is rather slow, according to the value $A_C\approx 0.015$ from Eq.~\eqref{eq: IR values of C}.
Similarly, the coupling function $D(\tilde\rho)$ for the squared Weyl tensor decreases logarithmically ($A_D\approx 0.031 >0$)
\al{
D(\tilde\rho) = \bar D - A_D \ln \tilde\rho\,.
\label{eq: asymptotic behavior of D}
} 
For $\tilde\rho\to \infty$ both $c$ and $d$ vanish $\sim \ln \tilde\rho/\tilde\rho$.

With $F=\xi_\infty \varphi^2 + 2w_\infty k^2$ the effective Planck mass depends on the scalar field $\varphi$.
For the cosmology of this type of models it becomes dynamical. For typical solutions $\varphi$ and $F$ diverge for infinitely increasing time~\cite{Wetterich:1987fm,Wetterich:2014gaa}.
The observed Planck mass can be associated with the present value of the scalar field, $M^2 =  F(t_0) =\xi_\infty \varphi^2(t_0)$.
If we choose the renormalization scale $k$ such that $u_\infty k^4$ coincides roughly with the present dark energy density, $u_\infty k^4 \approx (2\cdot 10^{-3}\,\text{eV})^4$, the increasing scalar field has reached today a large value $\varphi^2(t_0)/k^2=\tilde\rho(t_0)\approx 10^{60}$. 
This large value is connected to the huge age of the universe in Planck units. The infrared limit for $\tilde\rho\to \infty$ becomes a very good approximation.

For processes involving finite length scales the associated non-zero momenta typically act as an additional infrared cutoff, replacing effectively in the logarithms of Eqs.~\eqref{eq: asymptotic behavior of C} and \eqref{eq: asymptotic behavior of D}, $\ln\tilde\rho\to \ln(\varphi^2/(k^2+q^2))$, resulting in
\al{
D=\bar D' + A_D \ln\left( \frac{k^2 +q^2}{M^2}\right)\,.
\label{eq: modified asymptotic behavior of D}
}
The corrections of higher derivative gravity to Einstein's gravity are suppressed by $DC_{\mu\nu\rho\sigma}C^{\mu\nu\rho\sigma}/M^2 R$ or $CR/M^2$.
Associating $R\sim q^2$, $C_{\mu\nu\rho\sigma}C^{\mu\nu\rho\sigma}\sim q^4$, and using $M^2/q^2\sim wk^2/q^2$, the suppression factor $Dq^2/M^2 \approx (D/w)(q^2/k^2)=dq^2/k^2\approx 10^{-30}D q^2/k^2$ is tiny for all momenta sufficiently below the Planck mass.
Modifications of the ghost pole for the graviton propagator in higher derivative gravity due to the logarithmic running of $D$ in Eq.~\eqref{eq: modified asymptotic behavior of D} have been discussed in Ref.~\cite{Wetterich:2019qzx}.
Instead of a ghost pole on the real axis for $q^2$ there is a pair of poles in the complex plane.
It is not clear if this is compatible with an acceptable graviton propagator.

Leaving the issue of the graviton propagator open, a model of quantum gravity based on a scaling solution with a flow in $\tilde\rho$ from the SFT-fixed point for $\tilde\rho\to 0$ to the infrared limit for $\tilde\rho\to \infty$ seems acceptable. 
This would realize ``fundamental scale invariance"~\cite{Wetterich:2020cxq}. The observable gravitational interactions on length scales smaller than the present horizon would agree with the predictions of Einstein gravity if the matter sector is also characterized by a scaling solution, with all particle masses proportional to the scalar field $\varphi$ and $\varphi$-independent dimensionless couplings.
Observable mass ratios are then time independent despite a cosmological time evolution of $\varphi$. On cosmological scales the scaling solution with $u=u_\infty$, $w\sim \tilde\rho$ gives rise to dynamical dark energy~\cite{Wetterich:1987fm}.
By a Weyl scaling to the Einstein frame the effective Planck mass and particle masses become constants and all dependence on the renormalization scale $k$ is eliminated~\cite{Wetterich:2020cxq}.

\section{Conclusions}
\label{sec: Conclusions}
Within functional renormalization we have computed the flow of higher derivative gravity with up to four derivatives of the metric.
We have coped with this technical challenge by use of the gauge invariant flow equation (or equivalent physical gauge fixing).
This has permitted us a decomposition into fluctuation modes for which the individual contributions can be separated, helping for a qualitative understanding of different effects.

In the double limit $C^{-1} \to 0$, $D^{-1}\to 0$ we recover the perturbative $\beta$-functions and the associated asymptotic freedom at the SFT-fixed point.
According to the perturbative $\beta$-functions the coupling $D^{-1}$ increases as the renormalization scale $k$ is lowered and runs outside the perturbative domain.
The non-perturbative character of the functional flow equation allows us to follow the flow of $D^{-1}$ and $C^{-1}$ in the non-perturbative domain where $D$ and $C$ are no longer large. We find that for $D>0$ nothing stops the decrease of $D$ with decreasing $k$. As a consequence, $D$ reaches zero at some $k_0$.

For negative $D$ the sign of the $\beta$-function for $D$ can change. The zero of $\beta_D$ at $D<0$ permits an additional fixed point with small finite $D_*<0$. This fixed point lies outside the perturbative domain of higher derivative gravity.
At the fixed point also $C$, the dimensionless effective Planck mass $w$ and the dimensionless effective cosmological constant $u$ take fixed values.
This additional R-fixed point can be associated with asymptotic safety.
It is possible to formulate a consistent quantum field theory for gravity based on the R-fixed point.

The negative value of $D$, as well as the effective mass term $\tilde m_t^2$ being not very far from the breakdown of the truncation at $\tilde m_t^2=-1$, may cast same doubts on the reliability of our truncation.
An extended truncation would be much welcome for a test of robustness. On the other hand, investigations of the momentum dependence of the graviton propagator by different functional renormalization approaches~\cite{Denz:2016qks,Bonanno:2021squ} suggest that the R-fixed point persists for an arbitrary momentum dependence of the inverse graviton propagator beyond the terms in fourth order in momentum included in the present paper.
It is an interesting question if a critical flow trajectory connects the SFT- and R-fixed points.

The existence of the R-fixed point seems not mandatory for a consistent quantum field theory of metric gravity.
The microscopic theory may be defined as well at the SFT-fixed point and be characterized by asymptotic freedom.
The trajectories away from the SFT-fixed point may join directly the ``infrared region" for which an effective theory of gravity based on the Einstein-Hilbert action becomes a very good approximation.

It is possible that the effective action for quantum gravity is described by a scaling solution for which the functions $u$, $w$, $D$ and $C$ depend on the dimensionless scalar invariant $\tilde \rho$, without explicit dependence on $k$.
In this case, the limit $\tilde \rho\to \infty$ defines an infrared fixed point, and flow trajectories could connect directly the SFT- and infrared fixed points. The behavior of this scaling solution for large $\tilde \rho$ is given by
\al{
&w= \frac{1}{2}\xi_\infty \tilde \rho + w_\infty\,,&
&u=u_\infty\,, 
}
with constant $\xi_\infty$ and $u_\infty$.
While $C$ and $D$ depend logarithmically on $\tilde \rho$ for large $\tilde \rho$
\al{
&C=\bar C + A_C \ln \tilde \rho\,,&
&D=\bar D -A_D \ln \tilde \rho\,,
}
the ratios $d=D/w$, $c=C/w$ vanish $\sim \ln \tilde \rho/\tilde \rho$.
These ratios are the relevant quantities to measure the deviations from an effective action based on the Einstein-Hilbert truncation.

We may replace for the logarithmic flow of $C$ and $D$ the renormalization scale $k^2$ by $k^2+q^2$, with $q^2$ a typical squared momentum $q^2$ or corresponding covariant (negative) Laplacian $\Delta=-D^2$.
Taking $\tilde \rho=\varphi^2/k^2$ with $\varphi$ a scalar singlet field, the gravitational effective action according to the scaling solution is given for $q^2\ll \varphi^2$ and $k^2 \ll \varphi^2$ by
\al{
\Gamma &= \int_x\sqrt{g} \left[ u_\infty k^4 - \left( \frac{\xi_\infty}{2}\varphi^2 + w_\infty k^2 \right) R 
 -\frac{1}{2}R\left( \bar C + A_C\ln \frac{\varphi^2}{\Delta+k^2} \right) R
 +\frac{1}{2} C_{\mu\nu\rho\sigma} \left( \bar D - A_D\ln \frac{\varphi^2}{\Delta+k^2} \right)  C^{\mu\nu\rho\sigma} \right]\,.
}
Together with a suitable kinetic term for $\varphi$, and particle masses $\sim \varphi$ as appropriate for the quantum scale symmetry at an IR-fixed point, this effective action seems compatible with observation.
The scale $k$ is the only overall scale and we may choose it as $k\approx 2\cdot 10^{-3}$\,eV.
The present effective value of the Planck mass is $\sim \xi_\infty^{1/2}\varphi$.
It is dynamical and reaches for present cosmology the value $2\cdot 10^{18}$\,GeV, such that $k^2/\varphi^2\approx 10^{-60}$ explains the weakness of gravity. The scalar field $\varphi$ can be associated with the cosmon, providing for dynamical dark energy.

\begin{center}
{\bf Acknowledgements}
\end{center}
We thank Benjamin Knorr and Jan M. Pawlowski  for valuable discussions.
This work is supported by the DFG Collaborative Research Centre ``SFB 1225 (ISOQUANT)" and Germany’s
Excellence Strategy EXC-2181/1-390900948 (the Heidelberg Excellence Cluster STRUCTURES).
The work of M.\,Y. is supported by the Alexander von Humboldt Foundation. 

\vspace{2ex}
\noindent {\bf Note added.}~~After submitting our paper on arXiv, a recent paper~\cite{Barvinsky:2021ijq} has computed the heat kernel coefficients for a more general non-minimal forth-order derivative operator, $D^4+\hat \Omega^{abc}D_aD_bD_c+\hat D^{ab}D_a D_b+H^aD_a+\hat P$. Because gravitational systems with higher derivative operators could contain such a type of derivative in the two-point functions, those results may make computations simple and then may be quite useful for investigations of asymptotically safe gravity and its extended systems.

\appendix

\section{Setup}
\label{App: setups}
Our main tool in this work is the functional renormalization group~\cite{Wetterich:1992yh,Morris:1993qb,Reuter:1993kw,Ellwanger:1993mw}.
A central object is the scale-dependent 1PI effective action (or effective average action) $\Gamma_k$.
For $k\to0$, one obtains the full 1PI effective action $\Gamma_{k=0}=\Gamma$.
The scale change of $\Gamma_k$ is described by the following functional differential equation~\cite{Wetterich:1992yh}:
\al{
\label{App: flow equation}
\p_t \Gamma_k=\frac{1}{2}\Tr\left[ \left( \Gamma_k^{(2)}+{\mathcal R}_k \right)^{-1}\p_t {\mathcal R}_k\right]\,.
}
Here, ${\mathcal R}_k$ is a regulator which suppress low momentum modes with $|p|<k$ such that high momentum modes with $|p|>k$ are integrated out, $\p_t = k\p_k$ is the dimensionless scale derivative, and $\Gamma_k^{(2)}$ is the Hessian, i.e.  the full two-point function defined by the second-order functional derivative with respect to field variables.
Tr denotes the functional trace acting on all internal spaces of fields such as momenta (eigenvalues of the covariant derivative), flavor and color. 
See Refs.~\cite{Morris:1998da,Berges:2000ew,Aoki:2000wm,Bagnuls:2000ae,%
  Polonyi:2001se,Pawlowski:2005xe,Gies:2006wv,Delamotte:2007pf,%
  Rosten:2010vm,Kopietz:2010zz,Braun:2011pp,Dupuis:2020fhh} on the derivation and technical aspects of the flow equation \eqref{App: flow equation} and its applications to various systems.
  
Although the flow equation \eqref{App: flow equation} is exact, namely it is derived without any approximations, one needs to make approximations to solve Eq.~\eqref{App: flow equation} even for a simple system. In general, the effective average action $\Gamma_k$ includes an infinite number of effective operators generated by quantum effects. Therefore, approximations can be made by restricting an infinite-dimensional theory space into a finite subspace.

In this appendix, we explain our setups and techniques to derive the flow generators or beta functions for the coupling functions in a derivative expansion for quantum gravity in fourth order in derivatives. 
Firstly, we give the specific form of the effective average action for the gravitational system in order to investigate the asymptotic freedom and asymptotic safety scenarios for quantum gravity. Secondly, properties of the decomposed metric fluctuation field are summarized. Thirdly, we list useful identities for the covariant derivatives acting on various spin fields which are used to evaluate the trace for spacetime indices in the flow equation \eqref{App: flow equation}. Finally, we show the expanded projectors into polynomials of curvature invariants. The decomposition of the metric fluctuation field is realized by acting appropriate projectors on the metric fluctuation field. Those involve generally an infinite number of the covariant derivatives and curvature operators. For our purpose in this work, it is convenient to expand the projectors in polynomials of curvature invariants up to its squared forms.
 
\subsection{Effective average action}
The starting point to derive the flow equations for the gravitational system is to split the metric field $g_{\mu\nu}$ into a background field $\bar g_{\mu\nu}$  and a fluctuation fields $h_{\mu\nu}$:
\al{
g_{\mu\nu} =\bar g_{\mu\nu }+ h_{\mu\nu}\,.
}
Hereafter, quantities, variables and operators contracted by the background field are presented by a bar on them.

The effective average action as a truncated system reads
\al{
\Gamma_k= \Gamma_k^\text{matter}+ \Gamma_k^\text{gravity} + \Gamma_\text{gf} +\Gamma_\text{gh}\,.
}
Here the action for the gravity part is parametrized with
\al{
  \Gamma_k^\text{gravity} &=  \int d^4 x\, \sqrt{g}
  \left[ U\fn{\rho} -\frac{F\fn{\rho}}{2} R +  \frac{\mathcal C\fn{\rho}}{2}R^2 + \frac{\mathcal D\fn{\rho}}{2}R_{\mu\nu}R^{\mu\nu} +{\mathcal E\fn{\rho}} R_{\mu\nu\rho\sigma}R^{\mu\nu\rho\sigma} \right]\nn[2ex]
  &=\int d^4 x\, \sqrt{g}
  \left[U\fn{\rho} -\frac{F\fn{\rho}}{2} R - \frac{C\fn{\rho}}{2}R^2 + \frac{D\fn{\rho}}{2} C_{\mu\nu\rho\sigma}C^{\mu\nu\rho\sigma} + E\fn{\rho}G_4 \right]\,  ,\label{action for higher derivative}
}
where the Gauss-Bonnet term $G_4=R^2-4R_{\mu\nu}R^{\mu\nu}+R_{\mu\nu\rho\sigma}R^{\mu\nu\rho\sigma}$ is topological in four dimensional space, and $C_{\mu\nu\rho\sigma}C^{\mu\nu\rho\sigma}=G_4+2R_{\mu\nu}R^{\mu\nu}-2/3R^2=R_{\mu\nu\rho\sigma}^2-2R_{\mu\nu}^2+1/3R^2$ is the squared Weyl tensor.
The gauge and ghost actions for diffeomorphisms are given respectively by
\al{
  \Gamma_\text{gf} &= 
    \frac{1}{2\alpha}\int d^4x \sqrt{ \bar{g} } \, \bar g^{\mu\nu} \Sigma_\mu \Sigma_\nu \,,\label{standard gauge fixing}\\[2ex]
  \Gamma_\text{gh} &= -\int d^4x \sqrt{ \bar{g} } \,\bar C_\mu \left[ \bar g^{\mu\nu}\bar D^2 +\frac{1-\beta}{2}\bar D^\mu \bar D^\nu +\bar R^{\mu\nu}\right] C_\nu\,,
  \label{standard ghost action}
}
where $C_\mu$ and $\bar C_\mu$ are the ghost and anti-ghost fields and the gauge fixing function obeys
\al{
\Sigma_\mu&= \bar D^\nu h_{\nu\mu}-\frac{\beta+1}{4}\bar D_\mu h\,,
\label{Eq: gauge fixing function}
}
with $h=\bar g^{\mu\nu}h_{\mu\nu}$ the trace mode of the metric fluctuation.
The constants $\alpha$ and $\beta$ are the gauge fixing parameters.
In this work, we use the physical gauge fixing $\beta=-1$ and $\alpha\to 0$ for which the gauge fixing action \eqref{standard gauge fixing} deals with the path integral for the gauge field on the gauge orbit satisfying $\bar D^\nu h_{\nu\mu}=0$. The physical gauge fixing acts only on the gauge modes which are generated by the action of a gauge transformation on the background metric $\bar g_{\mu\nu}$.
For the matter part, we give the action for free $N_S$-scalars, $N_F$-Weyl fermions,  $N_V$-gauge bosons,
\al{
\Gamma_k^\text{matter}=
\int_x \sqrt{\bar g}\left[ 
\frac{1}{2}(\p_\mu \varphi)^2 + m^2 \varphi^2 \right]
+\frac{1}{4}\int_x\sqrt{\bar g}\,F_{\mu\nu}^aF^{a\mu\nu}+\Gamma_\text{gf}^{(V)}+\Gamma_\text{gh}^{(V)}
+\int_x \sqrt{\bar e}\left[ 
\bar\psi i\bar{\Slash D}\psi + y \varphi   \bar\psi \gamma^5\psi
\right]\,,
}
where $\sqrt{e}$ denotes the determinant of vierbein $e_\mu{}^a$.
The gauge fixing and the ghost action for the gauge symmetries read 
\al{
&\Gamma_\text{gf}^{(V)} =\frac{1}{2\alpha_V} \int \df^4 x \sqrt{\bar g}\, (\bar D_\mu A^\mu)^2\,,&
&\Gamma_\text{gh}^{(V)} = \int \df^4 x \sqrt{\bar g}\, \bar c\, \p_\mu \bar D^\mu c\,,
}
where $\alpha_V$ is the (dimensionless) gauge fixing parameter for the additional gauge symmetries in the matter sector beyond diffeomorphisms.

\subsection{Physical metric decomposition}
\label{App sec: Physical metric decomposition}
In general, the fluctuating metric $h_{\mu\nu}$ is a second rank symmetric tensor, namely it has 10 degrees of freedom in four dimensional spacetime.
This tensor can be decomposed into physical and gauge degrees of freedom.
In this work, we employ the physical metric decomposition~\cite{Wetterich:2016vxu} which is given by
\al{
\label{App: physical metric decomposition}
h_{\mu\nu}	&=	f_{\mu\nu} + a_{\mu\nu}\,,&
a_{\mu\nu}&=\bar D_\mu a_\nu +\bar D_\nu a_\nu \,,
} 
where $f_{\mu\nu}$ is the physical metric satisfying the transverse condition, i.e. $\bar D^\mu f_{\mu\nu}=0$, and $a_\mu$ is the spin-1 vector gauge mode.
Thus the physical metric $f_{\mu\nu}$ has 6 degrees of freedom, while there are 4 degrees of freedom in the gauge mode $a_\mu$, corresponding to infinitesimal diffeomorphism transformations.

We introduce the trace mode (the physical spin-0 scalar mode), $\sigma:={\bar g}^{\mu\nu}f_{\mu\nu}$, and split the physical metric fluctuations, 
\al{
&f_{\mu\nu}
= T_{\mu\nu}{}^{\rho\tau}f_{\rho\sigma}+ I_{\mu\nu}{}^{\rho\tau} f_{\rho\sigma}
=b_{\mu\nu} + \frac{1}{4}\bar g_{\mu\nu} \sigma\,.
}
The two parts read
\al{
T_{\mu\nu}{}^{\rho\tau}f_{\rho\tau}=(E_{\mu\nu}{}^{\rho\tau} - I_{\mu\nu}{}^{\rho\tau})f_{\rho\tau}
=f_{\mu\nu}-\frac{1}{4}g_{\mu\nu}\sigma =: b_{\mu\nu}\,,\qquad\quad
I_{\mu\nu}{}^{\rho\tau}f_{\rho\tau}=\frac{1}{4}g_{\mu\nu}\sigma\,,
\label{App: Decomposition no1}
}
where we have introduced the unit matrix acting on symmetric tensors and the projection tensors,
\al{
&E_{\mu\nu}{}^{\rho\tau}=\frac{1}{2}(\delta_\mu^\rho \delta_\nu^\tau +\delta_\mu^\tau \delta_\nu^\rho)\,,&
&T_{\mu\nu}{}^{\rho\tau}=E_{\mu\nu}{}^{\rho\tau}-I_{\mu\nu}{}^{\rho\tau}\,,&
&I_{\mu\nu}{}^{\rho\tau}=\frac{1}{4}g_{\mu\nu}g^{\rho\tau}\,.
\label{App: the symmetric unit matrix and the projection tensors}
}
The projection tensors satisfy the conditions as the orthogonal basis,
\al{
&T_{\mu\nu}{}^{\alpha\beta}T_{\alpha\beta}{}^{\rho\tau}=T_{\mu\nu}{}^{\rho\tau}\,,&
&I_{\mu\nu}{}^{\alpha\beta}I_{\alpha\beta}{}^{\rho\tau}=I_{\mu\nu}{}^{\rho\tau}\,,&
&T_{\mu\nu}{}^{\rho\tau}I_{\rho\tau}{}^{\alpha\beta}=0\,.
}
Using these projections, one can define the projection operators $P_{a}$ and $P_f$ on the gauge and physical modes,
\al{
\left( P_a h\right)_{\mu\nu} &= \left( \bar{\mathbb D}_1\right)_{\mu\nu}{}^\alpha \left({\bar{\mathcal D}_1}^{-1} \right)_\alpha{}^{\beta} (-\bar D^\gamma) E_{\gamma\beta}{}^{\rho\sigma}  h_{\rho\sigma}=\bar D_\mu a_\nu +\bar D_\nu a_\nu = a_{\mu\nu}\,, 
\label{App: projector for a}
\\[1ex]
\left( P_f h\right)_{\mu\nu} &= (E-P_a)_{\mu\nu}{}^{\rho\sigma}h_{\rho\sigma} = f_{\mu\nu}\,,
\label{App: projector for f}
}
where
\al{
&\left( \bar{\mathbb D}_1\right)_{\mu\nu}{}^\alpha=\bar D_{\mu}\delta^\alpha_\nu+\bar D_{\nu}\delta^\alpha_\mu\,,&
&\left({\bar{\mathcal D}_1}^{-1}\right)_\mu{}^\nu=\left(\delta_{\mu}^\nu \bar\Delta_V - \bar D_\mu \bar D^\nu -\bar R_{\mu}{}^\nu \right)^{-1}\,.
\label{App: differential operators in gauge mode}
}
These projectors satisfy
\al{
&(P_f)_{\mu\nu}{}^{\alpha\beta}(P_f)_{\alpha\beta}{}^{\rho\sigma}=(P_f)_{\mu\nu}{}^{\rho\sigma}\,,&
&(P_a)_{\mu\nu}{}^{\alpha\beta}(P_a)_{\alpha\beta}{}^{\rho\sigma}=(P_a)_{\mu\nu}{}^{\rho\sigma}\,,&
&(P_a)_{\mu\nu}{}^{\alpha\beta}(P_f)_{\alpha\beta}{}^{\rho\sigma}=0\,,
}
and, at the lowest order, 
\al{
&\tr P_a=4\,,&
& \tr P_f=6\,.
}
In Eq.~\eqref{App: differential operators in gauge mode} and the following we denote the Laplacian $\bar\Delta=-\bar D^\mu\bar D_\mu$ acting on scalars, vectors or tensors by $\bar\Delta_S$, $\bar\Delta_V$ and $\bar\Delta_T$, respectively.

The physical metric fluctuations and the gauge modes are further decomposed into irreducible representations of the Lorentz group,
\al{
f_{\mu\nu}&=t_{\mu\nu} + \hat S_{\mu\nu} \sigma=t_{\mu\nu}+\hat P_{\mu\nu}({\mathcal N}^{-1}\sigma)\,,&
a_{\mu\nu}&= \bar D_\mu \kappa_{\nu}	+\bar D _{\nu}\kappa_\mu +  \bar D_\mu \bar D_\nu u\,,
\label{eq: metric decomposition}
}
where $t_{\mu\nu}$ is the transverse-traceless (TT) tensor and $\kappa_\mu$ a transverse vector $\bar D^\mu \kappa_\mu=0$, while $\sigma(=\bar g^{\mu\nu}f_{\mu\nu})$ is the spin-0 scalar field defined above.
The projector on the physical fluctuations is defined by
\al{ 
\label{projector for spin 0}
&\hat S_{\mu\nu}=\left( \bar g_{\mu\nu} \bar\Delta_{S} + \bar D_\mu \bar D_\nu - \bar R_{\mu\nu}\right){\mathcal N}^{-1}  =\hat P_{\mu\nu}{\mathcal N}^{-1}\,, &
&\mathcal N=3\left(\bar\Delta_{S}-\frac{\bar R}{3}\right)\,.
}
This operator satisfies $\tr \hat S=\bar g^{\mu\nu} \hat S_{\mu\nu}=1$.
The decomposition \eqref{App: Decomposition no1} can be written as
\al{
f_{\mu\nu}=b_{\mu\nu} +\frac{1}{4}{\bar g}_{\mu\nu}\sigma =t_{\mu\nu} +\tilde s_{\mu\nu}\sigma + \frac{1}{4}{\bar g}_{\mu\nu}\sigma\,,
\label{App: decomposition of bmunu}
} 
where $\tilde s_{\mu\nu}$ satisfies the traceless condition, i.e.. ${\bar g}^{\mu\nu} \tilde s_{\mu\nu}=0$ and is explicitly given by
\al{
\tilde s_{\mu\nu}\sigma =\left[\left( \bar D_\mu \bar D_\nu -\frac{1}{4}\bar g_{\mu\nu}\bar D^2  \right) - \left( \bar R_{\mu\nu} -\frac{1}{4}{\bar g}_{\mu\nu} \bar R \right) \right] {\mathcal N}^{-1} \sigma
=: \left( \widetilde{\mathcal D}_{\mu\nu} - \widetilde R_{\mu\nu}  \right) {\mathcal N}^{-1}\sigma
=: {\tilde p}_{\mu\nu} {\mathcal N}^{-1}\sigma \,.
}

We will later need further identities for the physical scalar metric fluctuation $\sigma$, as
\al{
\hat S_{\mu\nu}\sigma
= \left(\tilde s_{\mu\nu} +\frac{1}{4}{\bar g}_{\mu\nu} \right)\sigma
= \left(\tilde p_{\mu\nu} + \frac{{\bar g}_{\mu\nu}}{4} {\mathcal N} \right){\mathcal N}^{-1} \sigma \,.
}
For an expansion linear in the curvature tensor one has
\al{
\widetilde{\mathcal D}_{\mu\nu}\sigma&= \frac{1}{4} \Big( \bar g_{\mu\nu}  + 4 \bar D_\mu\bar D_\mu (\bar\Delta_S)^{-1}  \Big)\bar\Delta_S\sigma \nn
&=\frac{1}{4} \left( \widetilde P_{\mu\nu} + 3\bar D_\mu (\bar\Delta_V)^{-1} \bar D_\nu \right)\bar\Delta_S\sigma 
-  \bar D_\mu (\bar\Delta_V)^{-2}\bar D_\alpha \bar R^\alpha{}_\nu \bar\Delta_S\sigma + \mathcal O(\bar R^2)\,,
}
and
\al{
\widetilde{\mathcal D}_{\mu\nu}\widetilde{\mathcal D}^{\mu\nu}\sigma
&= \frac{1}{4}\left(\bar\Delta_S \right){\mathcal N}\sigma + \bar R^{\mu\nu} \widetilde{\mathcal D}_{\mu\nu}\sigma\,,
}
where we define the transverse projector,
\al{
\widetilde P_{\mu\nu}=\bar g_{\mu\nu} + \bar D_\mu (\bar \Delta_V)^{-1} \bar D_\nu\,.
\label{App: transverse projection operator}
}

The projection operators which project out $t_{\mu\nu}$ and $\hat S_{\mu\nu}\sigma$ from $h_{\mu\nu}$ are
\al{
(P_t)_{\mu\nu}{}^{\rho\sigma}&= \left(E_{\mu\nu}{}^{\alpha\beta}-\hat S_{\mu\nu}\bar g^{\alpha \beta} \right) (P_f)_{\alpha\beta}{}^{\rho\sigma}
= \left(E_{\mu\nu}{}^{\alpha\beta}  -  I_{\mu\nu}{}^{\alpha \beta} - \tilde s_{\mu\nu} \bar g^{\alpha \beta}  \right) (P_f)_{\alpha\beta}{}^{\rho\sigma}
\,,
\label{App: projector for t}
\\
(P_\sigma)_{\mu\nu}{}^{\rho\sigma} &= \hat S_{\mu\nu}\bar g^{\alpha\beta} (P_f)_{\alpha\beta}{}^{\rho\sigma}\,,
\label{App: projector for sigma}
}
where
\al{
&\tr P_t =(P_t)_{\mu\nu}{}^{\mu\nu} =5\,,&
&\tr P_\sigma = (P_\sigma)_{\mu\nu}{}^{\mu\nu} =1\,.
}

Under infinitesimal diffeomorphism transformations for the metric fluctuations, the TT tensor $t_{\mu\nu}$ and the physical spin-0 scalar field $\sigma$ are invariant, whereas the transverse vector field $\kappa_\mu$ and the gauge spin-0 scalar field $u$ are not.
Here explicitly the transformations of the infinitesimal diffeomorphisms metric fluctuations are given by
\al{
\label{App: infinitesimal diffeomorphisms}
h_{\mu\nu}\to h_{\mu\nu}+\bar D_\mu \xi_\nu+\bar D_\nu \xi_\mu\,,
}
where the gauge parameters $\xi_\mu$ are decomposed as $\xi_\mu=\xi_\mu^\perp +\bar D_\mu \xi$ with $\bar D^\mu \xi_\mu^\perp=0$ and $\bar D^\mu \bar D_\mu \xi=\bar D^\mu \xi_\mu$.
The physical metric fluctuations are invariant under the transformation \eqref{App: infinitesimal diffeomorphisms}, i.e., $t_{\mu\nu}\to t_{\mu\nu}$ and $\sigma \to \sigma$, whereas the gauge modes are transformed as $\kappa_\mu \to \kappa_\mu+\xi_\mu^\perp$ and $u\to u +2\xi$.
Hence, the physical metric fluctuations are gauge invariant.

\subsection{Identities for covariant derivatives}
We summarize some identities for covariant derivative which are needed to evaluate the flow generators.
We start with the commutator of two covariant derivatives acting on arbitrary tensor $\phi_{\alpha_1\alpha_2\ldots \alpha_n}$,
\al{
[\bar D_\mu, \bar D_\nu] \phi_{\alpha_1\alpha_2\ldots \alpha_n}=\sum_{i=1}^n \bar R_{\mu\nu\alpha_i}{}^\beta \phi_{\alpha_1\ldots \alpha_{i-1}\beta \alpha_{i+1}\ldots \alpha_n}\,.
}
From this, one obtains
\al{
[\bar D_\mu , \bar\Delta]\phi &= \bar R_\mu{}^\nu \bar D_\nu \phi\,,\\[1ex]
[\bar D_\mu , \bar\Delta]\phi_\rho &= \bar R_\mu{}^\nu \bar D_\nu \phi_\rho -2\bar R^\nu{}_\mu{}^\lambda{}_\rho \bar D_\nu \phi_\lambda -(\bar D_\nu \bar R^\nu{}_\mu{}^\lambda{}_\rho)\phi_\lambda\,,\\[1ex]
[\bar D_\mu , \bar\Delta]\phi_{\rho\sigma} &= \bar R_\mu{}^\nu \bar D_\nu \phi_{\rho\sigma} -4\bar R_\nu{}_\mu{}^\lambda{}_{(\rho} \bar D^\nu \phi_{\sigma)\lambda} -(\bar D_\nu \bar R^\nu{}_\mu{}^\lambda{}_{(\rho})\phi_{\sigma)\lambda}\,,
}
where $A_{(\mu}B_{\nu)}=(A_\mu B_\nu - B_\nu A_\mu)/2$.
In this work we assume covariantly constant background curvature and drop the term $\bar D_\nu \bar R^\nu{}_\mu{}^\lambda{}_\rho$.

From the identity
\al{
 [\bar\Delta, \bar D_\mu \bar D_\nu]\phi 
 = \left(  \bar D_\mu [\bar\Delta,  \bar D_\nu] +   [\bar\Delta, \bar D_\mu ]\bar D_\nu \right) \phi 
  =  \left( -  \delta^\alpha_\mu \bar R_\nu{}^\beta   - \bar R_\mu{}^\alpha \delta_\nu^\beta  + 2\bar R^\alpha{}_\mu{}^\beta{}_\nu \right)\bar D_\alpha \bar D_\beta \phi
  =: -\Phi_{\mu\nu}{}^{\alpha\beta}\bar D_\alpha \bar D_\beta \phi \,,
 \label{App: tensor of Phi}
}
one obtains
\al{
[\bar\Delta, \hat S_{\mu\nu}]\sigma
&=    -\Phi_{\mu\nu}{}^{\alpha\beta} \bar D_\alpha \bar D_\beta {\mathcal N}^{-1} \sigma \,,\\[2ex]
[\bar\Delta^2, \hat S_{\mu\nu}]\sigma
&=
\Phi_{\mu\nu}{}^{\alpha\beta}\Phi_{\alpha\beta}{}^{\rho\sigma}
  \bar D_\rho \bar D_\sigma{\mathcal N}^{-1}\sigma
 - 2\Phi_{\mu\nu}{}^{\alpha\beta}\bar D_\alpha \bar D_\beta  \bar\Delta{\mathcal N}^{-1} \sigma\,,
\\[2ex]
[\bar R^{\alpha\beta}\bar D_\alpha \bar D_\beta, \hat S_{\mu\nu}]\sigma
&= \bar R^{\alpha\beta} \left( 
\bar g_{\mu\nu}\Phi_{\alpha\beta}{}^{\rho\sigma}
+\delta_\mu^\rho\bar R_{\alpha\nu\beta}{}^\sigma   
  + \bar R_{\alpha\mu\beta}{}^\sigma \delta_\nu^\rho 
 - 2\delta_\alpha^\rho\bar R_{\mu\beta\nu}{}^\sigma  \right)\bar D_\rho \bar D_\sigma{\mathcal N}^{-1}\sigma\,.
}
The tensor $\Phi_{\mu\nu}{}^{\alpha\beta}$ satisfies
\al{
\bar g^{\mu\nu}\Phi_{\mu\nu}{}^{\alpha\beta}
= \bar R^{\alpha\beta} + \bar R^{\beta\alpha} -2\bar R^{\alpha\beta}=0\,.
}
Then, one can define useful Laplacians, called ``Lichnerowicz Laplacians"~\cite{2018GReGr..50..145L} acting on a spin-2 tensor field,
\al{
\bar\Delta_{L2}h_{\mu\nu} \equiv (\bar\Delta_{L2})_{\mu\nu}{}^{\rho\sigma}h_{\rho\sigma} 
&=-\bar D^2 E_{\mu\nu}{}^{\rho\sigma} \,h_{\rho\sigma} + \bar R_\mu{}^{\rho}\delta^{\sigma}_\nu h_{\rho\sigma}+\bar R_\nu{}^{\rho} \delta^\sigma_\mu h_{\rho\sigma}-2\bar R_{\mu}{}^{\rho}{}_\nu{}^{\sigma}h_{\rho \sigma}\nn
&=\left( \bar\Delta_T E_{\mu\nu}{}^{\rho\sigma} - \Phi_{\mu\nu}{}^{\rho\sigma}\right) h_{\rho\sigma} \,,
\label{App: Lichnerowicz Laplacian for spin-2}
}
which obeys
\al{
\bar\Delta_{L2}(\bar D_\mu \bar D_\nu \phi)&=\bar D_\mu \bar D_\nu \bar\Delta_{S}\phi\,,&
\bar\Delta_{L2}\bar g_{\mu\nu}\phi&=\bar g_{\mu\nu}\bar\Delta_{S}\phi\,,
\label{properties for Lichnerowicz Laplacians for spin 2}
}
while in a general background one has
\al{
\bar\Delta_{L2}(\bar R_{\mu\nu}\sigma)
&=\bar R_{\mu\nu}\bar\Delta_{S}\sigma 
-\Phi_{\mu\nu}{}^{\rho\sigma} \bar R_{\rho\sigma}\,,
}
and then
\al{
\bar\Delta_{L2}(\hat S_{\mu\nu}\sigma)&=\hat S_{\mu\nu}\bar\Delta_{S}\sigma + \Phi_{\mu\nu}{}^{\rho\sigma} \bar R_{\rho\sigma} {\mathcal N}^{-1}\sigma\,.
}

Now we see that
\al{
\int_x \sqrt{\bar g}\,\sigma\hat S^{\mu\nu}\bar\Delta_T \hat S_{\mu\nu}\sigma
&= \int_x \sqrt{\bar g}\,\sigma \left( \hat S^{\mu\nu}\hat S_{\mu\nu} \bar\Delta_S  - \hat S^{\mu\nu}\Phi_{\mu\nu}{}^{\alpha\beta} \bar D_\alpha \bar D_\beta {\mathcal N}^{-1}  \right) \sigma \nn
 &=\int_x \sqrt{\bar g}\,\sigma \left( \hat S^{\mu\nu}\hat S_{\mu\nu} \bar\Delta_S  - \hat S^{\mu\nu} \bar R_\mu{}^\alpha{}_\nu{}^\beta  \bar D_\alpha \bar D_\beta {\mathcal N}^{-1}  \right) \sigma\,,
}
where we have used $\bar D^\mu \hat S_{\mu\nu}=0$.

\subsection{Expansion of projectors}
To calculate the traces in the flow generators, we expand the projectors into polynomials of curvature operators.
In this work, we calculate contributions up to the quadratic order of the curvature invariants, i.e. $\bar R^2$, $\bar R_{\mu\nu}\bar R^{\mu\nu}$ and $\bar R_{\mu\nu\rho\sigma}\bar R^{\mu\nu\rho\sigma}$.
As one can see from Eq.~\eqref{App: tt mode interaction} and \eqref{App: sigma mode interaction}, the interaction parts involve at least one curvature scalar.
Therefore, it is enough to have the linear order of curvature operators.

We start by expanding the inverse of ${\bar{\mathcal D}_1}$ defined in Eq.~\eqref{App: differential operators in gauge mode}:
\al{
\left({\bar{\mathcal D}_1}^{-1}\right)_\mu{}^\nu \xi_\nu&=
\left[
\frac{1}{\bar\Delta_V}\delta_\mu^\nu +\frac{1}{2}\frac{1}{\bar \Delta_V}\bar D_\mu \bar D^\nu \frac{1}{\bar \Delta_V}
+\frac{1}{\bar \Delta_V^2}\bar R_\mu{}^\nu  + \frac{1}{\bar\Delta_V^3}\bar R_{\mu\alpha} \bar D^\alpha \bar D^\nu  +\frac{1}{2}\frac{1}{\bar \Delta_V^4}
 \bar R^{\alpha\beta} \bar D_\alpha \bar D_\beta \bar D_\mu \bar D^\nu \right] \xi_\nu + {\mathcal O}(\bar R^2)\nn
 &=\left[
\frac{1}{\bar\Delta_V}\delta_\mu^\nu +\frac{1}{2}\frac{1}{\bar \Delta_V} \bar D^\nu \bar D_\mu\frac{1}{\bar \Delta_V}
+\frac{1}{2\bar \Delta_V^2}\bar R_\mu{}^\nu  + \frac{1}{\bar\Delta_V^3}\bar R_{\mu\alpha} \bar D^\alpha \bar D^\nu  +\frac{1}{2}\frac{1}{\bar \Delta_V^4}
 \bar R^{\alpha\beta} \bar D_\alpha \bar D_\beta \bar D_\mu \bar D^\nu \right] \xi_\nu + {\mathcal O}(\bar R^2)\,.
}
Then, from Eqs.\,\eqref{App: projector for a} and \eqref{App: projector for f}, one has
\al{
(P_a)_{\mu\nu}{}^{\rho\sigma} &= -\frac{1}{2}\left( \bar D_\mu \frac{1}{\bar\Delta_V} \bar D^\rho \delta_\nu^\sigma + \bar D_\mu \frac{1}{\bar\Delta_V} \bar D^\sigma \delta_\nu^\rho + \bar D_\nu \frac{1}{\bar\Delta_V} \bar D^\rho \delta_\mu^\sigma +\bar D_\nu \frac{1}{\bar\Delta_V} \bar D^\sigma \delta_\mu^\rho \right) \nn
&\qquad -\frac{1}{2}\left( \bar D_\mu \frac{1}{\bar\Delta_V} \bar D^\rho \bar D_\nu \frac{1}{\bar\Delta_V} \bar D^\sigma + \bar D_\mu  \frac{1}{\bar\Delta_V} \bar D^\sigma \bar D_\nu \frac{1}{\bar\Delta_V} \bar D^\rho  \right) \nn[1ex]
&\qquad
+ \left( \bar D_{\mu}\delta^\alpha_\nu+\bar D_{\nu}\delta^\alpha_\mu \right)
\left( 
\frac{1}{2\bar \Delta_V^2}\bar R_\alpha{}^\beta 
+\frac{1}{\bar\Delta_V^3}\bar R_{\alpha\tau} \bar D^\tau \bar D^\beta  +\frac{1}{2}\frac{1}{\bar \Delta_V^4} \bar R^{\tau\kappa} \bar D_\tau \bar D_\kappa \bar D_\alpha \bar D^\beta 
 \right) (-\bar D^\gamma) E_{\gamma\beta}{}^{\rho\sigma}
 + \mathcal O(\bar R^2) \nn[2ex]
&=: \left(P_a^{(0)} \right)_{\mu\nu}{}^{\rho\sigma}   + \left(P_a^{(1)}\right)_{\mu\nu}{}^{\rho\sigma} + \mathcal O(\bar R^2)\,, \\[3ex]
(P_f)_{\mu\nu}{}^{\rho\sigma} &= E_{\mu\nu}{}^{\rho\sigma} - (P_a)_{\mu\nu}{}^{\rho\sigma} 
=: E_{\mu\nu}{}^{\rho\sigma}  -\left( P_a^{(0)} \right)_{\mu\nu}{}^{\rho\sigma}   - \left( P_a^{(1)} \right)_{\mu\nu}{}^{\rho\sigma}  + \mathcal O(R^2)\nn
&=: \left( P_f^{(0)} \right)_{\mu\nu}{}^{\rho\sigma} + \left(P_f^{(1)}\right)_{\mu\nu}{}^{\rho\sigma}+ \mathcal O(\bar R^2) \,,
}
where superscripts, $(0)$ and $(1)$, on the projectors denote the order of curvature operators, and we have defined $P_f^{(0)}=E-P_a^{(0)}$ and $P_f^{(1)}=- P_a^{(1)}$.
Here and hereafter, it is supposed that all projectors act on $h_{\rho\sigma}$, i.e., indices $\rho$ and $\sigma$ are contracted and thus those are not open-indices.
In particular, the projector $P_f^{(0)}$ reads
\al{
\left(P_f^{(0)} \right)_{\mu\nu}{}^{\rho\sigma}& = \frac{1}{2}\left( \widetilde P_\mu{}^{\rho} \widetilde P_\nu{}^{\sigma} + \widetilde P_\mu{}^{\sigma} \widetilde P_\nu{}^{\rho} \right)\,, \\[2ex]
\left(P_f^{(1)} \right)_{\mu\nu}{}^{\rho\sigma} &= -\left( \bar D_{\mu}\delta^\alpha_\nu+\bar D_{\nu}\delta^\alpha_\mu \right)
\left( 
\frac{1}{2\bar \Delta_V^2}\bar R_\alpha{}^\beta 
+\frac{1}{\bar\Delta_V^3}\bar R_{\alpha\tau} \bar D^\tau \bar D^\beta  +\frac{1}{2}\frac{1}{\bar \Delta_V^4} \bar R^{\tau\kappa} \bar D_\tau \bar D_\kappa \bar D_\alpha \bar D^\beta 
 \right) (-\bar D^\gamma) E_{\gamma\beta}{}^{\rho\sigma}\,,
}
where $\widetilde P_\mu{}^{\nu}$ is defined in Eq.~\eqref{App: transverse projection operator}.
The physical scalar projector is expanded as
\al{
\hat S_{\mu\nu} 
&=\frac{1}{3}\widetilde P_{\mu\nu} -\frac{1}{3}  \bar D_\mu (\bar\Delta_V)^{-2}\bar D_\alpha \bar R^\alpha{}_\nu  -\frac{1}{3} \bar R_{\mu\nu} \frac{1}{\bar\Delta_S}+ \frac{1}{9}\widetilde P_{\mu\nu}\bar R \frac{1}{\bar\Delta_S} + {\mathcal O}(\bar R^2)\,,
}
from which the projector for $\sigma$ given in Eq.~\eqref{App: projector for sigma}, reads
\al{
(P_\sigma)_{\mu\nu}{}^{\rho\sigma} &= \hat S_{\mu\nu}\bar g^{\alpha\beta} (P_f)_{\alpha\beta}{}^{\rho\sigma}
= \left(\tilde s_{\mu\nu}\bar g^{\alpha\beta}  + I_{\mu\nu}{}^{\alpha\beta} \right)  (P_f)_{\alpha\beta}{}^{\rho\sigma}
\nn
&= \frac{1}{3}\widetilde P_{\mu\nu} \widetilde P^{\rho\sigma}
+\left(  -\frac{1}{3}\bar D_\mu (\bar\Delta_V)^{-2}\bar D_\alpha \bar R^\alpha{}_\nu  -\frac{1}{3} \bar R_{\mu\nu} \frac{1}{\bar\Delta_S} + \frac{1}{9}\bar R\widetilde P_{\mu\nu} \frac{1}{\bar\Delta_S} \right)\widetilde P^{\rho\sigma}  
 \nn
&\quad
-\frac{2}{3}\widetilde P_{\mu\nu} \bar D^{\alpha}
\left( 
\frac{1}{2\bar \Delta_V^2}\bar R_\alpha{}^\beta 
+\frac{1}{\bar\Delta_V^3}\bar R_{\alpha\tau} \bar D^\tau \bar D^\beta  +\frac{1}{2}\frac{1}{\bar \Delta_V^4} \bar R^{\tau\kappa} \bar D_\tau \bar D_\kappa \bar D_\alpha \bar D^\beta 
 \right) (-\bar D^\gamma) E_{\gamma\beta}{}^{\rho\sigma}
+ {\mathcal O}(\bar R^2)\,.
\label{App: projector of sigma}
}
Thus one has the traceless-transverse projector so that
\al{
(P_t)_{\mu\nu}{}^{\rho\sigma}
=\left(E_{\mu\nu}{}^{\alpha\beta}  -  I_{\mu\nu}{}^{\alpha \beta} - \tilde s_{\mu\nu} \bar g^{\alpha \beta}  \right) (P_f)_{\alpha\beta}{}^{\rho\sigma}
=:  \left( P_t^{(0)} \right)_{\mu\nu}{}^{\rho\sigma} + \left( P_t^{(1)} \right)_{\mu\nu}{}^{\rho\sigma} + {\mathcal O}(\bar R^2)\,,
\label{App: projector of tt}
}
where
\al{
  \left( P_t^{(0)} \right)_{\mu\nu}{}^{\rho\sigma}&=\frac{1}{2}\left( \widetilde P_\mu{}^{\rho} \widetilde P_\nu{}^{\sigma} + \widetilde P_\mu{}^{\sigma} \widetilde P_\nu{}^{\rho} \right) - \frac{1}{3}\widetilde P_{\mu\nu} \widetilde P^{\rho\sigma}\,, \\[2ex]
\left( P_t^{(1)} \right)_{\mu\nu}{}^{\rho\sigma} &=
-\left[ \left( \bar D_{\mu}\delta^\alpha_\nu + \bar D_{\nu}\delta^\alpha_\mu \right)+\frac{2}{3}\widetilde P_{\mu\nu} \bar D^{\alpha}\right]\bar R^{\tau\kappa}\left( 
\frac{1}{2\bar \Delta_V^2}\bar g_{\alpha \kappa} \delta^\beta_\tau  
+\frac{1}{\bar\Delta_V^3}\bar g_{\alpha\kappa} \bar D_\tau \bar D^\beta  +\frac{1}{2}\frac{1}{\bar \Delta_V^4}\bar D_\tau \bar D_\kappa \bar D_\alpha \bar D^\beta 
 \right) (-\bar D^\gamma) E_{\gamma\beta}{}^{\rho\sigma} \nn
 &\qquad
 +\bar D_\mu (\bar\Delta_V)^{-2}\bar D_\alpha \bar R^\alpha{}_\nu  \left( \frac{1}{3} \widetilde P^{\rho\sigma} \right)
+   \bar R_{\mu\nu}\left( \frac{1}{3} \widetilde P^{\rho\sigma} \right)\frac{1}{\bar\Delta_S}
 - \bar R \left( \frac{1}{3}\widetilde P_{\mu\nu} \right) \left(\frac{1}{3}\widetilde P^{\rho\sigma}\right)\frac{1}{\bar\Delta_S}\,.
}
We should note here that although the lowest order projectors $P_f^{(0)}$ and $P_t^{(0)}$ have no apparent dependence on curvatures, commutators between the covariant derivatives and Laplacians could yield terms with curvatures.

 \section{Inverse two-point functions}
\label{sect: Inverse two-point functions}
To derive the beta functions using the flow equation \eqref{App: flow equation}, we need the inverse two-point functions which are second functional derivatives with respect to metric fluctuation fields.
We first summarize the relations between the three couplings in the higher derivative operators for different bases. 
Using the fact that the Gauss-Bonnet term is topological we will only need the second functional derivative for the squared Ricci scalar curvature and the squared Ricci tensor.
For these invariants we will list the explicit forms of Hessians for the physical metric fluctuations $f_{\mu\nu}$ and the decomposed ones $t_{\mu\nu}$, $\sigma$, $a_{\mu\nu}$.

\subsection{Basis of gravitational interactions}
\label{sect: Basis of gravitational interactions}
For the higher derivative operators, one can write different bases. 
As given in Eq.~\eqref{Eq: effective action for gravity part}, one of the bases for the higher derivative terms is the Weyl basis which reads
\al{
\label{QG action in R2 C2 G4}
S_\text{HD}=\int_x\sqrt{g}\left[- \frac{C}{2} R^2 +  \frac{D}{2} C_{\mu\nu\rho\sigma} C^{\mu\nu\rho\sigma} + E G_4 \right]\,.
}
On the other hand, the calculations of the beta functions using the heat kernel method yield results in the basis spanned by $R^2$, $R_{\mu\nu}^2$ and $R_{\mu\nu\rho\sigma}^2$ rather than the operator basis in Eq.~\eqref{QG action in R2 C2 G4}. 
Therefore, one recasts \eqref{QG action in R2 C2 G4} as
\al{
\label{QG action in R2 Rmunu Rmunurhosigma}
S_\text{HD}=\int_x\sqrt{g}\left[ \frac{\mathcal C}{2} R^2 +   \frac{\mathcal D}{2}R_{\mu\nu}R^{\mu\nu} + {\mathcal E}R_{\mu\nu\rho\sigma}  R^{\mu\nu\rho\sigma} \right]\,.
}
In order to obtain the loop contributions for the squared Weyl tensor and the Gauss-Bonnet term \eqref{eq: Gauss-Bonnet term}, we compare between the actions \eqref{QG action in R2 C2 G4} and \eqref{QG action in R2 Rmunu Rmunurhosigma} and the read the relations between these coupling constants such that
\al{
\label{combinations between QG operators}
&{\mathcal C}=-C+\frac{D}{3}+2E\,,&
&{\mathcal D}=-2D-8E\,,&
&{\mathcal E}=\frac{D}{2}+E\,,
}
or equivalently
\al{
\label{eq: relation between CDE and maCmaDmaE}
&C=-{\mathcal C}-\frac{1}{3}{\mathcal D}-\frac{2}{3}{\mathcal E}\,,&
&D=\frac{1}{2}{\mathcal D}+4{\mathcal E}\,,&
&E=-\frac{{\mathcal D}}{4}-{\mathcal E}\,.
}
Once the beta function for ${\mathcal C}$, ${\mathcal D}$ and ${\mathcal E}$ are computed, we can obtain the beta functions for $C$, $D$ and $E$ using these relations.

Let us consider the variations for the effective action \eqref{action for higher derivative}.
The term ${\mathcal E}\fn{\rho}R_{\mu\nu\rho\sigma}R^{\mu\nu\rho\sigma}$ is written in term of the Gauss-Bonnet term such that
\al{
\int_x\sqrt{g}\,{\mathcal E}\fn{\rho}R_{\mu\nu\rho\sigma}R^{\mu\nu\rho\sigma}
&=\int_x\sqrt{g}\,{\mathcal E}\fn{\rho} \left[ 4R_{\mu\nu}R^{\mu\nu}-R^2+G_4 \right] \nn
&=\int_x\sqrt{g}\,{\mathcal E}\fn{\rho} \left[ 4R_{\mu\nu}R^{\mu\nu}-R^2\right] +\int_x\sqrt{g}\,{\mathcal E}\fn{\rho}\p_\mu {\mathcal J}_\text{GB}^\mu \nn
&=\int_x\sqrt{g}\,{\mathcal E}\fn{\rho} \left[ 4R_{\mu\nu}R^{\mu\nu}-R^2\right] -\int_x\sqrt{g}\,(\p_\mu{\mathcal E}\fn{\rho} ){\mathcal J}_\text{GB}^\mu +\text{(total derivative term)}\,,
}
where we used the fact that the Gauss-Bonnet term is topological, i.e. this term can be written as a total derivative.
The second term on the right-hand side does not contribute to the Hessians for a constant ${\mathcal E}\fn{\rho}$.
The effective action \eqref{action for higher derivative} is written as
\al{
  \Gamma_k^\text{grav} &=  \int_x\, \sqrt{g}
  \left[ U\fn{\rho} -\frac{F\fn{\rho}}{2} R +  \frac{1}{2}\left(\mathcal C\fn{\rho} -2{\mathcal E}\fn{\rho} \right)R^2 + \frac{1}{2}\left( \mathcal D\fn{\rho} + 8 {\mathcal E}\fn{\rho}\right)R_{\mu\nu}R^{\mu\nu}  \right] +\text{(total derivative term)}\nn[1ex]
  &=\int_x\, \sqrt{g}
  \left[ U\fn{\rho} -\frac{F\fn{\rho}}{2} R - \frac{1}{2}\left(C\fn{\rho} + \frac{2}{3}D\fn{\rho} \right)R^2 +D\fn{\rho}R_{\mu\nu}R^{\mu\nu}  \right] +\text{(total derivative term)}\,,\nn[1ex]
  &=\int_x\, \sqrt{g}
  \left[ U\fn{\rho} -\frac{F\fn{\rho}}{2} R - \frac{C\fn{\rho}}{2}R^2 +\frac{D\fn{\rho}}{2}C_{\mu\nu\rho\sigma}C^{\mu\nu\rho\sigma}  \right] +\text{(total derivative term)}\,,
}
where we used the relations \eqref{combinations between QG operators} in the second equality, and the squared Weyl tensor is given as
\al{
\frac{1}{2}C_{\mu\nu\rho\sigma}C^{\mu\nu\rho\sigma}=  R_{\mu\nu}R^{\mu\nu}-\frac{1}{3}R^2 + \frac{1}{2}G_4\,.
\label{App: approximate Weyl tensor squared}
}
Therefore, we do not have to calculate the variation for $R_{\mu\rho\nu\sigma} R^{\mu\rho\nu\sigma}$.  It is sufficient to evaluate variations for the effect action,
\al{
  \Gamma_k^\text{grav} 
 &\simeq \int_x\, \sqrt{g}
  \left[ U\fn{\rho} -\frac{F\fn{\rho}}{2} R - \frac{H\fn{\rho}}{2}R^2 +D\fn{\rho}R_{\mu\nu}R^{\mu\nu}  \right]\,,
  \label{eq: rewritten action}
}
where we define 
\al{
H\fn{\rho}= C\fn{\rho} + \frac{2}{3} D\fn{\rho}\,.
}

\subsection{Variations}
We list the second variations for the effective action in terms of the physical metric fluctuations $f_{\mu\nu}$ around a general background ${\bar g}_{\mu\nu}$, i.e. for the action \eqref{eq: rewritten action} we show 
\al{
\Gamma_{k,(ff)}^{\text{grav},(2)}&=\int_x\left[ U\left(\sqrt{g}\right)_{(2)}-\frac{F}{2}\left(\sqrt{g}R\right)_{(2)} -\frac{H}{2} \left(\sqrt{g}R^2\right)_{(2)} +D\left(\sqrt{g}R_{\mu\nu}^2\right)_{(2)} \right]\bigg|_{ff}\nn
&=:\Gamma_{(ff)}^{(U)}+\Gamma_{(ff)}^{(R)}+\Gamma_{(ff)}^{(R^2)}+\Gamma_{(ff)}^{(R_{\mu\nu}^2)}\,.
}
The second variations for each part are computed as
\al{
\Gamma_{(ff)}^{(U)}&= \frac{1}{4}U\int_x \sqrt{\bar g}\left[-f^{\mu\nu} E_{\mu\nu}{}^{\rho\tau}f_{\rho\tau} +\frac{1}{2}\sigma^2  \right] \nn
&= \frac{1}{4}U\int_x \sqrt{\bar g}\left[-t^{\mu\nu} E_{\mu\nu}{}^{\rho\tau}t_{\rho\tau} + \sigma\left\{\frac{1}{6} + \left(\frac{1}{3}-\hat S^{\mu\nu}\hat S_{\mu\nu} \right) \right\}\sigma - \sigma  \hat S^{\rho\tau}t_{\rho\tau}  - t^{\mu\nu}  \hat S_{\mu\nu}\sigma  \right]\,,
\label{Hessian of physical mode in potential}
\\[2ex]
\Gamma_{(ff)}^{(R)}
&=\frac{1}{4}\frac{F}{2}\int_x \sqrt{\bar g}\bigg[f^{\mu\nu}\left\{ \left( {\bar\Delta}_{T} +{\bar R}\right)  E_{\mu\nu}{}^{\rho\tau} - 2 {\bar R}_\mu{}^\rho \delta_\nu^\tau  - 2{\bar R}_\mu{}^\rho{}_\nu{}^\tau \right\} f_{\rho\tau}-\sigma \left({\bar \Delta}_{S} +\frac{{\bar R}}{2}\right)\sigma+ 2\sigma {\bar R}^{\rho\tau} f_{\rho\tau} \bigg] \nn[1ex]
&= \frac{1}{4}\frac{F}{2}\int_x \sqrt{\bar g}\bigg[t^{\mu\nu}\left\{ \left( {\bar\Delta}_{T} +{\bar R}\right)  E_{\mu\nu}{}^{\rho\tau} - 2{\bar R}_\mu{}^\rho \delta_\nu^\tau   - 2{\bar R}_\mu{}^\rho{}_\nu{}^\tau \right\} t_{\rho\tau}
-\sigma \bigg\{ \frac{2}{3} {\bar \Delta}_{S}  + \left( \frac{1}{3} - \hat S^{\mu\nu}\hat S_{\mu\nu}\right) {\bar \Delta}_{S}  \nn
&\qquad
+\left( \frac{1}{2} - \hat S^{\mu\nu}\hat S_{\mu\nu}\right) \bar R 
 + 2\hat S^{\mu\nu} \left({\bar R}_\mu{}^\rho \delta_\nu^\tau \hat S_{\rho\tau} -\bar R_{\mu\nu} \right) 
+ 2\hat S^{\mu\nu} {\bar R}_\mu{}^\rho{}_\nu{}^\tau  \left(  \hat S_{\rho\tau} - \bar D_\rho \bar D_\tau {\mathcal N}^{-1} \right) \bigg\}\sigma
\nn
&\quad
+\sigma \left( \hat S^{\rho\tau}\bar \Delta_T + \bar R \hat S^{\rho\tau}+ {\bar R}^{\rho\tau} - 2 \hat S^{\mu\nu}{\bar R}_\mu{}^\rho \delta_\nu^\tau   - 2\hat S^{\mu\nu}{\bar R}_\mu{}^\rho{}_\nu{}^\tau  \right)t_{\rho\tau}  
\nn
&\quad
+ t^{\mu\nu}  \left( \bar \Delta_T \hat S_{\mu\nu} + \bar R \hat S_{\mu\nu}+ {\bar R}_{\mu\nu} - 2 {\bar R}_\mu{}^\rho \delta_\nu^\tau \hat S_{\rho\tau}   - 2{\bar R}_\mu{}^\rho{}_\nu{}^\tau \hat S_{\rho\tau}  \right)  \sigma 
  \bigg]\,,
\\[2ex]
\Gamma_{(ff)}^{(R^2)}
&=\frac{1}{4} H\int_x \sqrt{\bar g}\Bigg[f^{\mu\nu}\left\{\left( \bar R\bar\Delta_T  + \frac{\bar R^2}{2}\right) E_{\mu\nu}{}^{\rho\tau} -2{\bar R} {\bar R}_{\mu}{}^\rho \delta_{\nu}^\tau -2{\bar R}_{\mu\nu}{\bar R}^{\rho\tau} -2\bar R{\bar R}_\mu{}^\rho{}_\nu{}^\tau\right\} f_{\rho\tau} \nn[1ex]
&\phantom{=\frac{C}{4}\int_x \sqrt{\bar g}\Bigg[f^{\mu\nu}}-2\sigma \left( \bar\Delta_S^2 + \frac{\bar R}{2}\bar\Delta_S + \frac{\bar R^2}{8} \right) \sigma+4 \sigma{\bar R}^{\rho\tau} \left(\bar\Delta_T +\frac{\bar R}{2} \right) f_{\rho\tau}\Bigg]\nn[1ex]
&=\frac{1}{4} H\int_x \sqrt{\bar g}\Bigg[t^{\mu\nu}\left\{\left( \bar R\bar\Delta_T  + \frac{\bar R^2}{2}\right) E_{\mu\nu}{}^{\rho\tau} -2{\bar R} {\bar R}_{\mu}{}^\rho \delta_{\nu}^\tau -2{\bar R}_{\mu\nu}{\bar R}^{\rho\tau} -2\bar R{\bar R}_\mu{}^\rho{}_\nu{}^\tau\right\} t_{\rho\tau} \nn[1ex]
&\quad
-\sigma \bigg\{ 2\bar\Delta_S^2 +\frac{1}{2}{\bar R} \bar\Delta_S + \left( \frac{1}{2} - \hat S^{\mu\nu}\hat S_{\mu\nu} \right){\bar R} \bar\Delta_S 
+ \left( \frac{1}{2} - \hat S^{\mu\nu}\hat S_{\mu\nu} \right) \frac{\bar R^2}{2}
+2 {\bar R}\hat S^{\mu\nu} \left(  {\bar R}_{\mu}{}^\rho \delta_{\nu}^\tau \hat S_{\rho\tau} - {\bar R}_{\mu\nu}  \right) 
\nn
&\qquad
+2 \bar R \hat S^{\mu\nu} {\bar R}_\mu{}^\rho{}_\nu{}^\tau \left(S_{\rho\tau}- \bar D_\rho \bar D_\tau {\mathcal N}^{-1}\right) 
+2 \hat S^{\mu\nu} {\bar R}_{\mu\nu}\left(  {\bar R}^{\rho\tau} \hat S_{\rho\tau}  -   2 \bar\Delta_S   \right)
 \bigg\} \sigma \nn
&\quad
+\sigma \left\{ 
\left(\hat S^{\rho\tau}\bar R +2{\bar R}^{\rho\tau}\right) \bar\Delta_T  
+ \left(  \hat S^{\rho\tau} {\bar R} +2{\bar R}^{\rho\tau}\right) \frac{\bar R}{2} 
-2\hat S^{\mu\nu}{\bar R} {\bar R}_{\mu}{}^\rho \delta_{\nu}^\tau
 -2\hat S^{\mu\nu}\bar R{\bar R}_\mu{}^\rho{}_\nu{}^\tau
 -2\hat S^{\mu\nu}{\bar R}_{\mu\nu}{\bar R}^{\rho\tau}
\right\}t_{\rho\tau}
 \nn
&\quad
+  t^{\mu\nu}\left\{ 
 \bar\Delta_T  \left(\hat S_{\mu\nu}\bar R +2{\bar R}_{\mu\nu}\right)
+  \frac{\bar R}{2} \left(  \hat S_{\mu\nu} {\bar R} +2{\bar R}_{\mu\nu}\right)
-2{\bar R} {\bar R}_{\mu}{}^\rho \delta_{\nu}^\tau\hat S_{\rho\tau}
 -2\bar R{\bar R}_\mu{}^\rho{}_\nu{}^\tau \hat S^{\rho\tau}
 -2{\bar R}_{\mu\nu}{\bar R}^{\rho\tau}\hat S_{\rho\tau}
\right\} \sigma
 \Bigg]
\,,\\[2ex]
\Gamma_{(ff)}^{(R_{\mu\nu}^2)}
&=\frac{1}{4}D\int_x \sqrt{\bar g}  \Bigg[ 
f^{\mu\nu}\bigg\{ \left(\bar\Delta_T^2 + 2\bar R^{\alpha\beta}\bar D_\beta\bar D_\alpha - \bar R^{\alpha\beta}\bar R_{\alpha\beta} \right) E_{\mu\nu}{}^{\rho\tau} - 4\bar\Delta_T {\bar R}_\mu{}^\rho{}_\nu{}^\tau \nn
&\qquad\qquad\qquad\qquad
+4 \bar R_\mu{}^{\alpha}{}_\nu{}^\beta \bar R_{\alpha}{}^\rho{}_\beta{}^\tau
+2 \bar R_{\mu}{}^{\rho} \bar R_\nu{}^\tau 
-2 \bar R_{\mu\alpha}\bar R^{\alpha\tau}\delta_{\nu}^\rho 
+4  \bar R^\rho{}_\mu{}^\tau{}_\alpha\bar R_\nu{}^\alpha
+4 \delta_\mu^\rho\bar R_{\nu\alpha}{}^\tau{}_\beta  \bar R^{\alpha\beta}
\bigg\} f_{\rho\tau}   \nn
&\quad
+ \sigma \left(\bar\Delta_{S}^2 -\bar R^{\alpha\beta} \bar D_\beta \bar D_\alpha +\frac{1}{2} \bar R^{\alpha\beta}\bar R_{\alpha\beta} \right)\sigma
+ \sigma \left( -4 \bar R^{\alpha\beta} \bar R_{\alpha}{}^\rho{}_\beta{}^\tau \right)f_{\rho\tau} \Bigg]\nn
&=\frac{1}{4}D\int_x \sqrt{\bar g}  \Bigg[ 
t^{\mu\nu}\bigg\{ \left(\bar\Delta_T^2 + 2\bar R^{\alpha\beta}\bar D_\beta\bar D_\alpha - \bar R^{\alpha\beta}\bar R_{\alpha\beta} \right) E_{\mu\nu}{}^{\rho\tau} 
- 4\bar\Delta_T {\bar R}_\mu{}^\rho{}_\nu{}^\tau\nn
&\qquad\qquad\qquad\qquad
+4 \bar R_\mu{}^{\alpha}{}_\nu{}^\beta\bar R_{\alpha}{}^\rho{}_\beta{}^\tau
+2 \bar R_{\mu}{}^{\rho} \bar R_\nu{}^\tau  
-2 \bar R_{\mu\alpha}\bar R^{\alpha\tau}\delta_{\nu}^\rho 
+4 \bar R_\nu{}^\alpha \bar R^\rho{}_\mu{}^\tau{}_\alpha
+4 \bar R^{\alpha\beta}\bar R_{\nu\alpha}{}^\tau{}_\beta \delta_\mu^\rho
\bigg\} t_{\rho\tau}   \nn
&\quad
+\sigma \bigg\{  
\frac{4}{3}\bar\Delta_{S}^2  - \left( \frac{1}{3} - \hat S^{\mu\nu}\hat S_{\mu\nu} \right) \bar\Delta_{S}^2 
-2\left(\frac{1}{2} -\hat S^{\mu\nu}\hat S_{\mu\nu} \right)\bar R^{\alpha\beta} \bar D_\beta \bar D_\alpha +\left(\frac{1}{2} - \hat S^{\mu\nu}\hat S_{\mu\nu}  \right)\bar R^{\alpha\beta}\bar R_{\alpha\beta}
   \nn
 &\quad
 +  \hat S^{\mu\nu}\Big(
 2 \Phi_{\mu\nu}{}^{\rho\tau} \left(  \hat S_{\rho\tau} + \bar R_{\rho\tau} {\mathcal N}^{-1} \right) \bar\Delta_S 
  + \Phi_{\mu\nu}{}^{\alpha\beta} \Phi_{\alpha\beta}{}^{\rho\tau} \hat S_{\rho\tau}
-\Phi_{\mu\nu}{}^{\alpha\beta}\Phi_{\alpha\beta}{}^{\rho\tau} \bar R_{\rho\tau}{\mathcal N}^{-1}
 + 2 \Phi_{\mu\nu}{}^{\alpha\beta}\Phi_{\alpha\beta}{}^{\rho\tau} \bar R_{\rho\tau} {\mathcal N}^{-1} \Big)
 \nn
 &\quad
-2 \hat S^{\mu\nu} \bar R^{\alpha\beta} \left( 
\delta_\mu^\rho\bar R_{\alpha\nu\beta}{}^\sigma   
  + \bar R_{\alpha\mu\beta}{}^\sigma \delta_\nu^\rho 
 - 2\delta_\alpha^\rho\bar R_{\mu\beta\nu}{}^\sigma  \right)\bar D_\rho \bar D_\sigma{\mathcal N}^{-1}
-4\hat S^{\mu\nu} \bar R_\mu{}^\rho{}_\nu{}^\tau \Big( \hat S_{\rho\tau} \bar\Delta_S  + \Phi_{\rho\tau}{}^{\alpha\beta} \left(  \hat S_{\alpha\beta} +\bar  R_{\alpha\beta} {\mathcal N}^{-1}\right) \Big)
\nn
 &\quad
+\hat S^{\mu\nu} (4 \bar R_\mu{}^{\alpha}{}_\nu{}^\beta \bar R_{\alpha}{}^\rho{}_\beta{}^\tau
+2 \bar R_{\mu}{}^{\rho} \bar R_\nu{}^\tau 
-2 \bar R_{\mu\alpha}\bar R^{\alpha\tau}\delta_{\nu}^\rho 
+4  \bar R^\rho{}_\mu{}^\tau{}_\alpha\bar R_\nu{}^\alpha
+4 \delta_\mu^\rho\bar R_{\nu\alpha}{}^\tau{}_\beta  \bar R^{\alpha\beta} \Big) \hat S_{\rho\tau}
\bigg\} \sigma
 \nn
&\quad
+ \sigma \bigg(
\hat S^{\rho\tau} \left( \bar\Delta_T^2 +2\bar R^{\alpha\beta}\bar D_\alpha \bar D_\beta \right)
-4 \hat S^{\mu\nu}\bar\Delta_T \bar R_\mu{}^\rho{}_\nu{}^\tau
-\hat S^{\rho\tau} \bar R^{\alpha\beta}\bar R_{\alpha\beta} 
 -2 \bar R^{\alpha\beta} \bar R_{\alpha}{}^\rho{}_\beta{}^\tau
 \nn
&\qquad
+  \hat S^{\mu\nu}\Big( 4 \bar R_\mu{}^{\alpha}{}_\nu{}^\beta\bar R_{\alpha}{}^\rho{}_\beta{}^\tau
+2 \bar R_{\mu}{}^{\rho} \bar R_\nu{}^\tau  
-2 \bar R_{\mu\alpha}\bar R^{\alpha\tau}\delta_{\nu}^\rho 
+4 \bar R_\nu{}^\alpha \bar R^\rho{}_\mu{}^\tau{}_\alpha
+4 \bar R^{\alpha\beta}\bar R_{\nu\alpha}{}^\tau{}_\beta \delta_\mu^\rho \Big)
 \bigg)t_{\rho\tau}
  \nn
&\quad
+ t^{\mu\nu} \bigg(
 \left( \bar\Delta_T^2 +2\bar R^{\alpha\beta}\bar D_\alpha \bar D_\beta \right)\hat S_{\mu\nu}
-4 \bar\Delta_T \bar R_\mu{}^\rho{}_\nu{}^\tau \hat S_{\rho\tau}
- \bar R^{\alpha\beta}\bar R_{\alpha\beta} \hat S_{\mu\nu}
 -2 \bar R_{\alpha\beta} \bar R_\mu{}^{\alpha}{}_\nu{}^\beta
 \nn
&\qquad
+  \Big( 4 \bar R_\mu{}^{\alpha}{}_\nu{}^\beta\bar R_{\alpha}{}^\rho{}_\beta{}^\tau
+2 \bar R_{\mu}{}^{\rho} \bar R_\nu{}^\tau  
-2 \bar R_{\mu\alpha}\bar R^{\alpha\tau}\delta_{\nu}^\rho 
+4 \bar R_\nu{}^\alpha \bar R^\rho{}_\mu{}^\tau{}_\alpha
+4 \bar R^{\alpha\beta}\bar R_{\nu\alpha}{}^\tau{}_\beta \delta_\mu^\rho \Big)\hat S_{\rho\tau}
 \bigg) \sigma 
  \Bigg]\,,
 \label{Hessian of physical mode in C2}
}
where the tensor $\Phi_{\mu\nu}{}^{\rho\sigma}$ is defined in Eq.~\eqref{App: tensor of Phi}.

\subsection{Hessians}
\label{App: Hessians in Einstein background spacetime}
\subsubsection{Metric fluctuations}
The Hessians $\Gamma^{(2)}_k$ for the action \eqref{eq: rewritten action}  is defined by
\al{
\Gamma_k^{(2)}=\frac{1}{2}\int_x\sqrt{\bar g} \pmat{t^{\mu\nu}  & \sigma & a^\gamma &\varphi}\left({\Gamma}^{(2)}_{\text{grav}}\right)\pmat{t_{\rho\sigma}\\[1ex] \sigma \\[1ex] a_\delta \\[1ex] \varphi}\,,
}
For the physical metric decomposition \eqref{physical metric decomposition}, it has a simple the structure,
\al{
\Gamma^{(2)}_{\text{grav}}=\frac{1}{2}\pmat{
\left(\Gamma^{(2)}_{(tt)}\right)_{\mu\nu}^{\phantom{\mu\nu}\rho\sigma} && \left(\Gamma^{(2)}_{(t\sigma)}\right)_{\mu\nu}^{\phantom{\mu\nu}\rho\sigma}  && 0 && \left(\Gamma^{(2)}_{(t\varphi)}\right)_{\mu\nu}\\[3ex]
\left(\Gamma^{(2)}_{(\sigma t)}\right)^{\rho\sigma} &&\left(\Gamma^{(2)}_{(\sigma\sigma)}\right) && 0 &&  \left(\Gamma^{(2)}_{(\sigma\varphi)}\right)  \\[3ex]
0 && 0 && \left(\Gamma^{(2)}_{(aa)}\right)_\gamma^{\phantom{\gamma}\delta} && \left(\Gamma^{(2)}_{(a\varphi)}\right)_{\gamma} \\[3ex]
\left(\Gamma^{(2)}_{(\varphi t)}\right)^{\rho\sigma}  &&  \left(\Gamma^{(2)}_{(\varphi \sigma)}\right) && \left(\Gamma^{(2)}_{(\varphi a)}\right)^{\delta} && \Gamma^{(2)}_{(\varphi\varphi)}
}
\,.
}

The two-point functions of the physical metric fluctuation read
\al{
&t^{\mu\nu}\left( \Gamma_{tt}^{(2)} \right)_{\mu\nu}{}^{\rho\sigma} t_{\rho\sigma}
-\frac{2}{3}\sigma\left( \Gamma_{\sigma\sigma}^{(2)} \right)\sigma 
+ t^{\mu\nu} \left( \Gamma_{t \sigma}^{(2)} \right)_{\mu\nu}  \sigma 
+  \sigma \left( \Gamma_{\sigma t}^{(2)} \right)^{\rho\sigma} t_{\rho\sigma}
\nn
&\qquad
=
t^{\mu\nu}  \Bigg[
{\mathbb K}_{(t)} E_{\mu\nu}{}^{\rho\sigma} +\Big({\mathbb M}_{(t)}(\bar R,\bar \Delta_T)\Big)_{\mu\nu}^{\phantom{\mu\nu}\rho\sigma}
\Bigg]t_{\rho\sigma} 
-\frac{2}{3} \sigma  \Bigg[
{\mathbb K}_{(\sigma)}  + {\mathbb M}_{(\sigma)}(\bar R,\bar \Delta_S)
\Bigg]  \sigma  \nn[1ex]
&\qquad\qquad+
 t^{\mu\nu} \Big({\mathbb M}_{(t\sigma)}(\bar R,\bar \Delta_S)\Big)_{\mu\nu}  \sigma 
+  \sigma \Big({\mathbb M}_{(\sigma t)}(\bar R,\bar \Delta_T)\Big)^{\rho\sigma} t_{\rho\sigma}
\,.
\label{App: two-point function of t and sigma}
} 
Here the regulated kinetic terms for the $t$-mode and the physical scalar $\sigma$-mode read, respectively,
\al{
{\mathbb K}_{(t)}&= \frac{F}{2}\bar\Delta_T +D \bar\Delta_T^2 -U\,,
\label{App: kinetic term of tt}
\\[2ex]
{\mathbb K}_{(\sigma)}&=\frac{F}{2}\bar\Delta_S + 3C \bar\Delta_S^2 -\frac{U}{4} \,.
\label{App: kinetic term of sigma in t sigma basis}
}

The interacting parts, denoted by ${\mathbb M}$ are computed up to the squared order of curvature operators.
For the $t$-mode we find
\al{
\Big({\mathbb M}_{(t)}(\bar R,\bar \Delta_T)\Big)_{\mu\nu}^{\phantom{\mu\nu}\rho\sigma}
&=
\left[ \frac{F}{2}\bar R +    H\bar R \bar\Delta_T +  2D \bar R^{\alpha\beta}\bar D_{\alpha}\bar D_\beta + \frac{H}{2} \bar R^2  - D \bar R^{\alpha\beta}\bar R_{\alpha\beta}  \right]  (P_t)_{\mu\nu}{}^{\rho\sigma}\nn[1ex]
&\quad
- 2 {\mathcal X} \left[ (P_t)_{\mu\nu}{}^{\alpha\beta}  \left( {\bar R}_\alpha{}^\gamma \delta_\beta^\delta  \right)(P_t)_{\gamma\delta}^{\phantom{\gamma\delta}\rho\sigma} \right]
-2 \left[  {\mathcal X}+ 2D \bar\Delta_T   \right] \left[ (P_t)_{\mu\nu}{}^{\alpha\beta} {\bar R}_\alpha{}^\gamma{}_\beta{}^\delta (P_t)_{\gamma\delta}^{\phantom{\gamma\delta}\rho\sigma} \right] \nn
&\quad
-2H\left[ (P_t)_{\mu\nu}{}^{\alpha\beta}  \left( {\bar R}_{\alpha\beta}{\bar R}^{\gamma\delta} \right)(P_t)_{\gamma\delta}^{\phantom{\gamma\delta}\rho\sigma}  \right] 
+D\bigg[  (P_t)_{\mu\nu}{}^{\alpha\beta}  \Big(
 4 \bar R_{\tau \alpha \kappa \beta}\bar R^{\tau \gamma \kappa\delta} 
 +2 \bar R_{\alpha}{}^{\gamma} \bar R_\beta{}^\delta 
   -2 \bar R_{\alpha\tau}\bar R^{\tau\delta}\delta_\beta^\gamma \nn
  &\qquad
 +4\bar R_\beta{}^\tau \bar R_\alpha{}^\gamma{}_\tau{}^\delta
  +4 \bar R^{\tau\kappa}\bar R_{\beta\tau}{}^\delta{}_\kappa \delta_\alpha^\gamma
\Big) (P_t)_{\gamma\delta}^{\phantom{\gamma\delta}\rho\sigma} \bigg] \,,
\label{App: tt mode interaction}
}
while for the $\sigma$-mode one finds
\al{
\frac{2}{3}{\mathbb M}_{(\sigma)}(\bar R,\bar \Delta_S)
&=
\left[\left( \frac{F}{2}\bar\Delta_S  + D \bar\Delta_S^2 - U \right)+  \frac{F}{2}\bar R +H\bar R \bar\Delta_S + 2D\bar R^{\alpha\beta}\bar D_\alpha \bar D_\beta  \right]\left(\frac{1}{3}-\hat S^{\mu\nu}\hat S_{\mu\nu} \right)
\nn
&\quad 
+\frac{1}{6}\left[ {\mathcal X}\bar R + 2D\bar R^{\alpha\beta}\bar D_\alpha \bar D_\beta  \right]
+\frac{2}{3}H\left[ \bar R  -6\hat S^{\mu\nu} {\bar R}_{\mu\nu}  \right]\bar\Delta_S
-\frac{1}{3}H\left[ \frac{1}{4}\bar R^2 - 6  \hat S^{\mu\nu} {\bar R}_{\mu\nu}{\bar R}^{\rho\tau}\hat S_{\rho\tau} \right] 
\nn
&\quad
+2{\mathcal X}\hat S^{\mu\nu} \left(  {\bar R}_{\mu}{}^\rho \delta_{\nu}^\tau \hat S_{\rho\tau} - {\bar R}_{\mu\nu}  \right) 
+2\left [{\mathcal X}  +2D\bar\Delta_S  \right]\hat S^{\mu\nu} \bar R_{\mu}{}^{\rho}{}_\nu{}^\tau \left(\hat S_{\rho\tau}- \bar D_\rho \bar D_\tau {\mathcal N}^{-1}\right)
   \nn
 &\quad
 -D\bigg[
\frac{1}{6} \bar R^{\alpha\beta}\bar R_{\alpha\beta}
+4   \bar R^{\alpha\beta} \delta_\mu^\rho\bar R_{\nu\alpha}{}^\tau{}_\beta \left( \hat S_{\rho\tau} - \bar D_\rho \bar D_\tau{\mathcal N}^{-1}  \right) 
+
 4 \hat S^{\mu\nu} \bar R^{\rho\alpha}  \bar R_{\mu\alpha\nu}{}^\tau \left(  \hat S_{\rho\tau} + \bar D_\rho \bar D_\tau{\mathcal N}^{-1} \right)
     \nn
&\quad
 +  \hat S^{\mu\nu}\Big(
2 \Phi_{\mu\nu}{}^{\rho\tau}\bar D_\rho \bar D_\tau \bar\Delta_S {\mathcal N}^{-1}
 + \Phi_{\mu\nu}{}^{\alpha\beta} \Phi_{\alpha\beta}{}^{\rho\tau}  \bar D_{\rho} \bar D_\tau {\mathcal N}^{-1} \Big)
\nn
&\quad
+\hat S^{\mu\nu} (4 \bar R_\mu{}^{\alpha}{}_\nu{}^\beta \bar R_{\alpha}{}^\rho{}_\beta{}^\tau
+2 \bar R_{\mu}{}^{\rho} \bar R_\nu{}^\tau 
-2 \bar R_{\mu\alpha}\bar R^{\alpha\tau}\delta_{\nu}^\rho 
\Big) \hat S_{\rho\tau}
\bigg]\,.
\label{App: sigma mode interaction}
}
In Eqs.~\eqref{App: tt mode interaction} and \eqref{App: sigma mode interaction} we have defined the shorthand
\al{
{\mathcal X}=\frac{F}{2}  + \left( C +\frac{2}{3}D \right)\bar R = \frac{F}{2} +H \bar R\,,
}
and $\mathcal N$ is given by Eq.~\eqref{projector for spin 0}.
The off-diagonal parts read
\al{
\left({\mathbb M}_{(\sigma t)}(\bar R,\bar \Delta_T)  \right)^{\rho\sigma}
&=\bigg[\frac{U}{3\bar\Delta_S} - \frac{2}{3}\frac{F}{2} +2H  \bar\Delta_S
-\frac{5}{3} D\bar\Delta_S 
 \bigg] \bar R^{\rho\sigma}\,,
 \\[2ex]
\left( {\mathbb M}_{(t\sigma)}(\bar R,\bar \Delta_S) \right)_{\mu\nu}
&= \left({\mathbb M}_{(\sigma t)}(\bar R,\bar \Delta_T)  \right)_{\mu\nu}\,.
\label{App: mixing term of Hessian}
}

In this work we employ the physical gauge fixing $\beta=-1$ and $\alpha\to0$.
For this choice, the gauge fixing action \eqref{standard gauge fixing} with \eqref{Eq: gauge fixing function} takes the form,
\al{
\Gamma_\text{gf} =  \frac{1}{2\alpha}\int d^4x \sqrt{ \bar{g} } \, (\bar D^\nu h_{\nu\mu})(\bar D^\rho h_{\rho}{}^\mu)=: \frac{1}{\alpha} \tilde\Gamma_\text{gf}\,,
}
so that the Hessian for the physical mode $f_{\mu\nu}$ does not involve the gauge parameter $\alpha$.
The Hessian for the gauge mode is given by
\al{
\left(\Gamma^{(2)}_{(aa)}\right)_\gamma^{\phantom{\gamma}\delta}= \left(\Gamma^{\text{grav},(2)}_{(aa)}\right)_\gamma^{\phantom{\gamma}\delta}+\frac{1}{\alpha}  \left(\tilde\Gamma^{(2)}_\text{gf}\right)_\gamma^{\phantom{\gamma}\delta}\,.
}
The propagator appearing in the flow equation involves the inverse of the (regulated ) Hessian. The Landau gauge $\alpha\to 0$ decouples the gauge mode in the propagator matrix. For $\alpha\to0$ we do not have to specify the explicit form of $\Gamma^{\text{gravity},(2)}_{aa}$.  Then the Hessian for the spin-1 gauge mode $a_\mu$ takes a simple form
\al{
\label{App: total spin-1 hessian}
\alpha\left(\Gamma^{(2)}_{(aa)}\right)_\mu^{\phantom{\mu}\nu}&\to \left(\tilde\Gamma_\text{gf}^{(2)}\right)_\mu^{\phantom{\mu}\nu}= \left(\delta_{\mu}^{\alpha}\bar \Delta_V -\bar D_\mu \bar D^\alpha -\bar R_\mu{}^{\alpha} \right)
\left( \delta_{\alpha}^{\nu} \bar\Delta_V -\bar  D_{\alpha}\bar D^\nu -\bar R_{\alpha}{}^\nu \right)
= \left( \bar {\mathcal D}_1\right)_\mu^{\phantom{\mu}\alpha}\left( \bar {\mathcal D}_1\right)_\alpha^{\phantom{\alpha}\nu}
\,.
}
The remaining Hessian for the physical modes is given by
\al{
\Gamma_\text{phys}^{(2)}=\pmat{
\left(\Gamma^{(2)}_{(tt)}\right)_{\mu\nu}^{\phantom{\mu\nu}\rho\sigma} && \left(\Gamma^{(2)}_{(t\sigma)}\right)_{\mu\nu} && \left(\Gamma^{(2)}_{(t \varphi)}\right)_{\mu\nu}  \\[4ex]
\left(\Gamma^{(2)}_{(\sigma t)}\right)^{\rho\sigma}&& \Gamma^{(2)}_{(\sigma\sigma)} &&  \Gamma^{(2)}_{(\sigma\varphi)}\\[3ex]
\left(\Gamma^{(2)}_{(\varphi t)}\right)^{\rho\sigma}  && \Gamma^{(2)}_{(\varphi \sigma)} && \Gamma^{(2)}_{(\varphi\varphi)}
}\,.
\label{app: physical Hessian}
}
In the limit of constant scalar fields considered in the present work the mixing terms between $\varphi$ and metric fluctuations ($t_{\mu\nu}$ and $\sigma$) vanish.
The Hessian \eqref{app: physical Hessian} becomes block-diagonal, with a separate two-point function for the scalar $\varphi$ given by  
\al{
\Gamma^{(2)}_{(\varphi\varphi)}&= Z_\varphi  \bar \Delta_S +m^2\,.
}
Here we define the effective derivative dependent scalar mass term
\al{
m^2=m_\varphi^2 -\frac{\tilde \xi_\varphi}{2}\bar R+ \frac{\gamma_\varphi}{2} \bar R^2 +\frac{\delta_\varphi}{2} \bar R_{\mu\nu}^2  +  \varepsilon_\varphi\bar R_{\mu\nu\rho\sigma}^2\,,
}
with
\al{
&m_\varphi^2=\frac{\p^2 U}{\p \varphi^2}\,,&
&\tilde \xi_\varphi=\frac{\p^2 F}{\p \varphi^2}\,,&
& \gamma_\varphi=\frac{\p^2 \mathcal C}{\p \varphi^2}\,,&
& \delta_\varphi=\frac{\p^2 \mathcal D}{\p \varphi^2}\,,&
& \varepsilon_\varphi=\frac{\p^2 \mathcal E}{\p \varphi^2}\,.
}

\subsubsection{Ghost}
We next consider the Hessians for the ghost fields. 
From the ghost action \eqref{standard ghost action} with $\beta=-1$, one finds
\al{
\left( \Gamma_\text{gh,$\bar C C$}^{(2)}\right)_\mu^{\phantom{\mu}\nu}&=\delta_\mu^{\nu}\bar \Delta_V - \frac{1-\beta}{2}\bar D_\mu \bar D^\nu -\bar R_\mu{}^{\nu} 
\xrightarrow[]{\beta =-1}\delta_\mu^{\nu}\bar \Delta_V - \bar D_\mu \bar D^\nu -\bar R_\mu{}^{\nu}
=\left( \bar {\mathcal D}_1\right)_\mu^{\phantom{\mu}\nu}\,.
\label{App: total ghost hessian}
}
This differential operator is the same as the gauge modes in the metric, see Eq.~\eqref{App: total spin-1 hessian}.

\subsubsection{Jacobian}
\label{App: Jacobian}
Finally we evaluate Jacobians arising from the physical metric decompositions \eqref{App: physical metric decomposition}.
To this end, we first compute the product of two metric fluctuations,
\al{
\langle h_{\mu\nu}^{(1)},h^{(2)\mu\nu} \rangle &:= \int_x \sqrt{\bar g}\, h_{\mu\nu}^{(1)} E^{\mu\nu}{}_{\rho\sigma} h^{(2)}{}^{\rho\sigma}\nn[1ex]
&=\int_x \sqrt{\bar g}\, \bigg[ 
\left(
f_{\mu\nu}^{(1)}  +{\bar D}_\mu a_\nu^{(1)} + {\bar D}_\nu a_\mu^{(1)} 
\right)
\left(
f^{(2)}{}^{\mu\nu}  +{\bar D}^\mu a^{(2)}{}^\nu + {\bar D}^\nu a^{(2)}{}^\mu
\right)
\bigg]\nn[1ex]
&=\int_x \sqrt{\bar g}\, \bigg[
f_{\mu\nu}^{(1)} f^{(2)}{}^{\mu\nu} 
+2a_\mu ^{(1)}\left( {\bar g}^{\mu\nu}\bar\Delta_V -\bar D^{\mu} \bar D^{\nu} -{\bar R}^{\mu\nu} \right)a_\nu ^{(2)}
\bigg]\nn
&=\int_x \sqrt{\bar g}\, \bigg[
t^{(1)}_{\mu\nu}t^{(2)\mu\nu}
+\sigma^{(1)} \hat S_{\mu\nu} \hat S^{\mu\nu} \sigma^{(2)}
+ t^{(1)}_{\mu\nu}\hat S^{\mu\nu}\sigma^{(2)}
+\sigma^{(1)} \hat S_{\mu\nu}  t^{(2)\mu\nu}
+2a_\mu ^{(1)} \left( \bar{\mathcal D}_\text{1} \right)^{\mu\nu}  a_\nu ^{(2)}
\bigg]\nn
&=\int_x \sqrt{\bar g}\,
\bigg[
\pmat{ t^{(1)}_{\mu\nu} & \sigma^{(1)}}\pmat{
E^{\mu\nu}{}_{\rho\sigma} & -\bar R^{\mu\nu}{\mathcal N}^{-1}\\[2ex]
-{\mathcal N}^{-1}\bar R_{\rho\sigma}& \hat S^{\alpha\beta}\hat S_{\alpha\beta}
}
\pmat{
t^{(2)}{}^{\rho\sigma}\\[0.5ex]
\sigma ^{(2)}
}
+2a_\mu ^{(1)} \left( \bar{\mathcal D}_\text{1} \right)^{\mu\nu} a_\nu ^{(2)}
\bigg]\,,
}
where $\left( \bar{\mathcal D}_\text{1} \right)^{\mu\nu}= {\bar g}^{\mu\nu}\bar\Delta_V -\bar D^{\mu} \bar D^{\nu} -{\bar R}^{\mu\nu}$, and we have used $t_{\mu\nu}\hat S^{\mu\nu}\sigma =-t_{\mu\nu} \bar R^{\mu\nu}{\mathcal N}^{-1}\sigma$ since $\bar D_\mu t^{\mu\nu}=0$ and $\bar g^{\mu\nu} t_{\mu\nu}=0$.
Note that in an Einstein spacetime, $\bar R_{\mu\nu} =\bar g_{\mu\nu}\bar R/4$, the mixing terms vanish.

In order to obtain the Jacobian arising from the metric decomposition, we consider the Gaussian path integral for the metric fluctuation
\al{
1&=\int {\mathcal D}h_{\mu\nu}\,\exp\fn{-\frac{1}{2}\langle h_{\mu\nu},h^{\mu\nu}\rangle}\nn
&=J_\text{grav}\int {\mathcal D}t_{\mu\nu} {\mathcal D}\sigma {\mathcal D a_\mu} \,\exp\bigg[ -\frac{1}{2}\int_x \sqrt{\bar g}\,
\bigg\{
\pmat{ t_{\mu\nu} & \sigma}\pmat{
E_{(t)}^{\mu\nu}{}_{\rho\sigma} & -\bar R^{\mu\nu}{\mathcal N}^{-1}\\[2ex]
-{\mathcal N}^{-1}\bar R_{\rho\sigma}& \hat S^{\alpha\beta}\hat S_{\alpha\beta}
}
\pmat{
t^{\rho\sigma}\\[0.5ex]
\sigma
}
+2a_\mu \left( \bar{\mathcal D}_\text{1} \right)^{\mu\nu} a_\nu
\bigg\}\bigg]\nn[1ex]
&=J_\text{grav}\left[ \det{}_{(1)}\left(\bar{\mathcal D}_\text{1} \right) \right]^{-1/2}
\left[ \det{}_{(2)}\pmat{
E_{(t)}^{\mu\nu}{}_{\rho\sigma} & -\bar R^{\mu\nu}{\mathcal N}^{-1}\\[2ex]
-{\mathcal N}^{-1}\bar R_{\rho\sigma}& \hat S^{\alpha\beta}\hat S_{\alpha\beta}
}\right]^{-1/2},
}
where $E_{(t)}$ is the identical matrix \eqref{App: the symmetric unit matrix and the projection tensors} acting on the space satisfying the TT condition.
One finds the Jacobian arising from the decomposition of the metric fluctuation,
\al{
J_\text{grav}&=\left[ \det{}_{(1)}\left(\bar{\mathcal D}_\text{1} \right) \right]^{1/2}
\left[ \det{}_{(2)}\pmat{
E_{(t)}^{\mu\nu}{}_{\rho\sigma} & -\bar R^{\mu\nu}{\mathcal N}^{-1}\\[2ex]
-{\mathcal N}^{-1}\bar R_{\rho\sigma}& \hat S^{\alpha\beta}\hat S_{\alpha\beta}
}\right]^{1/2}
=J_\text{grav1} \cdot J_\text{grav2}
\,.
\label{App: Jacobian from metric decomposition}
}
The differential operator has already appeared in the Hessians for the gauge modes \eqref{App: total spin-1 hessian} in the metric and the ghost field \eqref{App: total ghost hessian}.
The Jacobian for spin 1 mode can be eliminated by the redefinitions of the vector fluctuation
\al{
\tilde a_\mu= (\bar{\mathcal D}_\text{1})^{1/2}a_\mu  \,.
\label{App: redefinitions of fields}
}
In the basis $\tilde a_\mu$, the Jacobian does not arise, while its Hessian has to be appropriately modified.
In this work, we do not use this redefined field \eqref{App: redefinitions of fields}.
In Section~\ref{evaluation of Jacobian}, we evaluate the flow contributions from the regulated Jacobian explicitly.

\section{Heat kernel method}
\label{App: heat kernel coefficients}
\subsection{Basics of heat kernel method}
We summarize the heat kernel techniques and coefficients in order to evaluate the flow generators.
The flow generator for a spin-$i$ mode takes the following form:
\al{
\zeta&=\frac{1}{2}\tr_{(i)} W\fn{\bar\Delta_i} 
= \int^\infty_{-\infty} \df s\, \widetilde W(s) \,\tr_{(i)}\left[ e^{-s\bar\Delta_i} \right]
\nn
&= \frac{1}{2(4\pi)^2}\int_x\sqrt{\bar g}\left[
b_0^{(i)}Q_{2}[W] +b_2^{(i)}\bar R \,Q_1[W] + \left(b_4^{(i)} \bar R^2+{\hat b}_4^{(i)} \bar R_{\mu\nu}\bar R^{\mu\nu}+{\tilde b}_4^{(i)} \bar R_{\mu\nu\rho\sigma}\bar R^{\mu\nu\rho\sigma} \right) Q_0[W]
\right]\nn[1ex]
&= \frac{1}{2(4\pi)^2}\int_x\sqrt{\bar g}\left[
b_0^{(i)}Q_{2}[W]+b_2^{(i)}\bar R \,Q_1[W] + \left( \underline{b}_4^{(i)} \bar R^2+\underline{\hat b}_4^{(i)} C_{\mu\nu\rho\sigma}C^{\mu\nu\rho\sigma}+\underline{\tilde b}_4^{(i)}G_4  \right) Q_0[W]
\right]\,,
\label{app: heat kernel general}
}
where the heat kernel trace is expanded as
\al{
\tr_{(i)}\left[ e^{-s\bar\Delta_i} \right]
&= \frac{1}{(4\pi s)^2} \int_x\sqrt{\bar g} \left[  b_0^{(i)}+ b_2^{(i)}s{\bar R} +s^2\left(b_4^{(i)} \bar R^2+{\hat b}_4^{(i)} \bar R_{\mu\nu}\bar R^{\mu\nu}+{\tilde b}_4^{(i)} \bar R_{\mu\nu\rho\sigma}\bar R^{\mu\nu\rho\sigma} \right)\right]\nn[1ex]
&= \frac{1}{(4\pi s)^2} \int_x\sqrt{\bar g} \left[  b_0^{(i)}+ b_2^{(i)}s{\bar R} +s^2\left( \underline{b}_4^{(i)} \bar R^2+ \underline{\hat b}_4^{(i)} C_{\mu\nu\rho\sigma}C^{\mu\nu\rho\sigma}+\underline{\tilde b}_4^{(i)} G_4 \right)\right]\,.
\label{App: heat kernel expansion}
}
Note that $\tr_{(i)}$ denotes
\al{
&\tr_{(0)}[O]= \tr[O]\,,&
&\tr_{(1)}[O_{\mu\nu}]= \tr_{(0)}[O_{\mu\nu}\bar g^{\mu\nu}]\,,&
&\tr_{(2)}[O_{\rho\sigma}{}^{\mu\nu}] =\tr_{(0)}[O_{\rho\sigma}{}^{\mu\nu} E_{\mu\nu}{}^{\rho\sigma}]\,.
}
The threshold functions $Q_n[W]$ are given by
\al{
Q_n[W]=\int^\infty_0\df s\,s^{-n}\widetilde W(s)\,,
}
for which the Mellin transformation yields
\al{
&Q_n[W]=\frac{1}{\Gamma\fn{n}}\int^\infty_0 \df z\,z^{n-1}W\fn{z}
\qquad
\text{for $n\geq 1$} \,,&
&Q_{-n}[W]=(-1)^n\frac{\p^n W}{\p z^n}\Bigg|_{z=0}
\qquad
\text{for $n\geq 0$} \,.
\label{App: Q functions}
}
In particular, when the flow kernel takes the form, $(\bar\Delta_i)^m W(\bar\Delta_i)$, the threshold function reads
\al{
Q_n[ z^mW] =\frac{\Gamma(n+m)}{\Gamma(n)} Q_{n+m}[W]\,, \qquad n+m>0\,.
\label{App: Delta W threshold functions}
}

The flow kernel in this work takes typically the form, 
\al{
W_{p,\epsilon}(z)=\frac{\p_t {\mathcal R}_k(z)}{(A P_k(z)^{1+2\epsilon}+ B P_k(z)^{2+2\epsilon} +C P_k(z)^{2\epsilon})^{p+1}}\,,
\label{App: flow kernel example}
}
with $k$-dependent constants $A$, $B$ and $C$ . 
Here $P_k$ is the regulated momentum, i.e. $P_k=z+R_k(z)$ and the regulator function in the numerator of Eq.~\eqref{App: flow kernel example} is introduced such that $\bar\Delta_i$ is replaced to $P_k$, namely one gives ${\mathcal R}_k = A \left( P_k(z)^{1+2\epsilon} -z ^{1+2\epsilon} \right)+ B \left( P_k(z)^{2+2\epsilon} -z^{2+2\epsilon}  \right) +C\left( P_k(z)^{2\epsilon} -z^{2\epsilon}  \right)$ for which the derivative with respect to $t$ yields
\al{
\p_t{\mathcal R}_k(z) &= (\p_t A) \left( P_k^{1+2\epsilon} -z^{1+2\epsilon}  \right)+ (\p_t B) \left( P_k^{2+2\epsilon} -z^{2+2\epsilon}  \right) +  (\p_t C) \left( P_k^{2\epsilon} -z^{2\epsilon}  \right) \nn
&\qquad
 + \Big((1+2\epsilon)A P_k^{2\epsilon}+(2+2\epsilon)B P_k^{1+2\epsilon}  + 2\epsilon C P_k^{2\epsilon-1} \Big) \p_t R_k\,.
}
Let us now calculate the threshold functions \eqref{App: Q functions}. To this end, we employ the Litim-type cutoff function~\cite{Litim:2001up}, i.e. $R_k(z)=(k^2-z)\theta(k^2-z)$.
Thus, for $n\geq 1$, one finds
\al{
Q_n[W_{p,\epsilon}]&=\frac{1}{\Gamma\fn{n}}\int^\infty_0 \df z\,z^{n-1}\frac{\p_t{\mathcal R}_k(z) }{(A P_k(z)^{1+2\epsilon}+ B P_k(z)^{2+2\epsilon} +C P_k(z)^{2\epsilon})^{p+1}} \nn
&=\frac{2}{n\Gamma\fn{n}}\frac{k^{4+2n-4p\epsilon}}{(A k^2+ B k^4 +C )^{p+1}} \left( (1+2\epsilon) Ak^{-2} +2(1+\epsilon) B  +2 \epsilon k^{-4}C+ \frac{(1+2\epsilon)k^{-2}\p_t A}{2(n+1 +2\epsilon)} + \frac{(1+\epsilon) \p_t B}{n+2 + 2\epsilon} + \frac{ \epsilon k^{-4} \p_t C}{n+2\epsilon} \right)\nn
&=\frac{2}{n\Gamma\fn{n}}\frac{k^{2n-4p-4p\epsilon}}{(a+b +c )^{p+1}} \left( (1+2\epsilon) a +2(1+\epsilon)b  +2 \epsilon c+ \frac{(1+2\epsilon)(\p_t a+2a)}{2(n+1 +2\epsilon)} + \frac{(1+\epsilon) \p_t b}{n+2 + 2\epsilon} + \frac{\epsilon (\p_t c+4c)}{n+2\epsilon} \right)\nn
&= \frac{(1+2\epsilon)(n+2+2\epsilon)}{n+1 +2\epsilon}a^{-p}  \Bigg( 1+ 2\frac{(1+\epsilon)(n+1 +2\epsilon)}{(1+2\epsilon)(n+2+2\epsilon)}\frac{b}{a}  +  2\frac{\epsilon \left( n+2+2\epsilon \right)(n+1 +2\epsilon)}{(n+2\epsilon)(1+2\epsilon)(n+2+2\epsilon)}\frac{c}{a}\Bigg)\nn
&\quad
\times \left( 1 -\frac{\eta_n}{2(n+2+2\epsilon)} \right) \left[2k^{2n-4p-4p\epsilon}\ell_{p}^{2n} (b/a+c/a) \right]
\label{App: threshold functions general}
}
where $a=Ak^{-2}$, $B=b$ and $c=Ck^{-4}$.
Here the anomalous dimension is defined as
\al{
\eta_n
&= -\frac{\p_t a}{a} -\frac{ \p_t\left( 1+2\frac{1+\epsilon}{1+2\epsilon} \frac{n+1 +2\epsilon}{n+2 + 2\epsilon}\frac{b}{a} + 2\frac{\epsilon}{1+2\epsilon}\frac{n+1 +2\epsilon}{n+2\epsilon} \frac{c}{a} \right) }{1+2\frac{1+\epsilon}{1+2\epsilon} \frac{n+1 +2\epsilon}{n+2+2\epsilon}\frac{b}{a}  +  2\frac{\epsilon}{1+2\epsilon}\frac{n+1 +2\epsilon}{n+2\epsilon}\frac{c}{a}}\,.
\label{App: general anomalous dimension}
}

We define the dimensionless threshold function as
\al{
\ell_p^{2n} (x) = \frac{1}{n!}\frac{1}{(1 + x)^{p+1}}\,,
}
where $n\Gamma(n)=\Gamma\fn{n+1}=n!$ is used. Note that the threshold functions for $p\geq1$ can be obtained from $\ell_0^d$ so that
\al{
\ell_p^d\fn{x}=(-1)^p\frac{1}{p!}\frac{\p^p}{\p x^p}\ell_0^d\fn{x}\,.
\label{eq: expanded threshold functions}
}
For $n=0$, one has
\al{
Q_0[W_{p,\epsilon}]&= \lim_{z\to 0}W_{p,\epsilon}(z) \nn
&=\frac{k^{4-4\epsilon p}}{(A k^{2}+ B k^{4} +C )^{p+1}}\left( (2+4\epsilon)Ak^{-2}+(4+4\epsilon)B + 4\epsilon C k^{-4} + (\p_t A) k^{-2} + (\p_t B) +  (\p_t C)k^{-4}  \right) \nn
&=\frac{k^{-4p-4\epsilon p}}{(a + b  +c)^{p+1}}\Big( (2+4\epsilon)a+(4+4\epsilon)b + 4\epsilon c + (2a+\p_t a) + (\p_t b) +  (4c+\p_t c) \Big) \nn
&= 2(1+\epsilon) a^{-p} \left(1+\frac{b}{a} +\frac{c}{a}\right)\left(1 - \frac{\eta_0}{4(1+\epsilon)} \right)\frac{2k^{-4p-4\epsilon p} }{(1 + b/a  +c/a)^{p+1}} \,,
}
with the anomalous dimension of field
\al{
\eta_0 = -\frac{\p_t a }{a}-  \frac{ \p_t \left(1  + \frac{b}{a} + \frac{c}{a} \right)}{1+\frac{b}{a} +\frac{c}{a}}\,.
}
Indeed, this result agrees with setting $n=0$ for Eq.~\eqref{App: threshold functions general}.

\subsection{Projected heat kernel}
We specify the heat kernel expansion \eqref{App: heat kernel expansion} for each spin mode,
We start with the symmetric spin-2-mode case:
\al{
\tr_\text{(2)}\left[ e^{-s\bar\Delta_{T}} \right]
& =\tr_\text{(2)}\left[ e^{-s\bar \Delta_{T}} E\right] \nn
&=\frac{1}{(4\pi s)^2} \int_x\sqrt{\bar g} \left[   b_0^{(2)}+ b_2^{(2)} s\,\bar R +s^2\left( b_4^{(2)}\bar R^2+{\hat b}_4^{(2)} \bar R_{\mu\nu}\bar R^{\mu\nu}+ {\tilde b}_4^{(2)}\bar R_{\mu\nu\rho\sigma}\bar R^{\mu\nu\rho\sigma} \right)\right]\,,
\label{App: heat kernel expansion for spin 2}
} 
with the heat kernel coefficients,
\al{
& b_0^{(2)}= 10\,,&
& b_2^{(2)} =\frac{5}{3}\,,&
& b_4^{(2)} =\frac{5}{36}\,,&
&{\hat b}_4^{(2)} = -\frac{1}{18}\,,&
& {\tilde b}_4^{(2)}=-\frac{4}{9}\,.
}
Let us now consider the case where the projection operator $P_a$ given in Eq.~\eqref{App: projector for a} is inserted, namely,  $\tr_\text{(2L)}\left[ e^{-s\bar \Delta_{T}} \right]=\tr_\text{(2)}\left[ e^{-s\bar \Delta_T} P_a  \right]$.
Here, the evaluation of the heat kernel coefficients is performed as follows:
\al{
\tr_\text{(2)}\left[ e^{-s\bar\Delta_T} P_a  \right] &=  \tr_\text{(2)}\left[ e^{-s\bar\Delta_T} \left\{ \left( \bar{\mathbb D}_1\right)_{\mu\nu}{}^\alpha \left({\bar{\mathcal D}_1}^{-1} \right)_\alpha{}^{\beta} (-\bar D^\gamma) E_{\gamma\beta}{}^{\rho\sigma} \right\}  \right]  \nn
&=\tr_\text{(2)}\left[ e^{-s\bar\Delta_T}   \right] 
+\tr_\text{(1)}\left[ \left[(-\bar D^\gamma),\, e^{-s\Delta_V}\right] \left\{\left( \bar{\mathbb D}_1\right)_{\gamma\beta}{}^\alpha \left({\bar{\mathcal D}_1}^{-1} \right)_\alpha{}^{\beta}\right\}  \right] \nn
&=\tr_\text{(2)}\left[ e^{-s\bar\Delta_T}   \right] 
+\tr_\text{(1)}\left[ e^{-s\bar\Delta_V}\left\{ -s [-\bar D^\gamma,\bar \Delta_V] +\frac{s^2}{2}[[-\bar D^\gamma,\bar\Delta_V],\bar\Delta_V] +\mathcal O(s^3) \right\} \left\{\left( \bar{\mathbb D}_1\right)_{\gamma\beta}{}^\alpha \left({\bar{\mathcal D}_1}^{-1} \right)_\alpha{}^{\beta}\right\}  \right] \nn
&= \frac{1}{(4\pi s)^2} \int_x\sqrt{\bar g} \left[   b_0^\text{(2L)}+ b_2^\text{(2L)} s \,\bar R+s^2\left( b_4^\text{(2L)} \bar R^2+{\hat b}_4^\text{(2L)} \bar R_{\mu\nu}\bar R^{\mu\nu}+ {\tilde b}_4^\text{(2L)} \bar R_{\mu\nu\rho\sigma}\bar R^{\mu\nu\rho\sigma} \right)\right]\,.
\label{app: 2T heat kernel expansion}
}
Once obtaining the heat kernel coefficients $b^{(2L)}_i$, we obtain the heat kernel expansion for the physical 2T-mode from
\al{
\tr_\text{(2T)}\left[ e^{-s\bar\Delta_{T}} \right]&=\tr_\text{(2)}\left[ e^{-s\bar\Delta_T} P_f  \right] 
= \tr_\text{(2)}\left[ e^{-s\bar\Delta_T}  \right] - \tr_\text{(2)}\left[ e^{-s\bar\Delta_T} P_a  \right]\nn
&=\frac{1}{(4\pi s)^2} \int_x\sqrt{\bar g} \left[   b_0^\text{(2T)}+ b_2^\text{(2T)} s \,\bar R+s^2\left( b_4^\text{(2T)} \bar R^2+{\hat b}_4^\text{(2T)} \bar R_{\mu\nu}\bar R^{\mu\nu}+ {\tilde b}_4^\text{(2T)}\bar R_{\mu\nu\rho\sigma} \bar R^{\mu\nu\rho\sigma} \right)\right]\,.
}
Here the first term corresponds to Eq.~\eqref{App: heat kernel expansion for spin 2} and the second to Eq.~\eqref{app: 2T heat kernel expansion}.
In the same manner, we can evaluate the heat kernel expansion for the spin 0 and TT cases,
\al{
\tr_\text{(0)}\left[ e^{-s\bar\Delta_{S}} \right]&=\tr_\text{(2)}\left[ e^{-s\bar\Delta_T} P_\sigma \right] \nn
&= \frac{1}{(4\pi s)^2} \int_x\sqrt{\bar g} \left[   b_0^\text{(0)}+ b_2^\text{(0)} s \,\bar R+s^2\left( b_4^\text{(0)} \bar R^2+{\hat b}_4^\text{(0)}\bar  R_{\mu\nu}\bar R^{\mu\nu}+ {\tilde b}_4^\text{0}\bar R_{\mu\nu\rho\sigma}\bar R^{\mu\nu\rho\sigma} \right)\right]\,,\\[2ex]
\tr_\text{(2TT)}\left[ e^{-s\bar\Delta_{T}} \right]&=\tr_\text{(2)}\left[ e^{-s\bar\Delta_T} P_t \right] 
=\tr_\text{(2)}\left[ e^{-s\bar\Delta_T} (P_f-P_\sigma) \right] 
= \tr_\text{(2T)}\left[ e^{-s\bar\Delta_T} \right] - \tr_\text{(0)}\left[ e^{-s\bar\Delta_S} \right]\nn
&= \frac{1}{(4\pi s)^2} \int_x\sqrt{\bar g} \left[   b_0^\text{(2TT)}+ b_2^\text{(2TT)} s \,\bar R+s^2\left( b_4^\text{(2TT)} \bar R^2+{\hat b}_4^\text{(2TT)} \bar R_{\mu\nu}\bar R^{\mu\nu}+ {\tilde b}_4^\text{(2TT)}\bar R_{\mu\nu\rho\sigma}\bar R^{\mu\nu\rho\sigma} \right)\right]\,.
} 
In the next subsection, we exhibit the heat kernel coefficients $b_i^\text{(2TT)}$ and $b_i^\text{(0)}$.
Using them, one can obtain also $b_i^\text{(2T)}$ and $b_i^\text{(2L)}$.

\subsection{Heat kernel coefficients}
\label{App: hear kernel technique}
\subsubsection{Coefficients for each mode in metric fluctuations}
The heat kernel coefficients for the Laplacian $\bar\Delta_T$ acting on the TT tensor in a general background spacetime was computed in Ref.\,\cite{Benedetti:2010nr}, and result in
\al{
&b_0^\text{(2TT)}=5\,,&
&b_2^\text{(2TT)}=-\frac{5}{6}\,,&
&b_4^\text{(2TT)}=-\frac{137}{216}+ \frac{N}{2\chi_E}\,,&
&\hat b_4^\text{(2TT)}=-\frac{17}{108}- \frac{2N}{\chi_E}\,,&
&\tilde b_4^\text{(2TT)}=\frac{5}{18}+ \frac{N}{2\chi_E}\,,
\tag{C21-I}
}
where the superscript (2TT) stands for the TT spin-2 field, $\chi_E$ is the Euler characteristic associated to zero modes involved in the metric decompositions, and $N$ is the sum of the number of Killing vectors and conformal ones.
In the Weyl bases \eqref{app: heat kernel general} and \eqref{App: heat kernel expansion} this yields
\al{
&\underline{b}_4^\text{(2TT)}=-\frac{385}{648}\,, &
&\underline{\hat b}_4^\text{(2TT)}=\frac{103}{216}\,, &
&\underline{\tilde b}_4^\text{(2TT)}=-\frac{43}{216}+ \frac{N}{2\chi_E}\,,
\tag{C21-II}
}
For the Laplacian $\bar\Delta_S$ acting on spin-0 scalar field, heat kernel coefficients are well-known as
\setcounter{equation}{21}
\al{
&b_0^\text{(0)}=1\,,&
&b_2^\text{(0)}=\frac{1}{6}\,,&
&b_4^\text{(0)}=\frac{1}{72}\,,&
&\hat b_4^\text{(0)}=-\frac{1}{180}\,,&
&\tilde b_4^\text{(0)}=\frac{1}{180}\,,
\tag{C22-I}
}
implying for the Weyl bases
\al{
&\underline{b}_4^\text{(0)}=\frac{1}{72}\,,&
&\underline{\hat b}_4^\text{(0)}=\frac{1}{120}\,,&
&\underline{\tilde b}_4^\text{(0)}=-\frac{1}{360}\,.
\tag{C22-II}
}
\setcounter{equation}{22}

Next, we give a formula in order to evaluate the flow contribution from the (spin-1) measure modes, i.e., $a_{\mu}$, $C_\mu$, $\bar C_\mu$ and the Jacobian.
As we have derived in the last section, the measure mode in the physical gauge fixing ($\beta=-1$, $\alpha\to0$) takes the following uniform differential operator:
\al{
\left(\bar{\mathcal D}_1 \right)_\mu^{\phantom{\mu}\nu}=\delta_{\mu}^\nu \bar\Delta_V - \bar D_\mu \bar D^\nu -\bar R_{\mu}{}^\nu\,.
\label{App: Differential op for measure mode}
}
For differential operators taking the typical form $\left(\bar{\mathcal D}_1 \right)_{\mu}^{\phantom{\mu}\nu}=\delta_{\mu}^\nu \bar\Delta_V + a\bar D_\mu \bar D^\nu -\bar R_{\mu}{}^\nu$ with a constant $a$, the heat kernel coefficients have been evaluated~\cite{Endo:1984sz,Gusynin:1988zt,Gusynin:1997dc} so that
\al{
&b_0^\text{(1M)}=\frac{4-6a+3a^2}{(1-a)^2}~~\stackrel{a=-1}{=}~~\frac{13}{4}\,,&
&b_2^\text{(1M)}=\frac{10-13a+6a^2}{6(1-a)^2}~~\stackrel{a=-1}{=}~~\frac{29}{24}\,,& \nn[1ex]
&b_4^\text{(1M)}=\frac{8-10a+5a^2}{36(1-a)^2}~~\stackrel{a=-1}{=}~~\frac{23}{144}\,,&
&\hat b_4^\text{(1M)}=\frac{43-26a-2a^2}{90(1-a)^2}~~\stackrel{a=-1}{=}~~\frac{67}{360}\,,&
&\tilde b_4^\text{(1M)}=-\frac{11}{180}\,.
\tag{C24-I}
\label{App: heat kernel coefficients for measure modes in normal}
}
Again, we can translate to the Weyl basis
\al{
&\underline{b}_4^\text{(1M)}=\frac{13 - 12a + 4a^2}{36(1-a)^2}~~\stackrel{a=-1}{=}~~\frac{29}{144}\,,&
&\underline{\hat b}_4^\text{(1M)}=\frac{7  +6 a+ 8a^2}{60(1-a)^2}~~\stackrel{a=-1}{=}~~-\frac{7}{240}\,,& \nn[0.5ex]
&\underline{\tilde b}_4^\text{(1M)}=-\frac{-32 + 4a +13a^2}{180(1-a)^2}~~\stackrel{a=-1}{=}~~-\frac{23}{720}\,.
\tag{C24-II}
\label{App: heat kernel coefficients for measure modes}
}
\setcounter{equation}{24}

\subsubsection{Heat kernel coefficients for Lichnerowicz Laplacians}
For free matter fields (vector, Weyl spinor and scalar), it is convenient to define the following Laplacians, called the  ``Lichnerowicz Laplacians"~\cite{2018GReGr..50..145L},
\al{
\bar\Delta_{L0}S&=-\bar D^2 \,S=\bar \Delta_S \,S\,,\\[1ex]
\bar\Delta_{L\frac{1}{2}}\psi&=\left( -\bar D^2+\frac{\bar R}{4} \right)\psi=-\bar{\mathcal D}^2\psi\,,\\[1ex]
\bar\Delta_{L1} V_\mu&=-\bar D^2\, V_\mu +\bar R_\mu{}^{\nu}V_\nu=\bar\Delta_V V_\mu +\bar R_\mu{}^{\nu}V_\nu\,.
\label{App: Lichnerowicz Laplacian for spin1}
}
These Laplacians in an Einstein spacetime obey
\al{
&\bar\Delta_{L1}\bar D_\mu S=\bar D_\mu \bar\Delta_{L0}S\,,&
&\bar D_\mu \bar \Delta_{L1}V^\mu = \bar\Delta_{L0} \bar D_{\mu}V^\mu\,.
\label{properties for Lichnerowicz Laplacians}
}
The heat kernel coefficients for the Lichnerowicz Laplacians are summarized in Table~\ref{hkcs}.

\begin{table*}
\begin{center}
\caption{Heat kernel coefficients for the Lichnerowicz Laplacians $\bar\Delta_{Li}$ acting on individual matter fields in a general four dimensional constant background. 
The abbreviation ``T'' denotes ``transverse".}
\label{hkcs}
\begin{tabular}{|c|c|c|c|c|}
\hline
    \makebox[2cm]{}  & \makebox[2.5cm]{Vector ($i=1$)} & \makebox[3.3cm]{T-vector ($i=1$T)} & \makebox[3.8cm]{Weyl spinor ($i=\frac{1}{2}$)} & \makebox[2.5cm]{Scalar ($i=0$)} \\
     & ($A_\mu$) & ($A_\mu^\text{T}$) & $(\psi)$ & ($a$, $\sigma$, $\varphi$) \\
\hline
\hline
 $\bar \Delta_{Li}$ & $-\delta^\mu_\nu \bar D^2 +\bar R^\mu{}_\nu$ & $-\delta^\mu_\nu \bar D^2 +\bar R^\mu{}_\nu$ & $-\bar D^2 +\frac{\bar R}{4}$ & $-\bar D^2$\\
\hline
\hline
$b_0^{(i)}$ & $4$ & $3$ & $2$ & $1$ \\
\hline
$b_2^{(i)}(R)$ & $-\frac{1}{3}$ &$-\frac{1}{2}$ & $-\frac{1}{6}$ & $\frac{1}{6}$ \\
\hline
\hline
$180 b_4^{(i)}(R^2)$ & $-20$ &$-\frac{45}{2}$ & $\frac{5}{4}$ & $\frac{5}{2}$ \\
\hline
$180 {\hat b}_4^{(i)}(R_{\mu\nu}^2)$ & $86$ &$87$ & $-2$ & $-1$ \\
\hline
$180{\tilde b}_4^{(i)}(R_{\mu\nu\rho\sigma}^2)$ & $-11$ &$-12$ & $-\frac{7}{4}$ & $1$ \\
\hline
\hline
$180 \underline{b}_4^{(i)}(R^2)$ & $5$ &$\frac{5}{2}$ & $0$ & $\frac{5}{2}$ \\
\hline
$180 \underline{\hat b}_4^{(i)}(C^2)$ & $21$ &$\frac{39}{2}$ & $-\frac{9}{2}$ & $\frac{3}{2}$ \\
\hline
$180\underline{\tilde b}_4^{(i)}(E)$ & $-32$ &$-\frac{63}{2}$ & $\frac{11}{4}$ & $-\frac{1}{2}$ \\
\hline
\end{tabular}
\end{center}
\end{table*}

\subsection{Off-diagonal heat kernel expansion}
The Hessians include the following operator:
\al{
\mathcal O=\sum_k^n M^{\alpha_1\cdots \alpha_{2k}} \bar D_{(\alpha_1} \cdots \bar D_{\alpha_{2k})}\,,
}
for which the off-diagonal heat kernel method~\cite{Benedetti:2010nr} is used to evaluate the flow generator,
\al{
\tr_i[W(\bar\Delta_i) \mathcal O]
&= \int^\infty_0 \df s\,\widetilde W(s) \tr_i \left[ e^{-s\bar\Delta_i}\sum_k^n M^{\alpha_1\cdots \alpha_{2k}} \bar D_{(\alpha_1} \cdots \bar D_{\alpha_{2k})} \right] \nn
&= \int_x \sqrt{\bar g} \int^\infty_0 \df s \,\widetilde W(s) \tr_i \left[ \sum_k^n M^{\alpha_1\cdots \alpha_{2k}} H_{\alpha_1\cdots \alpha_{2k}} \right]\,.
}
Here we have
\al{
H_{\alpha_{1} \ldots \alpha_{2 n}}=& \frac{1}{(4 \pi s)^{2}}\left\{(-2 s)^{-n}\left(\bar g_{\alpha_{1} \alpha_{2}} \cdots \bar g_{\alpha_{(2 n-1)} \alpha_{2 n}}+\left(\frac{(2 n) !}{2^{n} n !}-1\right) \text { perm. }\right)\left(1+\frac{1}{6} s \bar  R\right)\right.\\ &\left.+\frac{1}{6}(-2 s)^{-(n-1)}\left(\bar g_{\alpha_{1} \alpha_{2}} \cdots \bar g_{\alpha_{2 n-3} \alpha_{2 n-2}} \bar R_{\alpha_{2 n-1} \alpha_{2 n}}+\left(\frac{(2 n) !}{2^{n}(n-1) !}-1\right) \text { perm. }\right)\right\} 
+\mathcal O(\bar R^2) \,.
}
The first two lowest order terms read
\al{
H_{\alpha\beta}
&=\frac{1}{(4\pi s)^{2}} \left\{ -\frac{1}{2s}\bar g_{\alpha\beta} \left( 1 +\frac{1}{6}s\bar R \right) +\frac{1}{6}\bar R_{\alpha\beta} \right\} \,,\\[2ex]
H_{\alpha \beta \mu \nu}&=\frac{1}{(4 \pi s)^{2}}\bigg\{ \frac{1}{4 s^{2}}\left(\bar g_{\alpha \mu} \bar g_{\beta \nu}+\bar g_{\alpha \nu} \bar g_{\beta \mu}+\bar g_{\alpha \beta} \bar g_{\mu \nu}\right)\left(1+\frac{1}{6} s\bar R\right) \nn
&\quad
\left.-\frac{1}{12 s}\left(\bar g_{\alpha \mu} \bar R_{\beta \nu}+ \bar g_{\alpha \nu} \bar R_{\beta \mu}+\bar g_{\beta \mu} \bar R_{\alpha \nu}+\bar g_{\beta \nu}  \bar R_{\alpha \mu}+\bar g_{\alpha \beta} \bar R_{\mu \nu}+\bar g_{\mu \nu} \bar R_{\alpha \beta}\right)\right\}\,.
}
For instance, one can compute the flow generators as follows:
\al{
\tr_{(0)}\left[ W(\bar\Delta_S) \bar R^{\mu\nu}\bar D_\mu \bar D_\nu \right]&=
 \int_x \sqrt{\bar g} \int^\infty_0\df s\,\widetilde W(s)\frac{1}{(4\pi s)^2}\left\{ -\frac{1}{2s}\bar R \left( 1 +\frac{1}{6}s\bar R \right) +\frac{1}{6}\bar R^{\alpha\beta}\bar R_{\alpha\beta} \right\} \nn
 &=  \frac{1}{(4\pi)^2}\int^\infty_0\df s\,{\widetilde W(s)}\left\{ -\frac{1}{2s^3}\int_x \sqrt{\bar g}\bar R - \frac{1}{12s^2}\int_x \sqrt{\bar g}\bar R^2 +\frac{1}{6s^2} \int_x \sqrt{\bar g}\bar R^{\alpha\beta}\bar R_{\alpha\beta} \right\} \nn
 &=\frac{1}{(4\pi)^2}\left\{ -\frac{1}{2}Q_3[W] \int_x \sqrt{\bar g}\bar R - \frac{1}{12}Q_2[W] \int_x \sqrt{\bar g}\bar R^2 +\frac{1}{6}Q_2[W]  \int_x \sqrt{\bar g}\bar R^{\alpha\beta}\bar R_{\alpha\beta} \right\}\,.
 }
One can see from this that the lowest order term is obtained by the replacement $\bar R^{\mu\nu}\bar D_\mu \bar D_\nu \to -  (\bar R\bar\Delta_S)/4$ (equivalently $\bar R_{\mu\nu}\to \bar g_{\mu\nu}\bar R/4$ or $-\bar D_\mu \bar D_\nu\to \bar g_{\mu\nu} \bar \Delta_S/4$).
Indeed, at the lowest order, one has
\al{
-\frac{\bar R}{4}\tr_{(0)} \left[ \bar\Delta_S W(\bar\Delta_S) \right]
\stackrel{\mathcal O(\bar R)}{=}
 -\frac{1}{4(4\pi)^2} \frac{\Gamma(3)}{\Gamma(2)} Q_3[W] \int_x \sqrt{\bar g}\bar R
=- \frac{1}{2(4\pi)^2}Q_3[W] \int_x \sqrt{\bar g}\bar R\,,
}
where we have used Eq.~\eqref{App: Delta W threshold functions}, and the values of the Gamma function, $\Gamma(3)=2$ and $\Gamma(2)=1$.

Using the off-diagonal heat kernel expansion for a curvature tensor $O^{\mu\nu\rho\sigma}$ of order of $\bar R^2$, e.g., $\bar R^{\mu\nu}\bar R^{\rho\sigma}$, one can calculate
\al{
&\tr_{(0)}\left[ W(\bar\Delta_S) O^{\mu\nu\rho\sigma} \widetilde P_{\mu\nu}\widetilde P_{\rho\sigma} \right]
\stackrel{\mathcal O(\bar R^2)}{=}
 \frac{1}{(4\pi)^2} Q_{2}[W]
\int_x\sqrt{\bar g}\left\{ \frac{13}{24}\bar g_{\mu\nu}\bar g_{\rho\sigma}  +  \frac{1}{24}(\bar g_{\mu\rho}\bar g_{\nu\sigma} + \bar g_{\mu\sigma}\bar g_{\nu\rho}) \right\}O^{\mu\nu\rho\sigma}  \,,
\label{App: flow generator with projected operator}
}
where we have performed the following computations:
\al{
\tr_{(0)}\left[  W(\bar\Delta_S) O^{\mu\nu\rho\sigma} \bar g_{\rho\sigma}\bar D_\mu (\bar\Delta_S)^{-1} \bar D_\nu \right] 
&\stackrel{\mathcal O(\bar R^2)}{=}
-\frac{Q_2[W]}{(4\pi)^2}\int_x \sqrt{\bar g}\,\frac{1}{4} \bar g_{\mu\nu}\bar g_{\rho\sigma} O^{\mu\nu\rho\sigma}\,,
}
and
\al{
\tr_{(0)}\left[  W(\bar\Delta_S) O^{\mu\nu\rho\sigma} \bar D_\mu (\bar\Delta_S)^{-1} \bar D_\nu \bar D_\rho (\bar\Delta_S)^{-1} \bar D_\sigma \right] 
&\stackrel{\mathcal O(\bar R^2)}{=}
\frac{Q_2[W]}{(4\pi)^2}\int_x \sqrt{\bar g}\,\frac{1}{24}(\bar g_{\mu\rho}\bar g_{\nu\sigma} + \bar g_{\mu\sigma}\bar g_{\nu\rho} + \bar g_{\mu\nu}\bar g_{\rho\sigma} )O^{\mu\nu\rho\sigma}  \,.
}
Using Eq.~\eqref{App: flow generator with projected operator}, the flow generator with the TT projected operator is given by
\al{
\tr_{(0)}\left[ W(\bar\Delta_T) O^{\mu\nu\rho\sigma} \left(P_t^{(0)}\right)_{\mu\nu\rho\sigma} \right]
&\stackrel{\mathcal O(\bar R^2)}{=}
\tr_{(0)}\left[ W(\bar\Delta_T) O^{\mu\nu\rho\sigma} \left\{\frac{1}{2}\left( \widetilde P_{\mu\rho} \widetilde P_{\nu\sigma} + \widetilde P_{\mu\sigma} \widetilde P_{\nu\rho} \right) - \frac{1}{3}\widetilde P_{\mu\nu} \widetilde P_{\rho\sigma} \right\} \right] \nn
&
~~=\frac{5}{(4\pi)^2} Q_2[W] \int \df^4x \sqrt{\bar g}\left\{ \frac{1}{36} \bar g_{\mu\nu}\bar g_{\rho\sigma} + \frac{1}{18} \bar g_{\mu\rho}\bar g_{\nu\sigma}  + \frac{1}{18} \bar g_{\mu\sigma}\bar g_{\nu\rho} \right\} O^{\mu\nu\rho\sigma}\,.
}

We have the projected curvature tensors by $P_t$ and $P_\sigma$ (or $\hat S_{\mu\nu}\hat S_{\rho\sigma}$) given in Eq.~\eqref{App: projector of sigma} and Eq.~\eqref{App: projector of tt}. 
Here we show explicit forms of tensor products up to of order of $R^2$ which appear in the flow generators.
The Hessian for the TT metric fluctuation involves the following terms:
\al{
\tr_{(0)}\left[ e^{-s \bar\Delta_T}  \left(P_t \right)_{\rho\sigma}{}^{\mu\nu}\delta_\nu^\sigma {\bar R}_\mu{}^\rho \right]
&\stackrel{\mathcal O(\bar R^2)}{=} \tr_{(0)}\left\{ e^{-s \bar\Delta_T}   \left[ \left( P_t^{(0)} \right)_{\mu\nu}{}^{\rho\sigma} + \left( P_t^{(1)} \right)_{\mu\nu}{}^{\rho\sigma}  \right]\delta_\nu^\sigma {\bar R}_\mu{}^\rho \right\} \nn
&=\frac{1}{(4\pi)^2}\int_x\sqrt{\bar g}\,\left[\frac{5}{4s} \bar R -\frac{5}{108} \bar R^2 + \frac{25}{54}\bar R_{\mu\nu}\bar R^{\mu\nu}  \right] \,,
\label{App: Projection operator multiplication1}
 \\[5ex]
\tr_{(0)}\left[ e^{-s \bar\Delta_T} \left(P_t \right)_{\rho\sigma}{}^{\mu\nu}{\bar R}_\mu{}^\rho{}_\nu{}^\sigma \right]
&\stackrel{\mathcal O(\bar R^2)}{=} \tr_{(0)}\left\{  e^{-s \bar\Delta_T}\left[ \left( P_t^{(0)} \right)_{\mu\nu}{}^{\rho\sigma} + \left( P_t^{(1)} \right)_{\mu\nu}{}^{\rho\sigma}  \right] {\bar R}_\mu{}^\rho{}_\nu{}^\sigma \right\} \nn
&=\frac{1}{(4\pi)^2}\int_x\sqrt{\bar g}\,\left[- \frac{5}{12 s}\bar R - \frac{5}{108}\bar R^2 + \frac{25}{54}\bar R_{\mu\nu}\bar R^{\mu\nu} \right]  \,.
}
We have also the following tensor products of order of $\bar R^2$:
\al{
&\tr_{(0)}\left[ e^{-s \bar\Delta_T} \, {\bar R}^{\rho\sigma}\left(P_t \right)_{\rho\sigma}{}^{\mu\nu}{\bar R}_{\mu\nu} \right]
\stackrel{\mathcal O(\bar R^2)}{=} \tr_{(0)}\left[ e^{-s \bar\Delta_T} \,{\bar R}^{\rho\sigma}\left(P_t ^{(0)}\right)_{\rho\sigma}{}^{\mu\nu}{\bar R}_{\mu\nu} \right]
=\frac{1}{(4\pi)^2}\int_x\sqrt{\bar g} \left[-\frac{5}{36}\bar R^2+\frac{5}{9}\bar R_{\mu\nu}\bar R^{\mu\nu} \right] \,,  \\[5ex]
&\tr_{(0)}\left[ e^{-s \bar\Delta_T}\,\bar R_{\alpha \mu\beta \nu} \left(P_t \right)_{\rho\sigma}{}^{\mu\nu}
 R^{\alpha\rho\beta\sigma} \right]
 \stackrel{\mathcal O(\bar R^2)}{=} \tr_{(0)}\left[ e^{-s \bar\Delta_T} \,\bar R_{\alpha \mu\beta \nu} \left(P_t^{(0)} \right)_{\rho\sigma}{}^{\mu\nu}
 R^{\alpha\rho\beta\sigma}  \right] \nn
 &\phantom{\tr_{(0)}\left[ e^{-s \bar\Delta_T}\,\bar R_{\alpha \mu\beta \nu} \left(P_t \right)_{\rho\sigma}{}^{\mu\nu}
 R^{\alpha\rho\beta\sigma} \right]~~}
 =\frac{1}{(4\pi)^2}\int_x\sqrt{\bar g}\left[ -\frac{5}{36}\bar R_{\mu\nu}\bar R^{\mu\nu}  + \frac{5}{12} \bar R_{\mu\nu\rho\sigma} \bar R^{\mu\nu\rho\sigma}\right] \,, \\[5ex]
&\tr_{(0)}\left[ e^{-s \bar\Delta_T}\,\bar R_{\mu}{}^{\rho}  \left(P_t \right)_{\rho\sigma}{}^{\mu\nu} \bar R_\nu{}^\sigma  \right]
\stackrel{\mathcal O(\bar R^2)}{=} \tr_{(0)}\left[ e^{-s \bar\Delta_T} \,\bar R_{\mu}{}^{\rho}  \left(P_t^{(0)} \right)_{\rho\sigma}{}^{\mu\nu} \bar R_\nu{}^\sigma \right]
=\frac{1}{(4\pi)^2} \int_x\sqrt{\bar g}\,\left[ \frac{5}{18} \bar R^2 + \frac{5}{36} \bar R_{\mu\nu}\bar R^{\mu\nu} \right] \,,\\[5ex]
&\tr_{(0)}\left[ e^{-s \bar\Delta_T} \,\bar R_{\mu\alpha}  \left(P_t \right)_{\rho\sigma}{}^{\mu\nu}\bar R^{\alpha\sigma}\delta_\nu^\rho  \right]
\stackrel{\mathcal O(\bar R^2)}{=} \tr_{(0)}\left[ e^{-s \bar\Delta_T} \,\bar R_{\mu\alpha}  \left(P_t^{(0)} \right)_{\rho\sigma}{}^{\mu\nu} \bar R^{\alpha\sigma}\delta_\nu^\rho \right]
=\frac{1}{(4\pi)^2}\int_x\sqrt{\bar g} \left[\frac{5}{4}\bar R_{\mu\nu}\bar R^{\mu\nu} \right] \,, \\[5ex]
&\tr_{(0)}\left[ e^{-s \bar\Delta_T}\, \left(P_t \right)_{\rho\sigma}{}^{\mu\nu}\bar R_\nu{}^\alpha \bar R_\mu{}^\rho{}_\alpha{}^\sigma\right]
\stackrel{\mathcal O(\bar R^2)}{=} \tr_{(0)}\left[ e^{-s \bar\Delta_T} \,\left(P_t ^{(0)}\right)_{\rho\sigma}{}^{\mu\nu}\bar R_\nu{}^\alpha \bar R_\mu{}^\rho{}_\alpha{}^\sigma   \right]
=\frac{1}{(4\pi)^2}\int_x\sqrt{\bar g}\left[-\frac{5}{12}\bar R_{\mu\nu}\bar R^{\mu\nu} \right]\,,\\[5ex]
&\tr_{(0)}\left[ e^{-s \bar\Delta_T}\,\bar R^{\alpha\beta}  \left(P_t \right)_{\rho\sigma}{}^{\mu\nu}\bar R_{\nu\alpha}{}^\sigma{}_\beta \delta_\mu^\rho \right]
\stackrel{\mathcal O(\bar R^2)}{=} \tr_{(0)}\left[ e^{-s \bar\Delta_T}\, \bar R^{\alpha\beta}  \left(P_t^{(0)} \right)_{\rho\sigma}{}^{\mu\nu}  \bar R_{\nu\alpha}{}^\sigma{}_\beta \delta_\mu^\rho \right]
=\frac{1}{(4\pi)^2}\int_x\sqrt{\bar g}\,\left[ \frac{5}{4}\bar R_{\mu\nu}\bar R^{\mu\nu} \right]\,.
 \label{App: Projection operator multiplication:last}
}

In the Hessian for the spin-0 mode, we have
\al{
\hat S^{\mu\nu} \hat S_{\mu\nu} 
&={\mathcal N}^{-1}\left[\frac{1}{3}{\mathcal N}^2  -\bar R^{\mu\nu}\bar D_\mu \bar D_\nu  -\frac{\bar R^2}{3}+\bar R^{\mu\nu}\bar R_{\mu\nu} \right]{\mathcal N}^{-1} \nn
&= \frac{1}{3} - \frac{1}{9\bar\Delta_S} \bar R^{\mu\nu}\bar D_\mu \bar D_\nu  \frac{1}{\bar\Delta_S}
+\frac{1}{9\bar\Delta_S^2}\bar R^{\mu\nu}\bar R_{\mu\nu} - \frac{2}{27\bar\Delta_S^3}\bar R\bar R^{\mu\nu}\bar D_\mu \bar D_\nu -\frac{1}{27\bar\Delta_S^2}\bar R^2 \,,\\[2ex]
\hat S^{\mu\nu} \bar R_{\mu\nu}
&= \frac{1}{3}\widetilde P^{\mu\nu} \bar R_{\mu\nu} 
-\frac{1}{3\bar\Delta_S}\bar R^{\mu\nu}\bar R_{\mu\nu}  +\frac{1}{9\bar\Delta_S}\widetilde P^{\mu\nu} \bar R_{\mu\nu} \bar R\,, \\[2ex]
\hat S^{\mu\nu}  \bar R_{\mu}{}^\rho \delta_{\nu}^\sigma \hat S_{\rho\sigma}  
&= \frac{1}{3}\widetilde P^{\mu\nu}\left(\bar R_{\mu}{}^\rho \delta_{\nu}^\sigma \right)\frac{1}{3} \widetilde P_{\rho\sigma} 
-\frac{2}{3\bar\Delta_S}\bar R^{\mu\nu} \left(  \bar R_\mu{}^\rho  \delta_\nu^\sigma\right) \frac{1}{3}\widetilde P_{\rho\sigma}
+\frac{2}{9\bar\Delta_S} \widetilde P^{\mu\nu}\bar R \left(   \bar R_\mu{}^\rho  \delta_\nu^\sigma\right) \frac{1}{3}\widetilde P_{\rho\sigma} \,. 
} 
These terms involve inverse Laplacians $(\bar\Delta_S)^{-n}$ ($n>0$). These inverse Laplacians cause an IR divergence (corresponding to zero eigenvalue of $\bar\Delta_S$) when performing the integration for $z\sim \bar\Delta_S$ within the flow equations. To avoid it, we will employ a field redefinition for $\sigma$ as given in Eq.~\eqref{sigma redefinition}.

In the flow kernel of $\sigma$, we calculate the following tensor-product terms of order of $\bar R^2$:
\al{
&\tr_{(0)}\left[ e^{-s \bar\Delta_S} \,\hat S^{\mu\nu} \bar R_{\mu\nu}  \hat S^{\rho\sigma} \bar R_{\rho\sigma} \right]
\stackrel{\mathcal O(\bar R^2)}{=} \tr_{(0)}\left[ e^{-s \bar\Delta_S}\, \left(\frac{1}{3}\widetilde P^{\mu\nu} \bar R_{\mu\nu} \right) \left(\frac{1}{3}\widetilde P^{\rho\sigma} \bar R_{\rho\sigma} \right) \right] 
=\frac{1}{(4\pi)^2}\int_x\sqrt{\bar g}\, \left[ \frac{13}{216}\bar R^2 + \frac{1}{108}\bar R_{\mu\nu}\bar R^{\mu\nu} \right]\,,\\[5ex]
&\tr_{(0)}\left[ e^{-s \bar\Delta_S}\,\hat S^{\mu\nu}  \bar R_{\alpha\mu} \delta^\tau_\nu \bar R^{\alpha\rho}  \hat S_{\rho\sigma} \right]
\stackrel{\mathcal O(\bar R^2)}{=}\tr_{(0)}\left[ e^{-s \bar\Delta_S} \, \left( \frac{1}{3}\widetilde P^{\mu\nu} \bar R_{\alpha\mu} \right)\delta^\sigma_\nu\left(\frac{1}{3}\widetilde P_{\rho\sigma} \bar R^{\alpha\rho} \right) \right]
=\frac{1}{(4\pi)^2}\int_x\sqrt{\bar g}\, \left[ \frac{1}{12}\bar R_{\mu\nu}\bar R^{\mu\nu}  \right] \,,\\[5ex]
&\tr_{(0)}\left[ e^{-s \bar\Delta_S}\,\hat S^{\mu\nu}\bar R_\mu{}^\alpha \bar R_{\alpha\nu} \right]
\stackrel{\mathcal O(\bar R^2)}{=}\tr_{(0)}\left[ e^{-s \bar\Delta_S}\, \left(\frac{1}{3}\widetilde P^{\mu\nu} \bar R_\mu{}^\alpha \bar R_{\alpha\nu} \right) \right]
=\frac{1}{(4\pi)^2}\int_x\sqrt{\bar g}\, \left[\frac{1}{4}\bar R_{\mu\nu}\bar R^{\mu\nu}  \right] \,,\\[5ex]
&\tr_{(0)}\left[ e^{-s \bar\Delta_S}\,\hat S^{\mu\nu} \Phi_{\mu\nu}{}^{\rho\sigma} \bar R_{\rho\sigma} \right]
\stackrel{\mathcal O(\bar R^2)}{=}\tr_{(0)}\left[ e^{-s \bar\Delta_S} \,\left(\frac{1}{3}\widetilde P^{\mu\nu}  \Phi_{\mu\nu}{}^{\rho\sigma} \bar R_{\rho\sigma} \right) \right] 
=0\,,\\[5ex]
&\tr_{(0)}\left[ e^{-s \bar\Delta_S} \,\hat S^{\mu\nu} \delta_\nu^\sigma \Phi_{\mu\rho}{}^{\alpha\beta} \bar R_{\alpha\beta}\hat S_{\rho\sigma} \right]
\stackrel{\mathcal O(\bar R^2)}{=}\tr_{(0)}\left[ e^{-s \bar\Delta_S} \, \left(\frac{1}{3}\widetilde P^{\mu\nu}   \Phi_{\mu\rho}{}^{\alpha\beta} \right)\delta_\nu^\sigma  \left(\frac{1}{3}\widetilde P_{\rho \sigma}   \bar R_{\alpha\beta} \right)  \right]
=0\,.  
}

\section{Summary of matter contributions}
We briefly summarize the matter contributions.
For massless $N_S$-scalars, $N_V$-vector bosons and $N_F$-Weyl fermions (corresponding to $N_D=N_F/2$ Dirac fermions), one finds
\al{
&\p_t\int_x\sqrt{\bar g}\left( U-\frac{F}{2}\bar R+ \frac{\mathcal C}{2}\bar R^2+\frac{\mathcal D}{2}\bar R_{\mu\nu}\bar R^{\mu\nu}+{\mathcal E}\bar R_{\mu\nu\rho\sigma}\bar R^{\mu\nu\rho\sigma} \right)
=N_S\pi^{(S)}_k
+N_V(\pi_k^{(V)}-\delta_k^{(V)})
+N_F \pi^{(F)}_k
\nn
&
=\frac{1}{16\pi^2}\int_x\sqrt{\bar g}\Bigg[  \left(N_S +2N_V-2N_F\right)k^4 \ell_0^4\fn{0} +\frac{1}{6}\left( N_S -4N_V +N_F \right)k^2\ell_0^2\fn{0} \bar R \nn
&\quad
+\frac{1}{360}\left\{ \left( 5N_S-50N_V -\frac{5N_F}{2} \right) \bar R^2
+\left(-2N_S+176N_V + 4N_F \right) \bar R_{\mu\nu}^2
+\left( 2N_S -26N_V +\frac{7N_F}{2}  \right)\bar R_{\mu\nu\rho\sigma}^2
\right\}\ell^0_0\fn{0}
\Bigg]\,.
}
Projecting it on each curvature operator and using the definition of the threshold functions \eqref{eq: threshold function with Litim cutoff}, one obtains
\al{
\p_t U\big|_\text{matter}&= \frac{k^4}{32\pi^2}\left(N_S +2N_V-2N_F\right)\,,\\[2ex]
\p_t F\big|_\text{matter}&=-\frac{k^2}{48\pi^2}\left( N_S -4N_V +N_F \right)\,,\\[2ex]
\p_t {\mathcal C}\big|_\text{matter}&= \frac{1}{2880\pi^2}\left( 5N_S-50N_V -\frac{5N_F}{2} \right)\,,\\[2ex]
\p_t {\mathcal D}\big|_\text{matter}&=\frac{1}{2880\pi^2}\left( -2N_S+176N_V + 4N_F\right)\,,\\[2ex]
\p_t {\mathcal E}\big|_\text{matter}&=\frac{1}{5760\pi^2}\left( 2N_S -26N_V +\frac{7N_F}{2} \right)\,.
}
For $U$ and $F$, their matter dependence agrees with Ref.\,\cite{Dona:2013qba} and our last paper~\cite{Wetterich:2019zdo}.
With the relations \eqref{eq: relation between CDE and maCmaDmaE} among coupling constants, one infers the beta functions for $R^2$, the squared Weyl tensor and the Gauss-Bonnet term,
\al{
\p_t C\big|_\text{matter}&=-\frac{1}{576\pi^2} N_S\,,\\[2ex]
\p_t D\big|_\text{matter}&=\frac{1}{960\pi^2}\left( N_S+12N_V+3N_F  \right)\,,\\[2ex]
\p_t E\big|_\text{matter}&=-\frac{1}{5760\pi^2}\left( N_S+62N_V+\frac{11}{2}N_F  \right)\,.
}
For the higher derivative terms, we obtain the same result in Refs.\,\cite{Donoghue:2014yha}.

\section{Contributions from metric fluctuations}
\label{App: Contributions from metric fluctuations}
For the gravitational system \eqref{action for higher derivative} with Eqs.\,\eqref{standard gauge fixing} and \eqref{standard ghost action}, the flow equation with the metric decomposition \eqref{App: physical metric decomposition} reads
\al{
\p_t\Gamma_k = \frac{1}{2}\tr_{(2)}\frac{\p_t \mathcal R_k}{\Gamma_k^{(2)}+\mathcal R_k}\bigg|_{ff} 
+ \frac{1}{2}\tr_{(1)}\frac{\p_t \mathcal R_k}{\Gamma_k^{(2)}+\mathcal R_k}\bigg|_{aa}
+ J_{\text{grav1},k} 
-\tr_{(1)}\frac{\p_t \mathcal R_k}{\Gamma_k^{(2)}+\mathcal R_k}\bigg|_{\bar C C} \,.
\label{App: full RG equation}
}
Here, the first term on the right-hand side is the contribution from the physical metric fluctuations, while the remaining terms are the measure contributions coming from the measure modes of metric fluctuations with the Jacobian and the ghost fields.
As one will see in Appendix~\ref{App: measure contributions}, the measure contribution takes a simple form
\al{
\eta_g:=-\frac{1}{2}\tr_{(1)}\frac{\p_t  P_k\fn{\bar{\mathcal D}_1}}{P_k\fn{\bar{\mathcal D}_1}}\bigg|_\text{measure}=\frac{1}{2}\tr_{(1)}\frac{\p_t \mathcal R_k}{\Gamma_k^{(2)}+\mathcal R_k}\bigg|_{aa}
+ J_{\text{grav1},k} 
-\tr_{(1)}\frac{\p_t \mathcal R_k}{\Gamma_k^{(2)}+\mathcal R_k}\bigg|_{\bar C C}\,,
}
with the differential operator $\bar{\mathcal D}_1$ defined in Eq.~\eqref{App: differential operators in gauge mode}.
Below, we evaluate the different contributions to Eq.~\eqref{App: full RG equation}.

\subsection{Physical metric contribution}
\label{App: Physical metric contribution}
We evaluate the physical metric contribution in Eq.~\eqref{App: full RG equation}, whose flow generator reads
\al{
\pi_k^{(f)}:= \frac{1}{2}\tr_{(2)}\frac{\p_t \mathcal R_k}{\Gamma_k^{(2)}+\mathcal R_k}\bigg|_{ff}
&=\frac{1}{2}\tr_{(2)}\frac{\p_t \mathcal R_k}{\Gamma_k^{(2)}+\mathcal R_k}\bigg|_{tt} + \frac{1}{2}\tr_{(2)}\frac{\p_t \mathcal R_k}{\Gamma_k^{(2)}+\mathcal R_k}\bigg|_{\sigma\sigma} + J_{\text{grav0},k}\nn[2ex]
&=\pi_k^{(t)} + \pi_k^{(\sigma)}+ J_{\text{grav0},k}
\,.
\label{App: flow generator from graviton contribution}
}
The explicit form of $J_{\text{grav0},k}$ is presented in Eq.~\eqref{App: regulated Jacobian for spin 0}.
The Hessian for the TT mode, $\Gamma_k^{(2)}|_{tt}$ is given in Eq.~\eqref{App: two-point function of t and sigma}.

\subsubsection{Full propagator and cutoff function}

As we will see later, the second term in Eq.~\eqref{App: kinetic term of sigma in t sigma basis} causes an IR divergence due to $\bar\Delta_S^{-1}$ in the denominator of $\hat S^{\mu\nu}\hat S_{\mu\nu}$.
To avoid this divergence, we adopt a redefinition 
\al{
\sigma=\left( \bar\Delta_S \right)^{\epsilon}\hat\sigma \,,
\label{sigma redefinition}
}
with $\epsilon$ a positive parameter. In the truncation which we employ in this work, the choice $\epsilon=1$ is enough to remove the IR divergence.

We now employ the regulator functions for the TT mode and the physical scalar mode such that Laplacians in Eqs.\,\eqref{App: kinetic term of tt} and \eqref{App: kinetic term of sigma in t sigma basis} are replaced to $P_k(\bar\Delta_i)=\bar\Delta_i+R_k(\bar\Delta_i)$,
\al{
\p_t \mathcal R_k^{(t)}&=
\frac{\p_t F}{2}\Big( P_k - \bar\Delta_T\Big)+\p_tD\Big(P_k^2-\bar\Delta_T^2 \Big) 
+\bigg( \frac{F}{2} +2DP_k \bigg) \p_t R_k\,,\nn[2ex]
\p_t {\mathcal R}_k^{(\sigma)} 
&=\frac{\p_t F}{2}\left(P_k^{1+2\epsilon}- (\bar\Delta_S)^{1+2\epsilon}\right) +3 \p_t C\left(P_k^{2+2\epsilon}-(\bar\Delta_S)^{2+2\epsilon}\right) 
-\frac{\p_t U}{4}\left( P_k^{2\epsilon}- (\bar\Delta_S)^{2\epsilon} \right) \nn
&\qquad
+ \left( \frac{F}{2}(1+2\epsilon) +3(2+2\epsilon)C P_k  -\frac{U}{4}(2\epsilon)P_k^{-1} \right)P_k^{2\epsilon}\p_t R_k\,.
}

\subsubsection{Expansion of flow generator}
Let us here calculate the propagator, i.e. the inverse of Eq.~\eqref{App: two-point function of t and sigma}.
One finds the propagators for each field,
\al{
&\frac{1}{2}\left( \Gamma_k^{(2)}+\mathcal R_k \right)^{-1}\bigg|_{tt}
= \frac{1}{{\mathbb P}_{(t)}}
-\frac{1}{{\mathbb P}_{(t)}}{\mathbb M}_{(t)}\frac{1}{{\mathbb P}_{(t)}}
+\frac{1}{{\mathbb P}_{(t)}}{\mathbb M}_{(t)}\frac{1}{{\mathbb P}_{(t)}} {\mathbb M}_{(t)} \frac{1}{{\mathbb P}_{(t)}}
+\frac{1}{{\mathbb P}_{(t)}}{\mathbb M}_{(t\sigma)}\frac{1}{{\mathbb P}_{(\sigma)}} {\mathbb M}_{(\sigma t)} \frac{1}{{\mathbb P}_{(t)}}
+\cdots\,,\\[2ex]
&\frac{1}{2}\left( \Gamma_k^{(2)}+\mathcal R_k \right)^{-1}\bigg|_{\sigma\sigma}
= \frac{1}{{\mathbb P}_{(\sigma)}}
-\frac{1}{{\mathbb P}_{(\sigma)}}{\mathbb M}_{(\sigma)}\frac{1}{{\mathbb P}_{(\sigma)}}
+\frac{1}{{\mathbb P}_{(\sigma)}}{\mathbb M}_{(\sigma)}\frac{1}{{\mathbb P}_{(\sigma)}} {\mathbb M}_{(\sigma)} \frac{1}{{\mathbb P}_{(\sigma)}}
+\frac{1}{{\mathbb P}_{(\sigma)}}{\mathbb M}_{(\sigma t)}\frac{1}{{\mathbb P}_{(t)}} {\mathbb M}_{(t\sigma)} \frac{1}{{\mathbb P}_{(\sigma)}}
+\cdots\,.
}

Then the flow generator \eqref{App: flow generator from graviton contribution} is evaluated so that
\al{
\pi_k^{(f)}
&= \pi_k^{(t)}+\pi_k^{(\sigma)} + J_{\text{grav},k}
=\frac{1}{2}\tr_{(2)}\frac{\p_t \mathcal R_k}{\Gamma_k^{(2)}+\mathcal R_k}\bigg|_{tt}+\frac{1}{2}\tr_{(0)}\frac{\p_t \mathcal R_k}{\Gamma_k^{(2)}+\mathcal R_k}\bigg|_{\sigma\sigma} + J_{\text{grav},k}\,.
\label{App: flow generators for physical metric fluctuations}
}
Here the flow generators for the TT mode and the spin 0 physical scalar mode are given, respectively, by
\al{
\pi_k^{(t)}&= \frac{1}{2}\tr_{(2)}\frac{\p_t {\mathcal R}_k^{(t)}}{{\mathbb P}_{(t)} }\bigg|_{tt} 
-\frac{1}{2}\tr_{(2)}\left[\p_t \mathcal R_k^{(t)} \cdot  {\mathbb P}_{(t)}^{-1}\cdot {\mathbb M}_{(t)} \cdot {\mathbb P}_{(t)}^{-1}\right] \bigg|_{tt} 
+\frac{1}{2}\tr_{(2)}\left[\p_t \mathcal R_k^{(t)} \cdot  {\mathbb P}_{(t)}^{-1}\cdot {\mathbb M}_{(t)} \cdot {\mathbb P}_{(t)}^{-1}\cdot {\mathbb M}_{(t)} \cdot {\mathbb P}_{(t)}^{-1}\right] \bigg|_{tt} \nn
&\quad
+\frac{1}{2}\tr_{(2)}\left[\p_t \mathcal R_k^{(t)} \cdot  {\mathbb P}_{(t)}^{-1}\cdot {\mathbb M}_{(t\sigma)} \cdot {\mathbb P}_{(\sigma)}^{-1}\cdot {\mathbb M}_{(\sigma t)} \cdot {\mathbb P}_{(t)}^{-1}\right] \bigg|_{tt} \nn[1ex]
&=:{\mathcal T}_{t}^{(1)}+{\mathcal T}_{t}^{(2)}+{\mathcal T}_{t}^{(3)}+{\mathcal T}_{t}^{(4)}\,,\\[2ex]
\pi_k^{(\sigma)}&= \frac{1}{2}\tr_{(0)}\frac{\p_t {\mathcal R}_k^{(\sigma)}}{{\mathbb P}_{(\sigma)} }\bigg|_{\sigma\sigma} 
-\frac{1}{2}\tr_{(0)}\left[\p_t {\mathcal R}_k^{(\sigma)} \cdot  {\mathbb P}_{(\sigma)}^{-1}\cdot {\mathbb M}_{(\sigma)} \cdot {\mathbb P}_{(\sigma)}^{-1}\right] \bigg|_{\sigma\sigma} 
+\frac{1}{2}\tr_{(0)}\left[\p_t {\mathcal R}_k^{(\sigma)} \cdot  {\mathbb P}_{(\sigma)}^{-1}\cdot {\mathbb M}_{(\sigma)} \cdot {\mathbb P}_{(\sigma)}^{-1}\cdot {\mathbb M}_{(\sigma)} \cdot {\mathbb P}_{(\sigma)}^{-1}\right] \bigg|_{\sigma\sigma} \nn
&\quad
+\frac{1}{2}\tr_{(0)}\left[\p_t {\mathcal R}_k^{(\sigma)} \cdot  {\mathbb P}_{(\sigma)}^{-1}\cdot {\mathbb M}_{(\sigma t)} \cdot {\mathbb P}_{(t)}^{-1}\cdot {\mathbb M}_{(t\sigma)} \cdot {\mathbb P}_{(\sigma)}^{-1}\right] \bigg|_{\sigma\sigma} \nn[1ex]
&=:{\mathcal T}_{\sigma}^{(1)}+{\mathcal T}_{\sigma}^{(2)}+{\mathcal T}_{\sigma}^{(3)}+{\mathcal T}_{\sigma}^{(4)}\,.
}

\subsubsection{Transverse-traceless mode}
The each contribution from the TT mode is calculated as follows:
The lowest order term is simply evaluated by using only the heat kernel technique such that
\al{
{\mathcal T}_{t}^{(1)}&=\frac{1}{2}\tr_{(2)}\frac{\p_t \mathcal R_k^{(t)}}{{\mathbb P}_{(t)} }\bigg|_{tt}
=:\frac{1}{2}\tr_{(2)} W_{t,0}(\bar\Delta_T)
 \nn
 &=\frac{1}{2}\frac{1}{16\pi^2} \int_x\sqrt{\bar g} \Bigg[ 5 Q_2[W_{t,0}] -\frac{5}{6}\bar R\, Q_1[W_{t,0}]  \nn
&\qquad
  + \left\{
\left( -\frac{137}{216} +\frac{N}{2\chi_E} \right){\bar R}^2 
+ \left( -\frac{17}{108} - \frac{2N}{\chi_E} \right) \bar R_{\mu\nu}\bar R^{\mu\nu}
+ \left( \frac{5}{18} +\frac{N}{2\chi_E} \right) \bar R_{\mu\nu\rho\sigma}\bar R^{\mu\nu\rho\sigma} \right\}Q_0[W_{t,0}]  \Bigg] \,.
}
Here we define the flow kernel of the TT mode, (with $\epsilon=0$ in Eq.~\eqref{App: flow kernel example})
\al{
W_{t,p}(\bar \Delta_T)=\frac{\p_t \mathcal R_k^{(t)}}{\left( \frac{F}{2}P_k +D P_k^2 -U \right)^{p+1}}\,.
}


In the next order, we need to calculate the tensor products between the projection operator and the curvature tensors.
One has
\al{
{\mathcal T}_{t}^{(2)}&=-\frac{1}{2}\tr_{(2)}\left[\p_t \mathcal R_k^{(t)} \cdot  {\mathbb P}_{(t)}^{-1}\cdot {\mathbb M}_{(t)} \cdot {\mathbb P}_{(t)}^{-1}\right] \bigg|_{tt} \nn
&= -\frac{1}{2}\tr \left[ W_{t,1}(\bar\Delta_T)\left\{ \frac{F}{2}\bar R +    H\bar R \bar\Delta_T +  2D \bar R^{\alpha\beta}\bar D_{\alpha}\bar D_\beta  + \frac{H}{2}\bar R^2  - D \bar R^{\alpha\beta}\bar R_{\alpha\beta}  \right\} \left(P_t \right)_{\mu\nu}{}^{\mu\nu} \right]  \nn
&\quad
 -\frac{1}{2}\tr\Bigg[W_{t,1}(\bar\Delta_T)\bigg\{
 - 2 \frac{F}{2} \left(  \delta_\nu^\sigma {\bar R}_\mu{}^\rho  + {\bar R}_\mu{}^\rho{}_\nu{}^\sigma \right)  
-4D \bar\Delta_T {\bar R}_\mu{}^\rho{}_\nu{}^\sigma
 - 2 H\bar R \left(  \delta_\nu^\sigma {\bar R}_\mu{}^\rho  + {\bar R}_\mu{}^\rho{}_\nu{}^\sigma \right) 
-2H{\bar R}_{\mu\nu}{\bar R}^{\rho\sigma} \nn
&\qquad 
+D\Big(
 4 \bar R_{\alpha \mu\beta \nu}\bar R^{\alpha\rho\beta\sigma} 
 +2 \bar R_{\mu}{}^{\rho} \bar R_\nu{}^\sigma 
 -2 \bar R_{\mu\alpha}\bar R^{\alpha\sigma}\delta_\nu^\rho 
 +4\bar R_\nu{}^\alpha \bar R_\mu{}^\rho{}_\alpha{}^\sigma
  +4 \bar R^{\alpha\beta}\bar R_{\nu\alpha}{}^\sigma{}_\beta \delta_\mu^\rho
\Big)
  \bigg\}\left(P_t \right)_{\rho\sigma}{}^{\mu\nu}\Bigg]  \nn
&=-\frac{1}{2}\frac{1}{16\pi^2}\int_x\sqrt{\bar g}\Bigg[
\left\{ \left( 10C + 5D  \right)Q_3[W_{t,1}]
+ \frac{10}{3}wk^2 Q_2[W_{t,1}] \right\} \bar R
+ \left\{ \left( \frac{5}{18}C +\frac{35}{54}D  \right) Q_2[W_{t,1}] -\frac{35}{54}wk^2Q_1[W_{t,1}]  \right\}\bar R^2 \nn
&\qquad
+ \left\{ \left( -\frac{10}{9}C - \frac{115}{27}D \right)Q_2[W_{t,1}] -\frac{50}{27}wk^2Q_1[W_{t,1}]  \right\} \bar R_{\mu\nu}\bar R^{\mu\nu}
+ \left\{ \frac{5}{3}D Q_2[W_{t,1}]   \right\}\bar R_{\mu\nu\rho\sigma}\bar R^{\mu\nu\rho\sigma}
\Bigg]\,,
}
 
 The next-to-next order reads
 \al{
 {\mathcal T}_{t}^{(3)}&= \frac{1}{2}\tr_{(2)}\left[\p_t \mathcal R_k^{(t)} \cdot  {\mathbb P}_{(t)}^{-1}\cdot {\mathbb M}_{(t)} \cdot {\mathbb P}_{(t)}^{-1}\cdot {\mathbb M}_{(t)} \cdot {\mathbb P}_{(t)}^{-1}\right] \bigg|_{tt} \nn
 &= \frac{1}{2} \tr\Bigg[W_{t,2}(\bar\Delta_T)
 \left\{ \left[ \frac{F}{2}\bar R +    H\bar R\bar\Delta_T   +D \left( 2\bar R^{\alpha\beta}\bar D_{\alpha}\bar D_\beta \right) \right]  E_{\mu\nu}{}^{\rho\sigma}
- \frac{F}{2} 2\delta_\nu^\sigma {\bar R}_\mu{}^\rho 
-2 \left[  \frac{F}{2} + 2D \bar\Delta_T   \right]{\bar R}_\mu{}^\rho{}_\nu{}^\sigma
\right\} \nn
&\quad
\times
 \left\{ \left[\frac{F}{2}\bar R +    H\bar R\bar\Delta_T   +D \left( 2\bar R^{\alpha\beta}\bar D_{\alpha}\bar D_\beta \right) \right]  E_{\rho\sigma}{}^{\gamma\delta}
- \frac{F}{2} 2\delta_\rho^\delta {\bar R}_\sigma{}^\gamma 
-2 \left[  \frac{F}{2} + 2D \bar\Delta_T   \right] {\bar R}_\rho{}^\gamma{}_\sigma{}^\delta
\right\} \left( P_t\right)_{\gamma\delta}{}^{\mu\nu}\Bigg]
\nn[2ex]
&= \frac{1}{2}\frac{1}{16\pi^2}\int_x\sqrt{\bar g}
\Bigg[
\Bigg\{ \left( 30C^2 +30CD -\frac{5}{3}D^2 \right)Q_4[W_{t,2}] + wk^2\left( \frac{40}{3}C + \frac{80}{9} D \right) Q_3[W_{t,2}] 
 +  \frac{5}{3}(wk^2)^2 Q_2[W_{t,2}]  \Bigg\} \bar R^2\nn
&\qquad
 + \left\{10D^2 Q_4[W_{t,2}] -\frac{160}{9}wk^2 D Q_3[W_{t,2}] + \frac{10}{9} (wk^2)^2 Q_2[W_{t,2}] \right\} \bar R_{\mu\nu}\bar R^{\mu\nu} \nn
 &\qquad
 +\left\{ 40D^2 Q_4[W_{t,2}] +\frac{40}{3}wk^2 D Q_3[W_{t,2}] +\frac{5}{3}(wk^2)^2 Q_2[W_{t,2}] \right\} \bar R_{\mu\nu\rho\sigma}\bar R^{\mu\nu\rho\sigma}
 \Bigg]\,.
}

The mixing term is given by
\al{
{\mathcal T}_t^{(4)}&=\frac{1}{2}\tr_{(2)}\left[\p_t \mathcal R_k^{(t)} \cdot  {\mathbb P}_{(t)}^{-1}\cdot {\mathbb M}_{(t\sigma)} \cdot {\mathbb P}_{(\sigma)}^{-1}\cdot {\mathbb M}_{(\sigma t)} \cdot {\mathbb P}_{(t)}^{-1}\right] \bigg|_{tt}  \nn
&=\frac{1}{2}\left(-\frac{3}{2}\right) \tr_{(0)}\Bigg[ W_{t \sigma t}[\bar\Delta_S] 
\bar\Delta_S\Bigg\{
{\bar R}^{\rho\sigma} \left( \frac{U}{3\bar\Delta_S} -\frac{2}{3}\frac{F}{2} +2 H \bar\Delta_S -\frac{5}{3} D\bar\Delta_S \right) 
 \Bigg\}\nn
&\phantom{\frac{1}{2}\left(-\frac{3}{2}\right) \tr_{(0)}\Bigg[ W_{t \sigma t}[\bar\Delta_S] 
\bar\Delta_S\Bigg\{
} \times
\Bigg\{
\left( \frac{U}{3\bar\Delta_S} -\frac{2}{3}\frac{F}{2} +2 H \bar\Delta_S -\frac{5}{3} D\bar\Delta_S \right) {\bar R}_{\alpha\beta}
 \Bigg\}\bar\Delta_S
\left( P_t \right)^{\alpha\beta}{}_{\mu\nu}\bar E^{\mu\nu}{}_{\rho\sigma} \Bigg] \nn
&=\frac{1}{2}\frac{1}{16\pi^2}\int_x\sqrt{\bar g}
\Bigg[ 
\Bigg\{ \left( 400C^2 - \frac{400}{3}CD +\frac{100}{9}D^2\right) Q_6[W_{\sigma t\sigma}] 
-\frac{80}{9}(6C-D)wk^2Q_5[W_{\sigma t\sigma}]\nn
& \quad
+ \frac{10}{9} \left( 2(w k^2)^2  + (6C  - D)  uk^4 \right)Q_4[W_{\sigma t\sigma}]
-\frac{20}{27}(wk^2)(uk^4)Q_3[W_{\sigma t\sigma}]
 +\frac{5}{54}(uk^4)^2 Q_2[W_{\sigma t\sigma}]
 \Bigg\}
 \left(\frac{\bar R^2}{4}-\bar R^{\mu\nu}\bar R_{\mu\nu} \right) 
\Bigg]\,,
}
where the flow kernel is defined as
\al{
W_{t\sigma t}(\bar \Delta_S)=\frac{\p_t \mathcal R^{(t)}_k}{\left( [\frac{F}{2}P_k+3CP_k^2 -\frac{U}{4}] (P_k)^{2\epsilon}\right)\left(\frac{F}{2}P_k+DP_k^2 -U\right)^2}\,.
}

\subsubsection{Physical spin-0 mode}
Let us evaluate the contribution from the spin-0 physical scalar mode.
To this end, we define the flow kernel as
\al{
W_{\sigma,p,\epsilon}(\bar\Delta_S)=\frac{\p_t {\mathcal R}_k^{(\sigma)}}{\left(\left[\frac{F}{2}P_k + 3C P_k^2 -\frac{U}{4} \right] P_k^{2\epsilon}\right)^{p+1}}\,.
\label{App: Wsigma}
}
$P_k^{2\epsilon}$ in the denominator of Eq.~\eqref{App: Wsigma} comes from the redefinition of $\sigma$ as given in Eq.~\eqref{sigma redefinition}.
In order to remove divergences arising from $1/\bar\Delta_S$ in the Hessian $\Gamma_{(\sigma\sigma)}^{(2)}$, setting $\epsilon=1$ is sufficient in the current truncation.

The lowest order term reads
\al{
{\mathcal T}_{\sigma}^{(1)}&=\frac{1}{2}\tr_{(0)}\frac{\p_t {\mathcal R}_k^{(\sigma)}}{{\mathbb P}_{(\sigma)} }\bigg|_{\sigma\sigma}
=:\frac{1}{2}\tr_{(0)} W_{\sigma,0,1}(\bar\Delta_S)
\nn
&= \frac{1}{2} \frac{1}{16\pi^2} \int_x\sqrt{\bar g} \Bigg[ 
Q_2[ W_{\sigma,0,1}] 
+\frac{\bar R}{6}Q_1[W_{\sigma,0,1}]
+\left(\frac{1}{72}{\bar R}^2- \frac{1}{180}\bar R_{\mu\nu}\bar R^{\mu\nu} + \frac{1}{180}\bar R_{\mu\nu\rho\sigma}\bar R^{\mu\nu\rho\sigma} \right)Q_0[W_{\sigma,0,1}]
\Bigg]\,.
}
We next calculate the next order contributions for which we need to use the formulae \eqref{App: Projection operator multiplication1}--\eqref{App: Projection operator multiplication:last}.
\al{
{\mathcal T}_{\sigma}^{(2)}&=-\frac{1}{2}\tr_{(0)}\left[\p_t {\mathcal R}_k^{(\sigma)} \cdot  {\mathbb P}_{(\sigma)}^{-1}\cdot {\mathbb M}_{(\sigma)}\cdot {\mathbb P}_{(\sigma)}^{-1}\right] \bigg|_{\sigma\sigma}  \nn
&= -\frac{1}{2}\frac{3}{2}\tr\left[W_{\sigma,1,1}(\bar\Delta_S)\bar\Delta_S^2
\left\{ 
\left[ \left( \frac{F}{2}\bar\Delta_S  + D \bar\Delta_S^2 - U \right)+\frac{F}{2}\bar R +H\bar R \bar\Delta_S + 2D\bar R^{\alpha\beta}\bar D_\alpha \bar D_\beta  \right]
\left(\frac{1}{3}-\hat S^{\mu\nu}\hat S_{\mu\nu} \right)
  \right\} 
   \right] \nn
&\quad
-\frac{1}{2}\frac{3}{2} \tr\Bigg[W_{\sigma,1,1}(\bar\Delta_S)\bar\Delta_S^2
\Bigg\{
\frac{1}{6}\left[ {\mathcal X}\bar R  + 2D\bar R^{\alpha\beta}\bar D_\alpha \bar D_\beta\right]
+\frac{2}{3}H\left[ \bar R  -6\hat S^{\mu\nu} {\bar R}_{\mu\nu}  \right]\bar\Delta_S
\Bigg\}  \Bigg] \nn
&\quad
 -\frac{1}{2} \frac{3}{2}\tr\Bigg[W_{\sigma,1,1}(\bar\Delta_S)\bar\Delta_S^2
\Bigg\{ 
2{\mathcal X}\hat S^{\mu\nu} \left(  {\bar R}_{\mu}{}^\rho \delta_{\nu}^\tau \hat S_{\rho\tau} - {\bar R}_{\mu\nu}  \right) 
+2\left [{\mathcal X}  +2D\bar\Delta_S  \right]\hat S^{\mu\nu} \bar R_{\mu}{}^{\rho}{}_\nu{}^\tau \left(\hat S_{\rho\tau}- \bar D_\rho \bar D_\tau {\mathcal N}^{-1}\right) \nn
&\qquad\qquad
-\frac{1}{3}H\left[ \frac{1}{4}\bar R^2 - 6\hat S^{\mu\nu} {\bar R}_{\mu\nu}{\bar R}^{\rho\tau}\hat S_{\rho\tau} \right]
-D\bigg[
\frac{1}{6} \bar R^{\alpha\beta}\bar R_{\alpha\beta}
+10 \hat S^{\mu\nu}  \bar R_{\alpha\mu} \delta^\tau_\nu \bar R^{\alpha\rho}  \hat S_{\rho\tau}
 -4 \hat S^{\mu\nu}\bar R_\mu{}^\alpha \bar R_{\alpha\nu}
 +2 \hat S^{\mu\nu} \Phi_{\mu\nu}{}^{\rho\tau} \bar R_{\rho\tau}
     \nn
 &\qquad\qquad
 + 2\hat S^{\mu\nu}\bar R_\mu{}^{\rho}\bar R_\nu{}^\tau \left( \hat S_{\rho\tau} - 2 \bar D_\rho  \bar D_\tau{\mathcal N}^{-1}  \right)
  -2 \hat S^{\mu\nu} \delta_\nu^\tau \Phi_{\mu\rho}{}^{\alpha\beta} \bar R_{\alpha\beta}\hat S_{\rho\tau}
  -2\hat S^{\mu\nu}   \Phi_{\mu\nu}{}^{\alpha\beta} \delta_\alpha^\rho\bar R_\beta{}^{\tau} \left( \hat S_{\rho\tau}  - \bar D_\rho  \bar D_\tau{\mathcal N}^{-1} \right)\bigg]
 \Bigg\}
 \Bigg] \nn[1ex]
 &=\frac{1}{2}\frac{1}{16\pi^2}\int_s\sqrt{\bar g}\Bigg[
 \left\{
 \frac{1}{4}wk^2 Q_4[W_{\sigma,1,1}]  -\frac{uk^4}{12} Q_3[W_{\sigma,1,1}]   + 12 C Q_5[W_{\sigma,1,1}]   \right\}\bar R
\nn
&\qquad
 + \bigg\{ -\frac{5}{36}wk^2 Q_3[W_{\sigma,1},1]+ \frac{35}{12} C Q_4[W_{\sigma,1,1}]
-\frac{5}{36} D Q_4[W_{\sigma,1,1}]+\frac{1}{72}uk^4 Q_2[W_{\sigma,1,1}] \bigg\}\bar R^2 \nn
&\qquad
+ \left\{ \frac{13}{18}wk^2 Q_3[W_{\sigma,1,1}]  - \frac{61}{6}CQ_4[W_{\sigma,1,1}] +\frac{5}{9}D Q_4[W_{\sigma,1,1}]  -\frac{5}{36} uk^4Q_2[W_{\sigma,1,1}]  \right\}\bar R_{\mu\nu}\bar R^{\mu\nu}
  \Bigg]\,.
}

The next-to-next order contribution reads
\al{
{\mathcal T}_{\sigma}^{(3)}&=\frac{1}{2}\tr_{(0)}\left[\p_t {\mathcal R}_k^{(\sigma)} \cdot  {\mathbb P}_{(\sigma)}^{-1}\cdot {\mathbb M}_{(\sigma)}  \cdot {\mathbb P}_{(\sigma)}^{-1}\cdot {\mathbb M}_{(\sigma)}  \cdot {\mathbb P}_{(\sigma)}^{-1}\right] \bigg|_{\sigma\sigma} \nn
&= \frac{1}{2}\left( \frac{3}{2}\right)^2\tr\Bigg[W_{\sigma,2,1}(\bar\Delta_S)\bar\Delta_S^4
\bigg\{ \left( \frac{F}{2}\bar\Delta_S  + D \bar\Delta_S^2 - U \right)\left(\frac{1}{3}-\hat S^{\mu\nu}\hat S_{\mu\nu} \right)
+H\left[ \frac{1}{2}\bar R  -4\hat S^{\mu\nu} {\bar R}_{\mu\nu}  \right]\bar\Delta_S \nn
&\qquad\qquad
+\left[ \frac{F}{2}\bar R +H\bar R \bar\Delta_S + 2D\bar R^{\alpha\beta}\bar D_\alpha \bar D_\beta \right]\left( \frac{1}{2}- \hat S^{\mu\nu}\hat S_{\mu\nu}\right) 
+2{\mathcal X}\hat S^{\mu\nu} \left(  {\bar R}_{\mu}{}^\rho \delta_{\nu}^\tau \hat S_{\rho\tau} - {\bar R}_{\mu\nu}  \right) 
\nn
&\qquad\qquad
+2\left [{\mathcal X}  +2D\bar\Delta_S  \right]\hat S^{\mu\nu} \bar R_{\mu}{}^{\rho}{}_\nu{}^\tau \left(\hat S_{\rho\tau}- \bar D_\rho \bar D_\tau {\mathcal N}^{-1}\right) \bigg\}^2
 \Bigg] \nn[1ex]
 &=\frac{1}{2}\frac{1}{16\pi^2}\int_x\sqrt{\bar g}\Bigg[
 \bigg\{ 
wk^2 \left( \frac{5}{36} wk^2 Q_6[W_{\sigma,2,1}]+20CQ_7[W_{\sigma,2,1}] -\frac{uk^4}{9} Q_5[W_{\sigma,2,1}] \right)
 +840C^2 Q_8[W_{\sigma,2,1}]
 \nn
 &\quad
+uk^4\left( \frac{uk^4}{144} Q_4[W_{\sigma,2,1}]  - \frac{20}{3} C Q_6[W_{\sigma,2,1}] \right)
 \bigg\} \bar R^2 
 + \bigg\{
wk^2 \left(  \frac{5}{18}wk^2 Q_6[W_{\sigma,2,1}] +40 C Q_7[W_{\sigma,2,1}]+\frac{uk^4}{9}Q_5[W_{\sigma,2,1}] \right) \nn
&\qquad
+1680 C^2Q_8[W_{\sigma,2,1}]
+uk^4\left( \frac{uk^4}{72} Q_4[W_{\sigma,2,1}]+\frac{20}{3} C Q_6[W_{\sigma,2,1}] \right)
 \bigg\} \bar R_{\mu\nu}\bar R^{\mu\nu}
 \Bigg]\,.
 }

Finally we evaluate the mixing effect which reads
\al{
{\mathcal T}_{\sigma}^{(4)}&=\frac{1}{2}\tr_{(0)}\left[\p_t {\mathcal R}_k^{(\sigma)} \cdot  {\mathbb P}_{(\sigma)}^{-1}\cdot {\mathbb M}_{(\sigma t)} \cdot {\mathbb P}_{(t)}^{-1}\cdot {\mathbb M}_{(t\sigma)} \cdot {\mathbb P}_{(\sigma)}^{-1}\right] \bigg|_{\sigma\sigma} \nn
&=\frac{1}{2}\left(-\frac{3}{2} \right)\tr_{(0)}\Bigg[ W_{\sigma t\sigma}[\bar\Delta_S] 
\bar\Delta_S\Bigg\{
\left( \frac{U}{3\bar\Delta_S} -\frac{2}{3}\frac{F}{2} +2 H \bar\Delta_S -\frac{5}{3} D\bar\Delta_S \right) 
{\bar R}^{\rho\sigma} \Bigg\}
\left( P_t \right)_{\rho\sigma}{}^{\alpha\beta} \nn
&\phantom{\frac{1}{2}\left(-\frac{3}{2} \right)\tr_{(0)}\Bigg[ W_{\sigma t\sigma}[\bar\Delta_S] 
\bar\Delta_S\Bigg\{} \times 
\Bigg\{
{\bar R}_{\alpha\beta}\left( \frac{U}{3\bar\Delta_S} -\frac{2}{3}\frac{F}{2} +2 H \bar\Delta_S -\frac{5}{3} D\bar\Delta_S \right) \Bigg\}\bar\Delta_S
\Bigg]\nn
&=\frac{1}{2}\frac{1}{16\pi^2}\int_x\sqrt{\bar g}
\Bigg[ 
\Bigg\{ \left( 400C^2 - \frac{400}{3}CD +\frac{100}{9}D^2\right) Q_6[W_{\sigma t\sigma}] 
-\frac{80}{9}(6C-D)wk^2Q_5[W_{\sigma t\sigma}]\nn
&\quad 
+ \frac{10}{9} \left( 2(w k^2)^2  + (6C  - D)  uk^4 \right)Q_4[W_{\sigma t\sigma}]
-\frac{20}{27}(wk^2)(uk^4)Q_3[W_{\sigma t\sigma}]
 +\frac{5}{54}(uk^4)^2 Q_2[W_{\sigma t\sigma}]
 \Bigg\} \left(\frac{\bar R^2}{4}-\bar R^{\mu\nu}\bar R_{\mu\nu} \right) 
\Bigg]\,,
}
where the flow kernel is given by
\al{
W_{\sigma t\sigma}[\bar\Delta_S]=\frac{\p_t {\mathcal R}_k^{(\sigma)}}{\left(\frac{F}{2}P_k+DP_k^2 -U\right)\left( [\frac{F}{2}P_k+3CP_k^2 -\frac{U}{4}] (P_k)^{2\epsilon}\right)^2  } \,.
}

\subsubsection{Jacobian}
\label{evaluation of Jacobian}
We have seen in Section~\ref{App: Jacobian} that the decomposition of the physical metric fluctuation \eqref{App: Jacobian from metric decomposition} and the field redefinition \eqref{sigma redefinition} with $\epsilon=1$ yield the Jacobian as
\al{
J_\text{grav2}=  \left[ \det{}_{(2)}\pmat{
E_{(t)}^{\mu\nu}{}_{\rho\sigma} &&& -\bar R^{\mu\nu}{\mathcal N}^{-1}\bar\Delta_S\\[2ex]
-\bar\Delta_S{\mathcal N}^{-1}\bar R_{\rho\sigma}&&&\bar\Delta_S \hat S^{\alpha\beta}\hat S_{\alpha\beta}\bar\Delta_S
}\right]^{1/2}
=:\left[\det{}_{(2)} \left( \Gamma_\text{Jac}^{(2)}\right)\right]^{1/2}.
\label{Jacobian as determinant}
}
Here the $(2,2)$-component and the off-diagonal parts are 
\al{
\bar\Delta_S \hat S^{\alpha\beta}\hat S_{\alpha\beta}\bar\Delta_S
&\simeq \frac{1}{3}\left( \bar\Delta_S^2 - \frac{1}{3} \bar R^{\mu\nu}\bar D_\mu \bar D_\nu  
 -\frac{1}{9}\bar R^2 +\frac{1}{3}\bar R^{\mu\nu}\bar R_{\mu\nu} + \frac{2}{9}\bar R\bar R^{\mu\nu}\frac{-\bar D_\mu \bar D_\nu}{\bar\Delta_S} \right)
 =: \frac{1}{3}(\bar\Delta_S^2 + M_\text{Jac0})\,,
 \label{App: DelSSDel}\\[2ex]
-\bar R_{\rho\sigma}{\mathcal N}^{-1}\bar\Delta_S&\simeq 
-\frac{1}{3}\bar R_{\rho\sigma}.
}
One can write the Jacobian \eqref{Jacobian as determinant} by using auxiliary fields as
\al{
\left[\det{}_{(2)} \left( \Gamma_\text{Jac}^{(2)}\right)\right]^{1/2}
&=\left[\det{}_{(2)} \left( \Gamma_\text{Jac}^{(2)}\right)\right]\left[\det{}_{(2)} \left( \Gamma_\text{Jac}^{(2)}\right)\right]^{-1/2} \nn
&=\int \mathcal D \bar{\bvec \chi} \mathcal D {\bvec \chi} \mathcal D {\bvec\theta}\, \exp\Bigg[
-\int_x\sqrt{\bar g}\, \bar{\bvec{\chi}}^T \left(\Gamma_\text{Jac}^{(2)} \right) {\bvec\chi}
-\frac{1}{2}\int_x\,\sqrt{\bar g}\, {\bvec\theta}^T\left( \Gamma_\text{Jac}^{(2)} \right) {\bvec\theta}
\Bigg],
}
where $\bar{\bvec\chi}$, ${\bvec\chi}$ are Grassmannian variables, while ${\bvec\theta}$ is an ordinary real variable.
These field can be decomposed into the TT mode and a scalar mode, i.e. ${\bvec\chi}^T=(\chi^\text{TT}_{\mu\nu}, \chi)$ and ${\bvec\theta}^T=(\theta^\text{TT}_{\mu\nu}, \theta)$.
We insert regulators for these fields as 
\al{
&\int_x\sqrt{\bar g} \bar{\bvec\chi}\left(\Gamma_\text{Jac}^{(2)} \right) {\bvec\chi}  + \frac{1}{2}\int_x\sqrt{\bar g} {\bvec\theta}\left( \Gamma_\text{Jac}^{(2)} \right) {\bvec\theta} 
\to \int_x\sqrt{\bar g} \bar{\bvec\chi} \left(\Gamma_\text{Jac}^{(2)} + {\mathcal R}_k^\text{Jac} \right) {\bvec\chi}  + \frac{1}{2}\int_x\sqrt{\bar g} {\bvec\theta}\left( \Gamma_\text{Jac}^{(2)} +{\mathcal R}_k^\text{Jac}\right){\bvec\theta}.
}
Here $\mathcal R_k^\text{Jac}$ replaces the squared Laplacian in Eq.~\eqref{App: DelSSDel} to $P_k(\bar\Delta_S)$. More specifically, we give
\al{
\mathcal R_k^\text{Jac}= \pmat{
(\mathcal R_k^\text{Jac2})^{\mu\nu}{}_{\rho\sigma}  & 0\\
0 & \frac{1}{3}\mathcal R_k^\text{Jac0}
}
=\pmat{
-\frac{1}{2}E_{(t)}^{\mu\nu}{}_{\rho\sigma} & 0\\
0 & \frac{1}{3}\left( P_k^2 -\bar\Delta_S^2 \right)
}.
\label{Regulator for Jacobian}
}
Then, the Jacobian \eqref{Jacobian as determinant} as a flow generator reads
\al{
J_{\text{grav2},k}
&=- \tr_{(2)} \frac{\p_t \mathcal R_k^\text{Jac}}{ \Gamma_\text{Jac}^{(2)}+ \mathcal R_k^\text{Jac}}
+\frac{1}{2} \tr_{(2)} \frac{\p_t \mathcal R_k^\text{Jac}}{ \Gamma_\text{Jac}^{(2)}+ \mathcal R_k^\text{Jac}}
=-\frac{1}{2} \tr_{(2)} \frac{\p_t \mathcal R_k^\text{Jac}}{ \Gamma_\text{Jac}^{(2)}+ \mathcal R_k^\text{Jac}}.
}

We evaluate the inverse matrix
\al{
\left( \Gamma_\text{Jac}^{(2)} + \mathcal R^\text{Jac}_k \right)^{-1}
=\pmat{
\frac{1}{2}E_{(t)}^{\mu\nu}{}_{\rho\sigma} &&& -\frac{1}{3}\bar R^{\mu\nu}\\[2ex]
-\frac{1}{3}\bar R_{\rho\sigma} &&&  \frac{1}{3} (P_k^2 + M_\text{Jac0})
}^{-1}.
}
As in Eq.~\eqref{Regulator for Jacobian}, the regulator matrix ${\mathcal R}_k^\text{Jac}$ has only a finite term in the (2,2)-component (i.e. scalar-mode component), so that it is enough to take the same component in the inverse matrix, i.e. up to the squared curvature operators, one has
\al{
\tr_{(2)}\frac{\p_t \mathcal R^\text{Jac}_k}{\Gamma_\text{Jac}^{(2)} + \mathcal R^\text{Jac}_k}
&= \tr_{(0)}\frac{\p_t \mathcal R^\text{Jac0}_k}{P_k^2 +M_\text{Jac0}}
+\tr_{(0)}\left[\frac{1}{\frac{1}{3}P_k^2} \Big[ (-\tfrac{1}{3}\bar R_{\rho\sigma}) (2P_t)^{\rho\sigma}{}_{\mu\nu} (-\tfrac{1}{3}\bar R^{\mu\nu}) \Big] \frac{1}{\frac{1}{3}P_k^2}\frac{1}{3}\p_t \mathcal R^\text{Jac0}_k \right]\nn[3ex]
&=\tr_{(0)}\frac{\p_t \mathcal R^\text{Jac0}_k}{P_k^2 + M_\text{Jac0}}
+\frac{2}{3}\tr_{(0)} \frac{\p_t \mathcal R^\text{Jac0}_k}{P_k^4} \Big[ \bar R_{\rho\sigma} (P_t)^{\rho\sigma}{}_{\mu\nu}\bar R^{\mu\nu}\Big] \,.
}
where the product between the off-diagonal parts and the unity matrix in the $(1,1)$-component has to be understood as that between the off-diagonal parts and the TT-mode projector.

Then, the flow equation from these contributions reads
\al{
J_{\text{grav0},k}&=-\frac{1}{2} \tr_{(0)}\frac{\p_t \mathcal R^\text{Jac0}_k(\bar\Delta_S) }{P_k^2 + M_\text{Jac0}}
 -\frac{1}{3}\tr_{(0)} \frac{\p_t\mathcal R^\text{Jac0}_k(\bar\Delta_S)}{P_k^4} (\bar R_{\mu\nu} (P_t)^{\mu\nu}{}_{\rho\sigma}\bar R^{\rho\sigma})
\nn
&= -\frac{1}{2}\frac{1}{16\pi^2}\int_x\sqrt{\bar g}
\Bigg[
Q_2[W_\text{Jac,0}]+\frac{1}{6} \left\{ Q_1[W_\text{Jac,0}]- Q_3[W_\text{Jac,1}] \right\} \bar R
+ \frac{1}{216}\left\{3Q_0[W_\text{Jac,0}] -4Q_2[W_\text{Jac,1}] +6Q_4[W_\text{Jac,2}] \right\} \bar R^2 \nn
&\qquad
+ \frac{1}{540} \left\{ -3Q_0[W_\text{Jac,0}] -50Q_2[W_\text{Jac,1}] + 30Q_4[W_\text{Jac,2}]  \right\} \bar R_{\mu\nu}\bar R^{\mu\nu} +\frac{1}{180}Q_0[W_\text{Jac,0}]\bar R_{\mu\nu\rho\sigma}\bar R^{\mu\nu\rho\sigma} \Bigg] \nn
&\quad
-\frac{1}{16\pi^2}\int_x\sqrt{\bar g}\, \frac{5}{27}Q_2[W_{\text{Jac},1}] \left(\bar R_{\mu\nu}\bar R^{\mu\nu} -\frac{\bar R^2}{4} \right)
\,.
\label{App: regulated Jacobian for spin 0}
}
where the flow kernel of the Jacobian is
\al{
W_{\text{Jac},p}[\bar \Delta_S]=\frac{\p_t \mathcal R_k^\text{Jac}(\bar\Delta_S)}{(P_k^2)^{p+1}}\,.
}

\subsection{Measure contribution}
\label{App: measure contributions}
Let us first evaluate the measure contribution without decomposing into the transverse spin-1 vector mode and spin-0 scalar mode, i.e.,
\al{
\eta_g=\delta^{(g)}_k-\epsilon^{(g)}_k=\frac{1}{2}\text{tr}_{(1)}\frac{\p_t \mathcal R_k}{\Gamma_k^{(2)}+\mathcal R_k}\Bigg|_{aa} +J_{\text{grav1},k} -\text{tr}_{(1)}\frac{\p_t \mathcal R_k}{\Gamma_k^{(2)}+\mathcal R_k}\Bigg|_{\bar C C}\,.
\label{eta_gk}
}
The contribution from the spin-1 metric fluctuation (first term on the right-hand side) has the squared differential operator, $(\bar{\mathcal D}_1)^2$, as given in Eq.~\eqref{App: total spin-1 hessian}, so that one finds
\al{
\delta^{(g)}_k\big|_a
=\frac{1}{2}\text{tr}_{(1)}\frac{\p_t \mathcal R_k}{\Gamma_k^{(2)}+\mathcal R_k}\Bigg|_{aa}+J_{\text{grav1},k}
=\frac{1}{2}\text{tr}_{(1)}\frac{\p_t \mathcal R_k}{ \bar{\mathcal D}_1+\mathcal R_k}\,,
} 
where the regularized Jacobian arising from the decomposition \eqref{App: Jacobian from metric decomposition} is evaluated as
\al{
J_{\text{grav1},k}=\frac{1}{2}\text{tr}_{(1)}\frac{\p_t \mathcal R_k}{\bar{\mathcal D}_1+\mathcal R_k}\,.
}
The spin-1 ghost contribution are given by the same form,
\al{
\epsilon^{(g)}_k =\text{tr}_{(1)}\frac{\p_t \mathcal R_k}{\Gamma_k^{(2)}+\mathcal R_k}\Bigg|_{\bar C C}
=\text{tr}_{(1)}\frac{\p_t \mathcal R_k}{\bar{\mathcal D}_1+\mathcal R_k}.
}

Thus, we see the relation $\epsilon_k^{(g)}=2\delta_k^{(g)}$, so that the measure contribution takes a simple form,
\al{
\eta_g=-\delta^{(g)}_k=-\frac{1}{2}\text{tr}_{(1)}\frac{\p_t \mathcal R_k}{\bar{\mathcal D}_1+\mathcal R_k}\,.
}
Let us choose the regulator such that the differential operator $\bar{\mathcal D}_1$ is replaced by $k^2$.
Using the heat kernel method with the coefficients \eqref{App: heat kernel coefficients for measure modes in normal} or \eqref{App: heat kernel coefficients for measure modes}, one can evaluate the measure contribution,
\al{
\eta_g &=-\frac{1}{2}\tr_{(1)}\frac{\p_t  P_k\fn{\bar{\mathcal D}_1}}{P_k\fn{\bar{\mathcal D}_1}} 
=: -\frac{1}{2}\tr_{(1)}W_g[\bar\Delta_V]
\nn[1ex]
&=-\frac{1}{2}\frac{1}{16\pi^2} \int_x\sqrt{\bar g} \left[ \frac{13}{4}Q_2[W_g]  + \frac{29}{24} {\bar R} \, Q_1[W_g] + \left(  \frac{23}{144}{\bar R}^2+ \frac{67}{360}{\bar R}_{\mu\nu}{\bar R}^{\mu\nu}  -\frac{11}{180}{\bar R}_{\mu\nu\rho\sigma}{\bar R}^{\mu\nu\rho\sigma} \right) Q_0[W_g]\right]\nn[2ex]
&=-\frac{1}{16\pi^2} \int_x\sqrt{\bar g} \left[ \frac{13}{4}k^4 \ell_0^4\fn{0}  + \frac{29}{24} k^2 \ell_0^2\fn{0}{\bar R} + \left(  \frac{23}{144}{\bar R}^2+ \frac{67}{360}{\bar R}_{\mu\nu}{\bar R}^{\mu\nu}  -\frac{11}{180}{\bar R}_{\mu\nu\rho\sigma}{\bar R}^{\mu\nu\rho\sigma} \right)\ell_0^0\fn{0}  \right]\nn[2ex]
&=-\frac{1}{16\pi^2} \int_x\sqrt{\bar g} \left[ \frac{13}{4}k^4 \ell_0^4\fn{0}  + \frac{29}{24} k^2 \ell_0^2\fn{0}{\bar R} + \left(  \frac{29}{144}{\bar R}^2 - \frac{7}{240}C_{\mu\nu\rho\sigma} C^{\mu\nu\rho\sigma}  -\frac{23}{720}G_4 \right)\ell_0^0\fn{0}  \right]\,.
\label{App: flow contribution from total measure mode}
}

We note a difference between the present result and Ref.~\cite{Wetterich:2019zdo} where we have considered the Einstein-Hilbert truncation in the maximally symmetric spacetime and have derived the beta functions of $U$ and $F$.
In Ref.~\cite{Wetterich:2019zdo} we have decomposed the gauge mode $a_\mu$ and the ghost field $C_\mu$ into the transverse modes and the scalar modes such that $V_\mu=V^\text{T}_\mu +\bar D_\mu V$ where $V^\text{T}_\mu \ni \{\kappa_\mu, C^\text{T}_\mu, \bar C^\text{T}_\mu\}$, $V\ni \{u, C,\bar C \}$, and $\bar D^\mu V^\text{T}_\mu=0$.
For these decompositions, the differential operator \eqref{App: Differential op for measure mode} acts on these fields as
\al{
V^\mu \left(\bar{\mathcal D}_1 \right)_\mu{}^\nu V_\nu=
V^\text{T}{}^\mu \left( \bar \Delta_V-\frac{\bar R}{4} \right) V^\text{T}_\mu + V \left[ 2\bar\Delta_S\left( \bar\Delta_S -  \frac{\bar R}{4} \right) \right] V\,,
}
where we have used $\bar R_{\mu\nu}=(\bar R/4)\bar g_{\mu\nu}$.
The differential operator $\bar\Delta_S$ in the Hessian for the spin-0 field is subtracted by the Jacobians arising from the field decomposition and the overall factor 2 in the spin-0 measure mode does not contribute in the flow generators. In Ref.~\cite{Wetterich:2019zdo} we have individually regularized $\bar{\mathcal D}_\text{1T}=\bar \Delta_V -{\bar R}/{4} $ and $\bar{\mathcal D}_0=\bar \Delta_S -{\bar R}/{4}$ and obtained measure contributions to the flow generators of the spin-1 and spin-0 measure modes, separately
\al{
\eta_1&=-\frac{1}{2}\tr_{(1)}\frac{\p_t  P_k\fn{\bar{\mathcal D}_\text{1T}}}{P_k\fn{\bar{\mathcal D}_\text{1T}}}
=-\frac{1}{16\pi^2} \int_x\sqrt{\bar g} \left[ 3k^4 \ell_0^4\fn{0}  +  k^2 \ell_0^2\fn{0}{\bar R}  \right]\,,
\label{App: spin-1 measure mode in the last work}
\\[2ex]
\eta_0&=-\frac{1}{2}\tr_{(0)}\frac{\p_t  P_k\fn{\bar{\mathcal D}_0}}{P_k\fn{\bar{\mathcal D}_0}}
=-\frac{1}{16\pi^2} \int_x\sqrt{\bar g} \left[ k^4 \ell_0^4\fn{0}  + \frac{5}{12} k^2 \ell_0^2\fn{0}{\bar R}  \right]\,.
\label{App: spin-0 measure mode in the last work}
}
Thus, the total measure contribution was obtained in Ref.~\cite{Wetterich:2019zdo} as
\al{
\eta_g=\eta_1+\eta_0=-\frac{1}{16\pi^2} \int_x\sqrt{\bar g} \left[ 4k^4 \ell_0^4\fn{0}  + \frac{17}{12} k^2 \ell_0^2\fn{0}{\bar R}  \right]\,.
}

This result differs from Eq.~\eqref{App: flow contribution from total measure mode}.
This disagreement arises from an overall factor 2 in the measure contribution of the spin-0 mode: In the present work we regularize the differential operator \eqref{App: Differential op for measure mode}, which corresponds to regularizing $2\bar{\mathcal D}_0$ rather than $\bar{\mathcal D}_0$ in Eq.~\eqref{App: spin-0 measure mode in the last work}.
Hence, the result in this study can be reproduced by replacing $k^2\to k^2/2$ in the spin-0 contribution $\eta_0$ in Eq.~\eqref{App: spin-0 measure mode in the last work}, namely,
\al{
\eta_0&=-\frac{1}{16\pi^2} \int_x\sqrt{\bar g} \left[ \left(\frac{k^2}{2}\right)^2 \ell_0^4\fn{0}  + \frac{5}{12} \left(\frac{k^2}{2}\right) \ell_0^2\fn{0}{\bar R}  \right]\,.
\label{App: spin-1 measure mode in the present work}
}
Then, the contribution from the total measure mode (sum of Eq.~\eqref{App: spin-1 measure mode in the last work} and Eq.~\eqref{App: spin-1 measure mode in the present work}) is modified to be
\al{
\eta_g=-\frac{1}{16\pi^2} \int_x\sqrt{\bar g} \left[  \frac{13}{4}k^4 \ell_0^4\fn{0}  + \frac{29}{24} k^2 \ell_0^2\fn{0}{\bar R} \right]\,.
}
This result agrees now with Eq.~\eqref{App: flow contribution from total measure mode}.
We conclude that the difference in results arises from different regularization procedures. The regularization of the operator \eqref{App: Differential op for measure mode} seems more universal and will be adopted here.

\bibliographystyle{JHEP} 
\bibliography{refs.bib}
\end{document}